\documentclass[twoside,12pt]{article}
\usepackage{epsfig}
\usepackage{subfigure,graphicx}
\usepackage{wrapfig,rotating}
\usepackage{amssymb}
\usepackage{amsmath}

\newcommand {\pom} {I\hspace{-0.2em}P}

\newcommand{\PO}{\rm l \! P }

\newcommand{\xpom}{x_{\PO} }

\newcommand{\be}{\begin{equation}}
\newcommand{\ee}{\end{equation}}

\newcommand {\apom} {\alpha_{\pomsub}}

\newcommand{\ftwod}{F_2^D}

\newcommand {\aprime} {\alpha^\prime_\pomsub}

\newcommand {\pomsub} {{\scriptscriptstyle \pom}}
\newcommand{\ftwopom}{F_2^{\pomsub}}

\def\rmDVCS{\mathrm{DVCS}}
\def\rmI{\mathrm{I}}

\def\AUTDVCS{A_{\mathrm{UT},\rmDVCS}}
\def\AUTI{A_{\mathrm{UT},\rmI}}

\newcommand\epsfigure[4][width=\hsize]{%
\begin{figure}[htbp]%
  \begin{center}%
     \IfFileExists{#2.eps.bb}%
       {\includegraphics[draft,#1]{#2}}%
       {\includegraphics[#1]{#2}}%
  \end{center}%
\caption{#3}\label{#4}%
\end{figure}%
}

\newcommand{\ffig}[4]{\begin{figure}[htbp]\vfill\begin{center}
\mbox{\epsfig{figure=#1,height=#2}}\caption{#3}\label{#4}
\end{center}\vfill\end{figure}}

\begin{document}
\title{\vspace{1cm} 
Advances in diffraction of subnuclear waves 
}
\author{Laurent SCHOEFFEL \\
CEA Saclay/Irfu-SPP, 91191 Gif-sur-Yvette, France
}
\maketitle
\begin{abstract}
In this review, we present and discuss the most recent results on
inclusive and exclusive diffractive processes at  
HERA and Tevatron colliders.
Measurements from fixed target experiments at HERMES and Jefferson
laboratory  are also reviewed. The complementarity of all
these results is analyzed in the context of perturbative QCD and
new challenging issues in nucleon tomography are studied.
A first understanding of how partons are localized in the nucleon
to build orbital momenta can be addressed with these
experimental results.
Some prospects are  outlined for new measurements in fixed target kinematic,
 at Jefferson
laboratory  and CERN, at COMPASS, or at the LHC.
Of special interest is the exclusive (coherent)
production of Higgs boson and heavy objects  at the LHC.
Based on the present knowledge, some perspectives are presented in this direction.

\end{abstract}

\begin{center}
\vspace{2cm}
To be published in {\it Progress in Particle and Nuclear Physics}
\end{center}
\begin{flushright}
\vspace{-17cm}
\end{flushright}

\newpage
\tableofcontents
\newpage



%
%
\section{  Introduction }

Understanding the fundamental structure of matter requires an understanding of how
quarks and gluons are assembled to form hadrons.
Of course, only when partons are the relevant degrees of freedom of the processes,
which we design in the following as perturbative processes.
The arrangement of quarks and gluons
inside nucleons can be probed by accelerating electrons, hadrons or nuclei to precisely 
controlled energies, smashing them into a target nucleus
and examining the final products.
Two kinds of reactions can be considered. 

The first one consists in
low momentum transfer processes with particles  that are hardly affected
in direction or energy by the scattering process. They provide a
low resolution image of the structure, which
allows to map the static, overall properties
of the proton (or neutron), such as  shapes, sizes, and response to
externally applied forces. This is the domain of form factors.
They depend on the three-momentum transfer to
the system. The Fourier transformation of  form factors
provides a direct information on the spatial distribution of charges
in the nucleon. 

A second type of reaction is designed to measure the population of the constituents as a
function of momentum, momentum distributions, through
deep inelastic scattering (DIS). 
It comes from higher energy processes
with particles
that have scored a near-direct hit on
a parton inside the nucleon, providing a higher resolution probe of the 
nucleon structure.
Such hard scattering events typically arise via electron-quark interactions
or quark-antiquark annihilation processes.
Nucleon can then be pictured as a large and ever-changing number of partons
having appropriate distributions of momentum and spin.

Many  experiments in the world located at
DESY (Hamburg), Jefferson Lab or JLab (Virginia), Brookhaven (New York),
Fermilab (Batavia) and CERN (Geneva) can measure these processes.
Both approaches described above 
are complementary, but bear some  drawbacks. The
form factor measurements do not yield any information about the
underlying dynamics of the system such as the momenta of the
constituents, whereas the momentum distributions do not give any
information on the spatial location of the constituents. In fact, more
complete information about the microscopic structure lies 
in the correlation between momenta
and transverse degrees of freedom.
New results in this direction are presented in this review
and the complementarity
of these  measurements, from all experiments listed above,  is discussed.

%
%
\section{  Basics of diffraction at HERA and Tevatron}

HERA was a collider where electrons 
or positrons of 27.6 GeV collided with protons of 920 GeV,
corresponding to a center of mass energy of about 300 GeV.
One of the most important experimental results from the DESY 
collider HERA
is the observation of a significant fraction, around $10\%$, of 
large rapidity gap events in deep inelastic scattering (DIS)
\cite{f2dearly,f2dall,marta,fldprel}. In these events,
the target proton emerges in the final state with a loss of a very small
fraction ($\xpom$) of its energy-momentum. 

\begin{figure}[hpt]
\begin{center}
\includegraphics[width=0.7\textwidth]{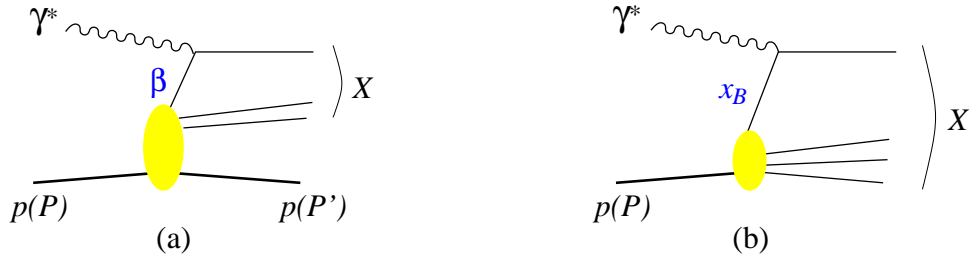}
\caption{Parton model diagrams for deep inelastic diffractive (a) and
  inclusive (b) scattering observed at lepton-proton collider HERA. The variable $\beta$ is the momentum
  fraction of the struck quark with respect to $P-P'$, and the Bjorken variable $x_{Bj}$ its
  momentum fraction with respect to $P$.}
\label{fig1}
\end{center}
\vspace{-0.5cm}
\end{figure}

In Fig. \ref{fig1}(a), we present this event topology,
$\gamma^* p \rightarrow X \ p'$, where the virtual photon $\gamma^*$ 
probes the proton structure and originates from the
electron. Then, the final hadronic state $X$ and the scattered proton are well separated in
space (or rapidity) and a gap in rapidity can be observed in 
the event with no particle produced 
between $X$ and the scattered proton. 
In the standard QCD description
of DIS,  such events are not expected in such an abundance since large
gaps are exponentially suppressed due to color strings formed between
the proton remnant and scattered  partons (see Fig. \ref{fig1}(b)). 
The theoretical description of such processes, also called diffractive processes, is 
 challenging since it must combine perturbative QCD effects of hard scattering with
non perturbative phenomena of rapidity gap  formation. 
The name diffraction in high-energy particle physics originates from the
analogy between optics and nuclear high-energy
scattering. In the Born approximation the equation for hadron-hadron elastic
scattering amplitude can be derived from the scattering of a plane wave
passing through and around an absorbing disk, resulting in an optic-like diffraction
pattern for hadron scattering. 
The quantum numbers of the initial beam particles are conserved during the reaction
and then 
the diffractive system is  well separated in rapidity from the scattered hadron.

\begin{figure}[hpt]
\vspace{-0.5cm}
\begin{center}
\includegraphics[width=1.\textwidth]{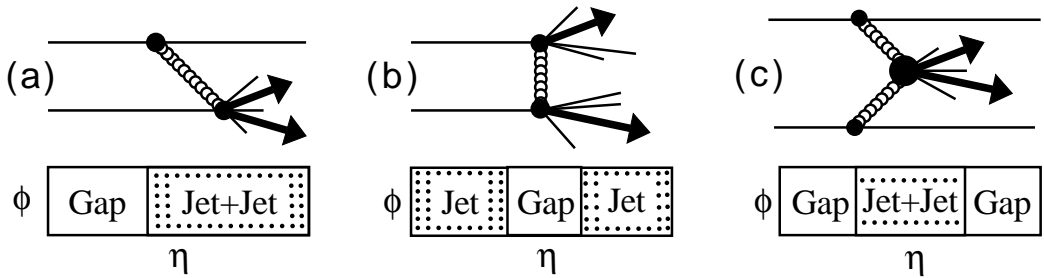}
\vspace*{-1cm}
\caption{ Schematic diagrams of topologies  representative of
hard diffractive processes studied by the proton-antiproton collider Tevatron.}
\label{fig2} 
\end{center}
\vspace{-0.5cm}
\end{figure}

The early discovery of large rapidity gap events at 
HERA \cite{f2dearly} has led to
a renaissance of the physics of diffractive scattering in an
entirely new domain, in which the large momentum transfer 
 provides a hard scale. 
This observation has then revived the 
rapidity gap physics with hard triggers, as large-$p_{\perp}$ jets, at the
proton-antiproton collider Tevatron (see Fig. \ref{fig2}). 
The Tevatron is a $p \bar{p}$ collider located close to Chicago at Fermilab,
USA. It is presently the collider with the highest center-of-mass energy of
about 2 TeV. Two main experiments are located around the ring, D\O\ and CDF.

\begin{figure}[htbp]
\centerline{\includegraphics[width=0.3\columnwidth]{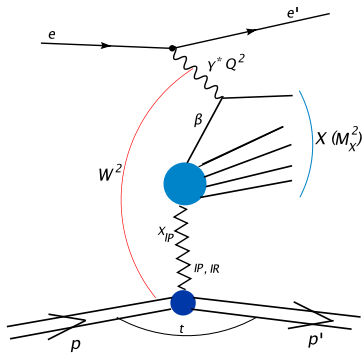}}
\caption{Diffractive kinematics.}\label{kin}
\end{figure}

In the single diffractive dissociation 
process in proton-proton scattering, $pp \rightarrow Xp$, 
at least one of the beam hadrons emerges intact from 
the collision, having lost only a small fraction of its energy 
and gained only a small transverse momentum. 
In the analogous process involving 
virtual photons, $\gamma^{*}p \to Xp$, 
an exchanged photon of virtuality $Q^2$ 
dissociates through its interaction with the proton at a 
squared four momentum transfer $t$ 
to produce a hadronic system $X$ with mass $M_X$.
The fractional longitudinal
momentum loss of the proton during the interaction is
denoted $x_{\pom}$, while the fraction of this momentum 
carried by the struck quark is denoted $\beta$. These variables 
are related to Bjorken $x$ by $x=\beta \, x_{\pom}$ (see Fig. \ref{kin}). 

Using the standard vocable, the vacuum/colorless exchange 
involved in the diffractive interaction
is called Pomeron in this review.
Whether
the existence of such  hard scales makes the diffractive processes
tractable within  perturbative QCD  or not has been a subject of
intense theoretical and experimental research during the past
decade.

%
%
\section{  Observation of diffractive events at HERA}

\subsection{The rapidity gap events}

\begin{figure}[t]
\begin{center}
\vspace{8.cm}
\hspace{-5.5cm}
\epsfig{file=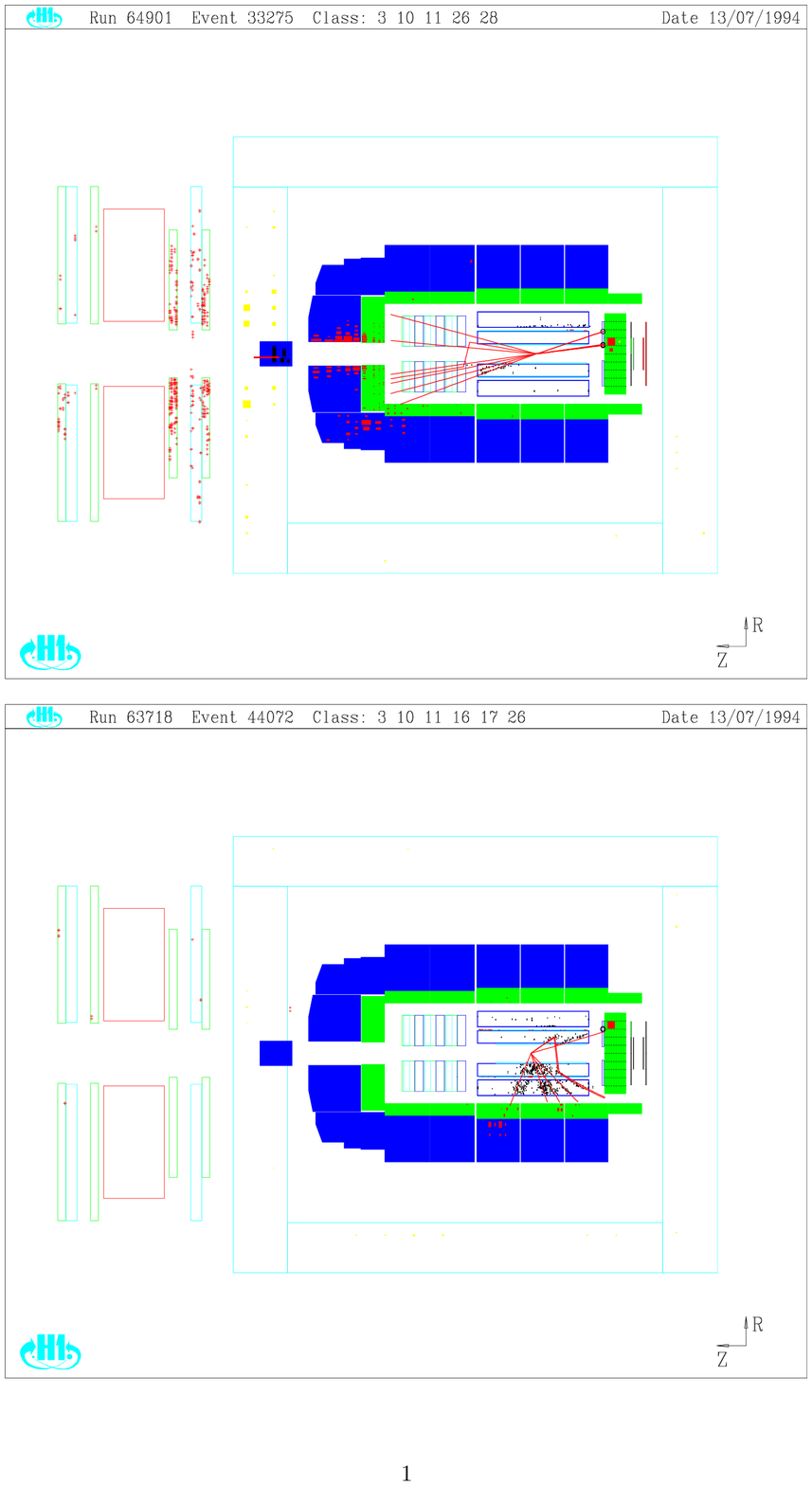,width=4.5cm,height=4.cm}
\vspace{-1cm}
\caption{Usual (top) and diffractive (bottom) events in the H1 experiment at HERA.
For a diffractive event, no hadronic activity is visible in the
 proton fragmentation region, as the proton remains intact
in the diffractive process. On the contrary, for a standard DIS event,
the proton is destroyed in the reaction and the flow of hadronic clusters
is clearly visible in the proton fragmentation region 
(+z direction, i.e. forward part of the detector).}
\label{fig1c}
\end{center}
\end{figure}

Let us start by giving 
a real example of a diffractive event in HERA experiments.
See  Fig.~\ref{fig1c}, which is the (exact) experimental reproduction of Fig.~\ref{fig1}.
A typical DIS event as shown in the upper plot of Fig.~\ref{fig1c} is $ep \rightarrow eX$
where electron and jets are produced in the final state. 
The electron is scattered in the  backward detector\footnote{At
HERA, the backward (resp. forward) directions are defined as the direction
of the outgoing electron (resp. proton).} (right of the figure)
whereas some hadronic activity is present in the forward region of the detector.
The proton is thus
completely destroyed and the interaction leads to jets and proton remnants directly observable
in the detector. 

The fact that much energy is observed in the forward region is
due to color exchange between the scattered jet and the proton remnants.
However, for events that we have called diffractive, the situation is completely
different. Such events appear like the one shown in the bottom  of Fig.~\ref{fig1c}.
The electron is still present in the backward detector, there is
still some hadronic activity (jets) in the LAr calorimeter, but no energy above
noise level is deposited in the forward part of the detectors. 
In other words, there is no color exchange between the
proton and the produced jets. The reaction can then be written as $ep \rightarrow epX$.
This is also called a Large Rapidity Gap  (LRG) event, and constitutes an
efficient experimental method to tag diffractive events.

\subsection{Proton tagging}

A second experimental technique to detect diffractive events is to tag the outgoing proton. 
The
idea is then to detect directly the intact proton in the final state. The proton
loses a small fraction of its energy and is thus scattered at very small angle
with respect to the beam direction. Some special detectors called roman pots can
be used to detect the protons close to the beam. 

The basic idea is simple. The roman pot
detectors are located far away from the interaction point and can move close to
the beam, when the beam is stable, to detect protons scattered at vary small
angles. 

\begin{figure}[t]
\begin{center}
\epsfig{file=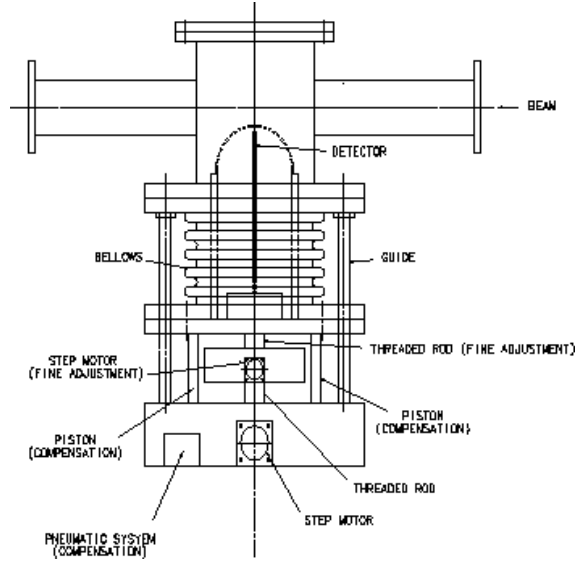,width=7.5cm}
\caption{Scheme of a roman pot detector.
The roman pot
detectors are located far away from the interaction point and can move close to
the beam, when the beam is stable, to detect protons scattered at vary small
angles. }
\end{center}
\label{fig2romanpot}
\end{figure}

The inconvenience is that the kinematical reach of those detectors is
much smaller than with the rapidity gap method. On the other hand,
the advantage is that it
gives a clear signal of diffraction since it measures the diffracted proton
directly.
A scheme of a roman pot detector as it is used by the H1 or ZEUS experiment is shown
in Fig. \ref{fig2romanpot}. The beam is the horizontal line at the upper part of the
figure. The detector is located in the pot itself and can move closer to the
beam when the beam is stable enough (during the injection period, the detectors
are protected in the home position). 

\subsection{The $M_X$ method}

\begin{figure}[t]
\begin{center}
\hspace{11cm}
\epsfig{file=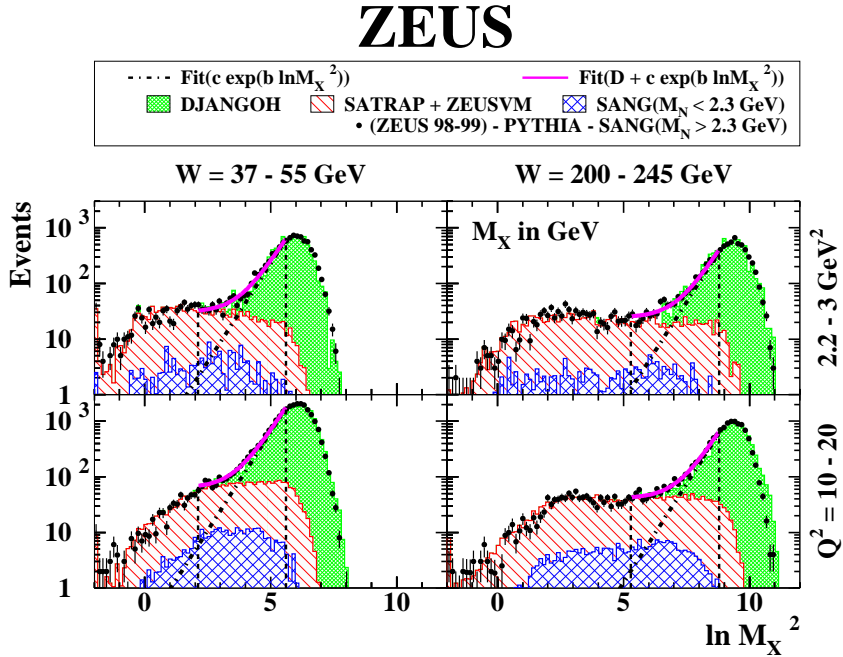,width=10cm, angle=270}
\caption{Illustration of the $M_X$ method used by the ZEUS collaboration to
define diffractive events.
The diffractive contribution corresponds to $d \sigma_{diff}/dM_X^2  \simeq constont$,
while the non diffractive part (at large $M_X$) leads to a shape in
$
\frac{d \sigma}{dM_X^2} \simeq \exp(b \log M_X^2)
$.}
\end{center}
\label{fig3mxmethod}
\end{figure}

The third method used at HERA mainly by the ZEUS experiment 
is based on the fact that there is a different behavior in $\log
M_X^2$, where $M_X$ is the total invariant mass produced in the event, either 
for diffractive or
non diffractive events. For diffractive events $d \sigma_{diff}/dM_X^2 =  
(s/M_X^2)  ^{\alpha -1} = const. ~~$if$ ~~ \alpha \sim 1$ (which is the case for
diffractive events). The ZEUS collaboration performs some fits of the 
$d\sigma/dM_X^2$ distribution:
\begin{eqnarray}
\frac{d \sigma}{dM_X^2} = D + c \exp(b \log M_X^2)
\end{eqnarray}
as illustrated in Fig. \ref{fig3mxmethod} The usual non diffractive events are
exponentially suppressed at high values of $M_X$. The difference between the
observed $d\sigma/dM_X^2$ data and the exponential suppressed distribution is
the diffractive contribution. 

\vfill
\clearpage

%
%
\section{  Measurement of the inclusive diffractive cross section HERA}

\subsection{  Inclusive diffraction as a leading twist process}

From  observation of diffractive events, using the different 
techniques exposed above,
the inclusive diffractive cross section has been measured at HERA by H1 and
ZEUS experiments over a wide kinematic range
\cite{f2dearly,f2dall,marta,fldprel}.
Similarly to inclusive DIS, 
cross section measurements for the reaction $ep \to eXp$ 
are conventionally expressed in terms of the
reduced diffractive cross section, $\sigma_r^{D(3)}$, which is related
to the measured cross section by
\begin{eqnarray}
\frac{{\rm d} \sigma^{ep \rightarrow eXp}}{
{\rm d} \beta {\rm d} Q^2 {\rm d} x_{\pom}} = 
\frac{4\pi\alpha^2}{\beta Q^4} \ \ \left[1-y+\frac{y^2}{2}\right] \ \ 
\sigma_r^{D(3)}(\beta,Q^2,x_{\pom}) \ .
\label{sigma-2}
\end{eqnarray}
At moderate inelasticities $y$, $\sigma_r^{D(3)}$
corresponds
to the diffractive structure function $F_2^{D(3)}$ to good approximation.

\begin{figure}[htbp]
\begin{center}
\includegraphics[width=10cm,height=8.5cm]{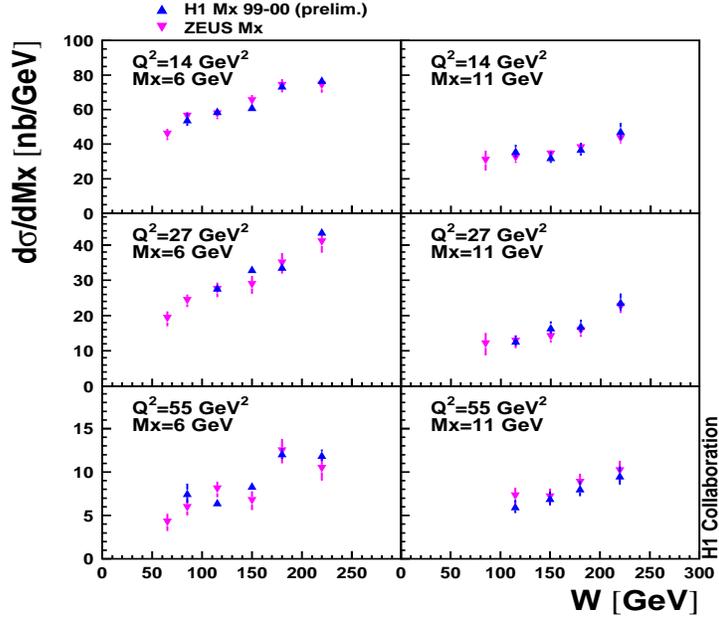}
\caption{The  cross section of the diffractive process $\gamma^* p \rightarrow p' X$, 
differential in the mass of the diffractively produced hadronic system $X$ ($M_X$),
is presented as a function of the center-of-mass energy of the $\gamma^*p$ system $W$.
Measurements at different values of the virtuality
$Q^2$ of the exchanged photon are displayed.
}
\label{figdata}
\end{center}
\vspace{-0.5cm}
\end{figure}

Fig. \ref{figdata} illustrates a first result for the diffractive cross section
as a function of $W$ for different $Q^2$ and $M_X$ values. 
We notice  that the diffractive cross section,
$ep \rightarrow epX$,
shows a hard dependence in the center-of-mass energy of the $\gamma^*p$ system $W$.
Namely, we measure a $W$ dependence of the form $\sim W^{ 0.6}$  for the diffractive cross section, 
compatible with the dependence expected
for a hard process. 

This  first observation is fundamental and allows further studies 
of the diffractive process in the context of
perturbative QCD (see next sections). 
The experimental selection  of
diffractive events is already a challenge but the discovery that 
these events build  a hard scattering process is a surprise and makes
the strong impact of HERA data into the field.
Indeed, the extent to which diffraction, even in the presence
of a hard scale, is a hard process, 
was rather unclear before HERA data. 
This has changed since then, with the 
arrival of accurate HERA data on diffraction in $ep$ scattering 
 and the realization that  
diffraction (measured to be a hard process)
in DIS can be described in close analogy with inclusive 
DIS \cite{f2dall,marta,fldprel}.

\begin{figure}[htbp]  
\begin{center}
     \epsfig{figure=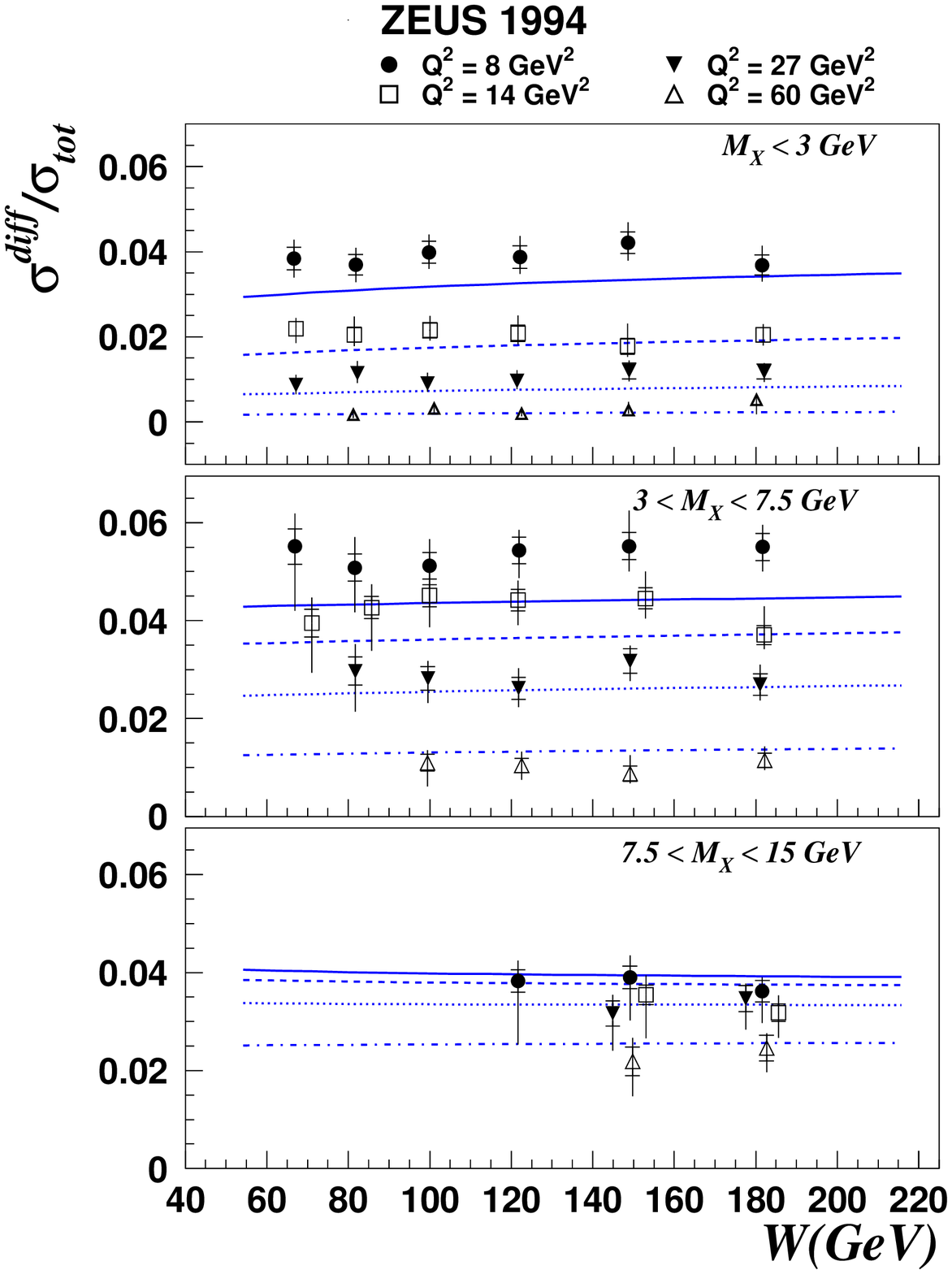,width=8cm}  
     \epsfig{figure=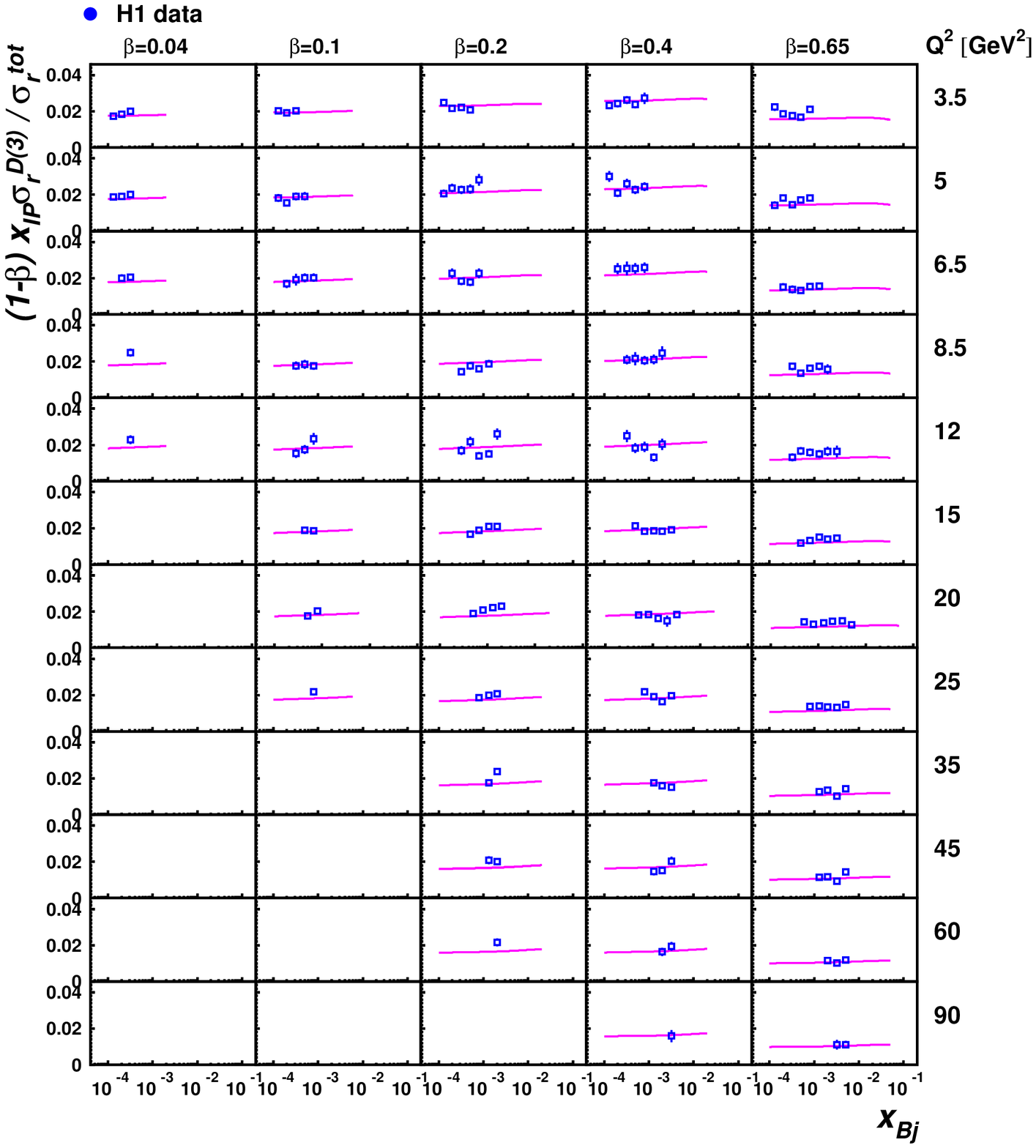,width=8cm}  
     \caption{  Left: Ratio of the diffractive versus the inclusive cross sections
as a function of $W$ for different values of  $Q^2$ and the diffractive mass $M_X$,
derived from early ZEUS  data.
Right: Ratio of the diffractive versus total cross sections, as a function of $x_{Bj}$,
derived from H1  data  
for different values of  $Q^2$ and $\beta$. A constant ratio of about
$0.02$ (2\%) is observed for each bin of measurements. If we add up the
five bins in $\beta$ (for the bulk of the $Q^2$  domain), 
we find immediately the
average number of 10\%. It gives the fraction of diffractive events
on the total DIS sample (see text).}  
\label{fig8}  
\end{center}
\end{figure}

This is also confirmed in Fig. \ref{fig8}, where the
ratio of diffractive to DIS cross sections is shown.
This ratio is found to depend weakly on the  Bjorken variable $x_{Bj}$ (or $W$)
at fixed values of the photon virtuality $Q^2$. Thus, 
we can conclude that
 diffraction in DIS is a leading twist effect with
logarithmic scaling violation in $Q^2$,
as for standard DIS. We discuss these results much further in the next sections.

\subsection{  Recent results on inclusive diffraction at HERA}

Extensive measurements of diffractive DIS cross sections have been made by both
the ZEUS and H1 collaborations at HERA,
using different experimental techniques \cite{f2dall,marta,fldprel}. 
Of course, the comparison of these techniques provides a rich
source of information to get a better understanding of their respective experimental gains
and prejudices.

\begin{figure}[tbp]
\begin{center}
\psfig{figure=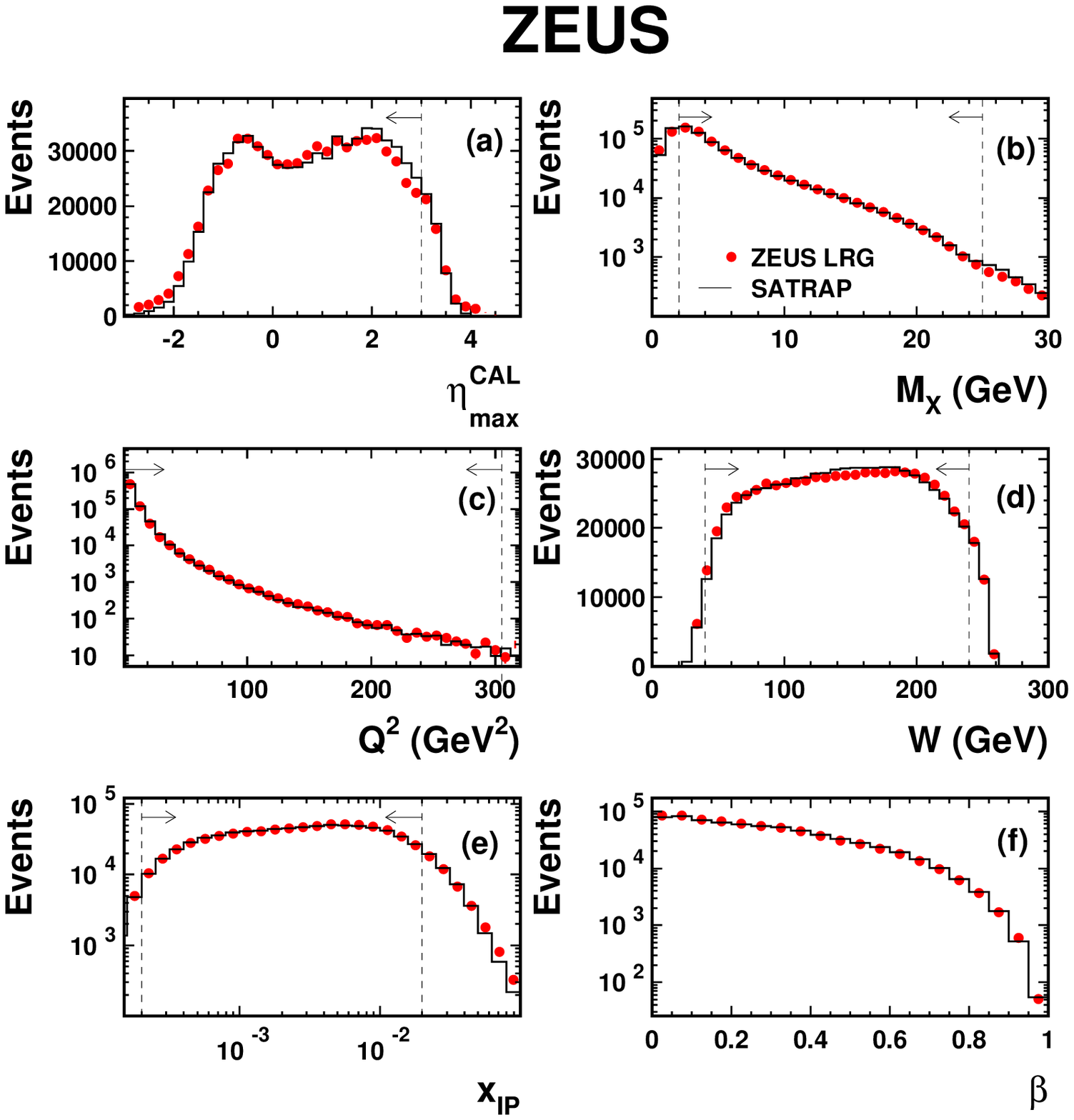,width=0.4\textwidth,angle=0}
\psfig{figure=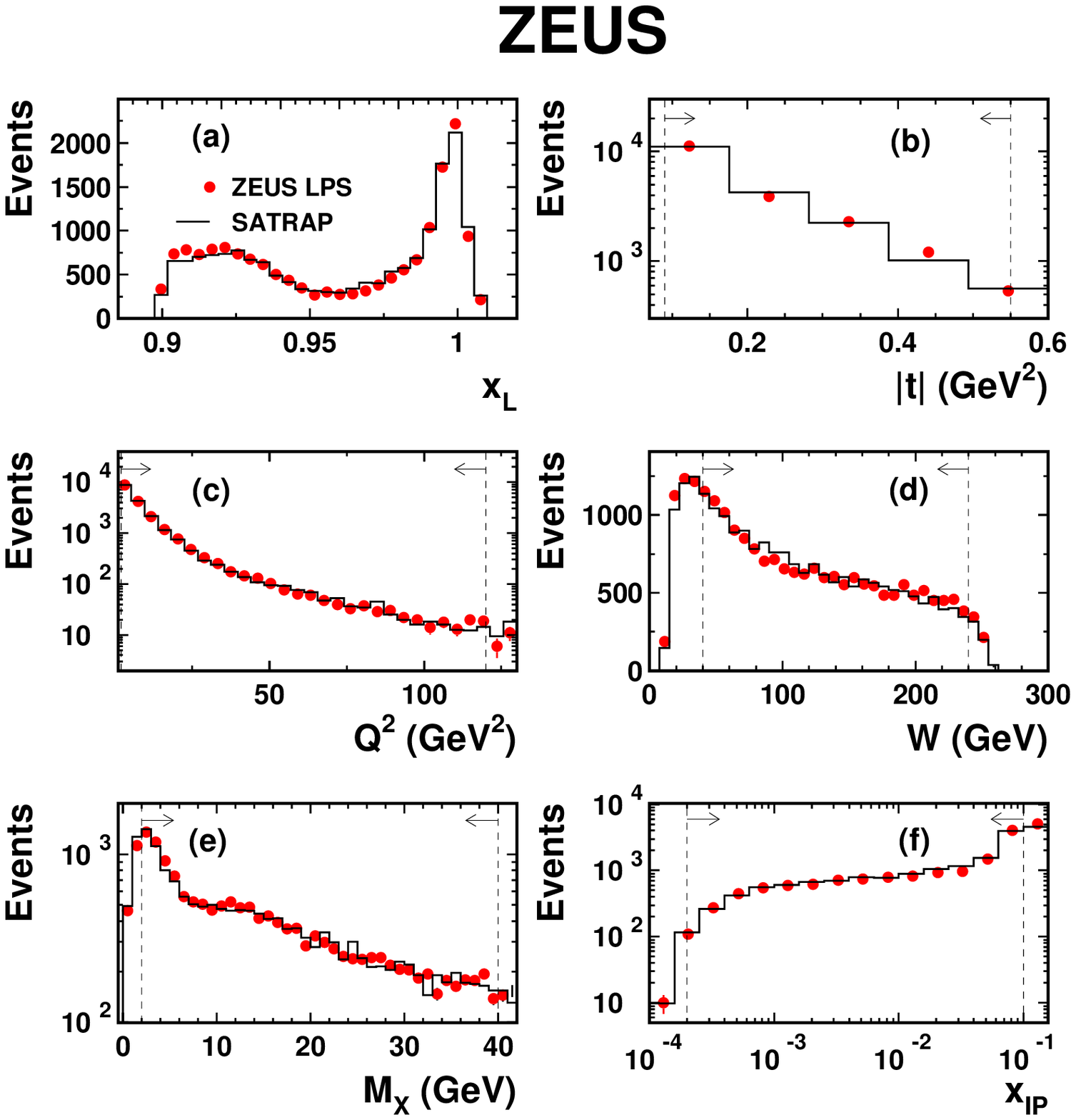,width=0.4\textwidth,angle=0}
\end{center}
\caption{Left: Comparison of the distributions of data (dots) to those obtained with
the Monte-Carlo (histograms) for typical variables in the LRG analysis.
Right: Comparison of the distributions of data (dots) to those obtained with
the Monte-Carlo (histograms) for typical variables in the LPS analysis.}
\label{datalrg}
\end{figure}

In Fig. \ref{datalrg}, the basis of the last ZEUS experimental analysis 
is summarized \cite{marta}. Data are compared to Monte-Carlo (MC) expectations for typical
variables. The MC is based on specific models for signal and backgrounds, 
and the good agreement with data is proof that the main ingredients of the
experimental analysis are under control.
These last sets of data (Fig. \ref{datalrg}) \cite{marta} contain
five to seven times more statistics than in preceding publications of diffractive
cross sections, and thus opens the way to new developments in data/models comparisons.

\begin{figure}[tbp]
\begin{center}
\psfig{figure=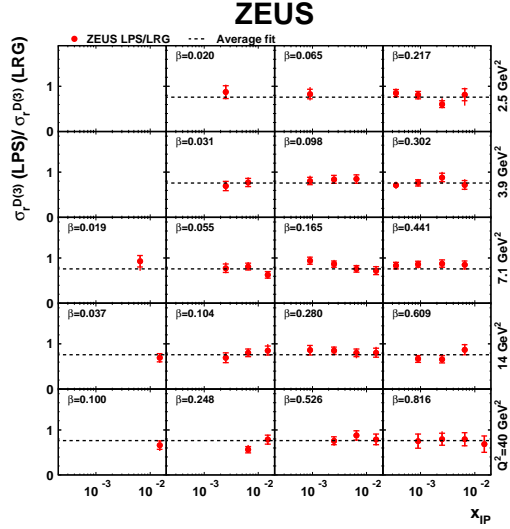,width=0.4\textwidth,angle=0}
\end{center}
\caption{Ratio of the diffractive cross sections, as obtained with the LPS and
the LRG experimental techniques. The lines indicate the average value of the ratio,
which is about 0.86. It implies that the LRG sample contains about
24\% of proton dissociation  events, corresponding to processes like $ep \rightarrow eXY$,
where $M_Y<2.3$ GeV. This fraction is approximately the same for H1 data (of course in the same $M_Y$ range).}
\label{lpsoverlrg}
\end{figure}

A first relative control of the  data samples is shown in Fig. \ref{lpsoverlrg}, where the
ratio of the diffractive cross sections is displayed,
as obtained with the LPS and
the LRG experimental techniques. The mean value of the ratio of $0.86$ indicates that
the LRG sample contains about 24\% of proton-dissociation background, which is not
present in the LPS sample. This background corresponds to events like
$ep \rightarrow e X Y$, where $Y$ is a low-mass excited state of the proton (with
$M_Y < 2.3$ GeV). 
It is obviously not present in the LPS analysis which  can select specifically a proton
in the final state.   This is the main background in the LRG analysis. Due to a lack
of knowledge of this background, it causes a large normalization uncertainty of 10  to 15 \% for the
cross sections extracted from the LRG analysis.

\begin{figure}[tbp]
\begin{center}
\psfig{figure=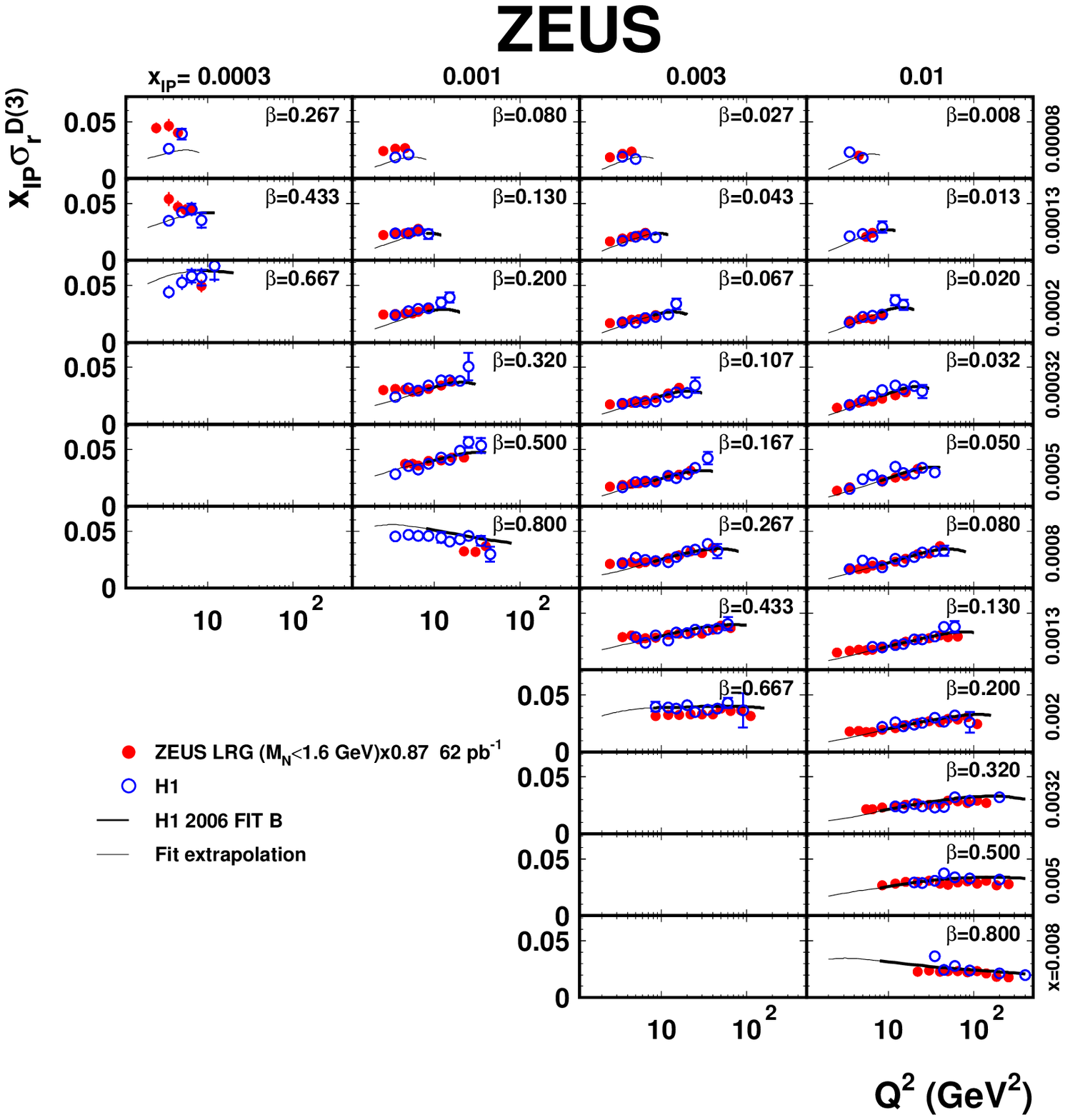,width=0.4\textwidth,angle=0}
\psfig{figure=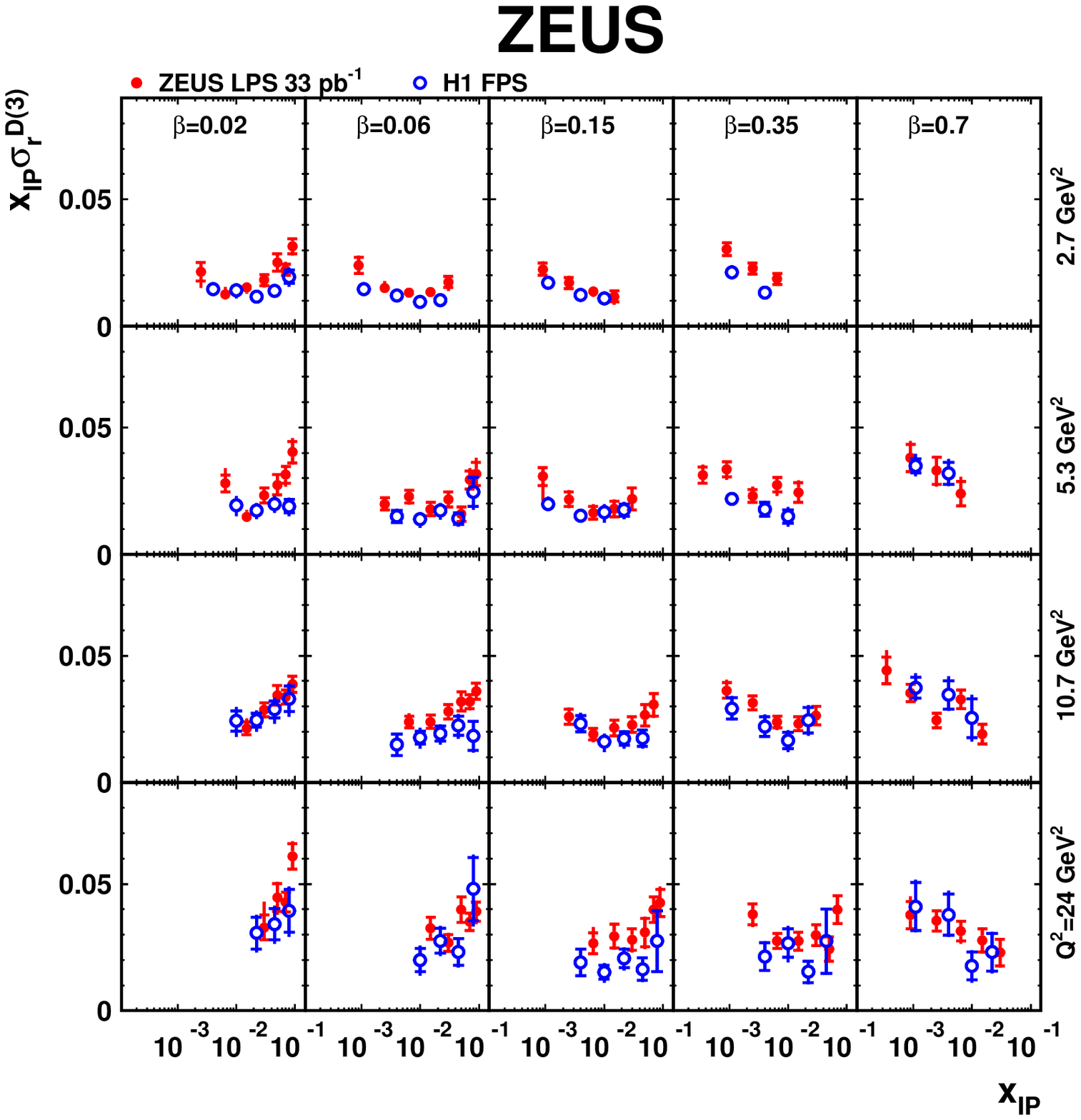,width=0.4\textwidth,angle=0}
\end{center}
\caption{Left: The diffractive cross sections obtained with the LRG method by the H1 and ZEUS experiments.
The ZEUS values have been rescaled (down) by a global factor of 13 \%. This value is compatible with the normalization uncertainty
of this sample. Right: The diffractive cross section obtained with the FPS (or LPS) method by the H1 and ZEUS experiments,
where the proton is tagged. The ZEUS measurements are above H1 by a global factor of about 10\%.}
\label{datah1zeus}
\end{figure}

We can then compare the results obtained by the H1 and ZEUS experiments for diffractive
cross sections (in Fig. \ref{datah1zeus}), using the LRG method.
A good compatibility of both data sets is observed, after rescaling the ZEUS points by 
a global factor of 13\%. This factor is compatible with the normalization uncertainty described above.

We can also compare the results obtained by the H1 and ZEUS experiments (in Fig. \ref{datah1zeus}),
 using the tagged proton  method (LPS for ZEUS and FPS for H1).
In this case, there is no proton dissociation background and the diffractive sample
is expected to be clean. It gives a good reference to compare both experiments. A global
normalization difference of about 10\% can be observed in Fig. \ref{datah1zeus},
which can be studied with more data. It remains compatible with the normalization
uncertainty for this tagged proton  sample.
It is interesting to note that the ZEUS measurements are globally above the H1 data by 
 about 10\% for both techniques, tagged proton or LRG.

\begin{figure}[tbp]
\begin{center}
\psfig{figure=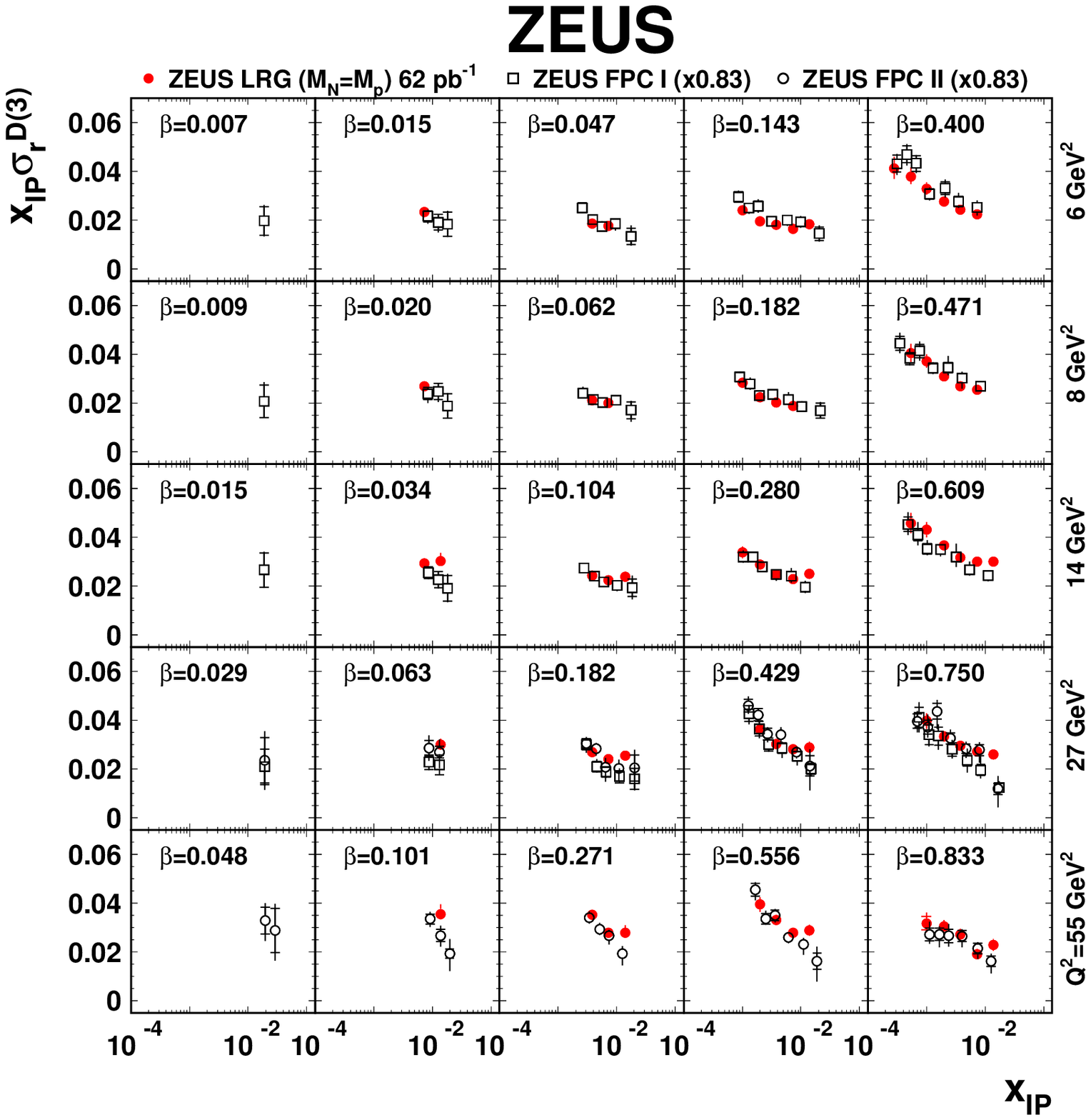,width=0.5\textwidth,angle=0}
\end{center}
\caption{The diffractive cross sections obtained with the LRG method (full dots) compared with the
results obtained with the $M_X$ method (open symbols: FPC I and FPC II). All values are converted to
$M_Y=M_p$.}
\label{datazeuslrgmx}
\end{figure}

In Fig. \ref{datazeuslrgmx}, we compare the results using the LRG and the $M_X$ methods,
for ZEUS data alone.
Both sets are in good agreement, which shows that there is no strong bias between these
experimental techniques.
The important message at this level is not only the observation of differences
as illustrated in Fig.  \ref{datah1zeus},
but the opportunity opened with the large statistics provided by the ZEUS measurements. 
Understanding discrepancies
between data sets is part of the experimental challenge of the next months. It certainly needs
 analysis of new data sets from the HERA experiments. However, already at the present
level, much can be done with existing data for the understanding of  diffraction at HERA.


\subsection{  Summary of recent results in one plot}

\begin{figure}[p] \unitlength 1mm
  \begin{center}
    \begin{picture}(100,200)
      \put(55,-5){\epsfig{file=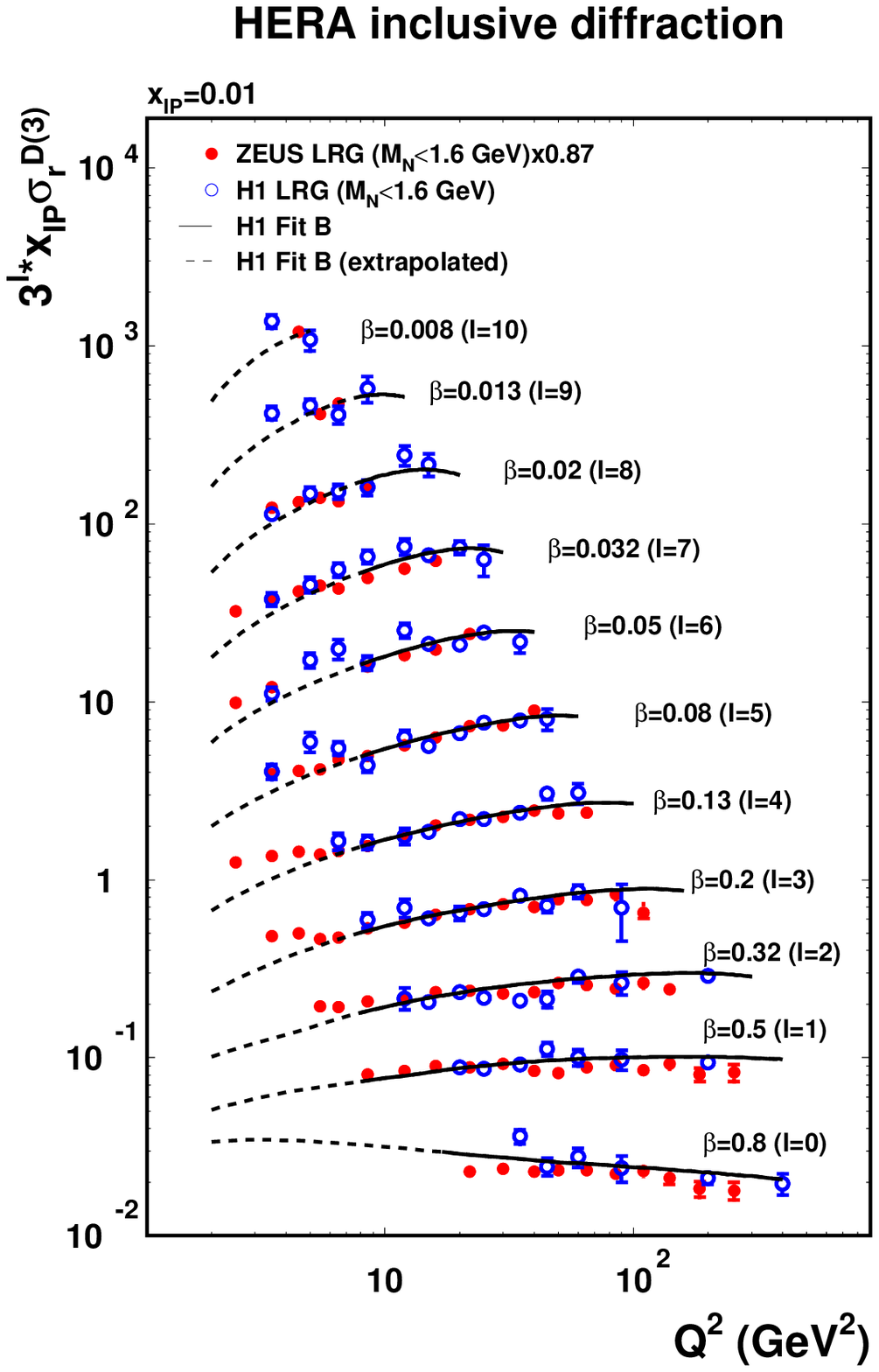,width=0.4\textwidth}}
      \put(55,105){\epsfig{file=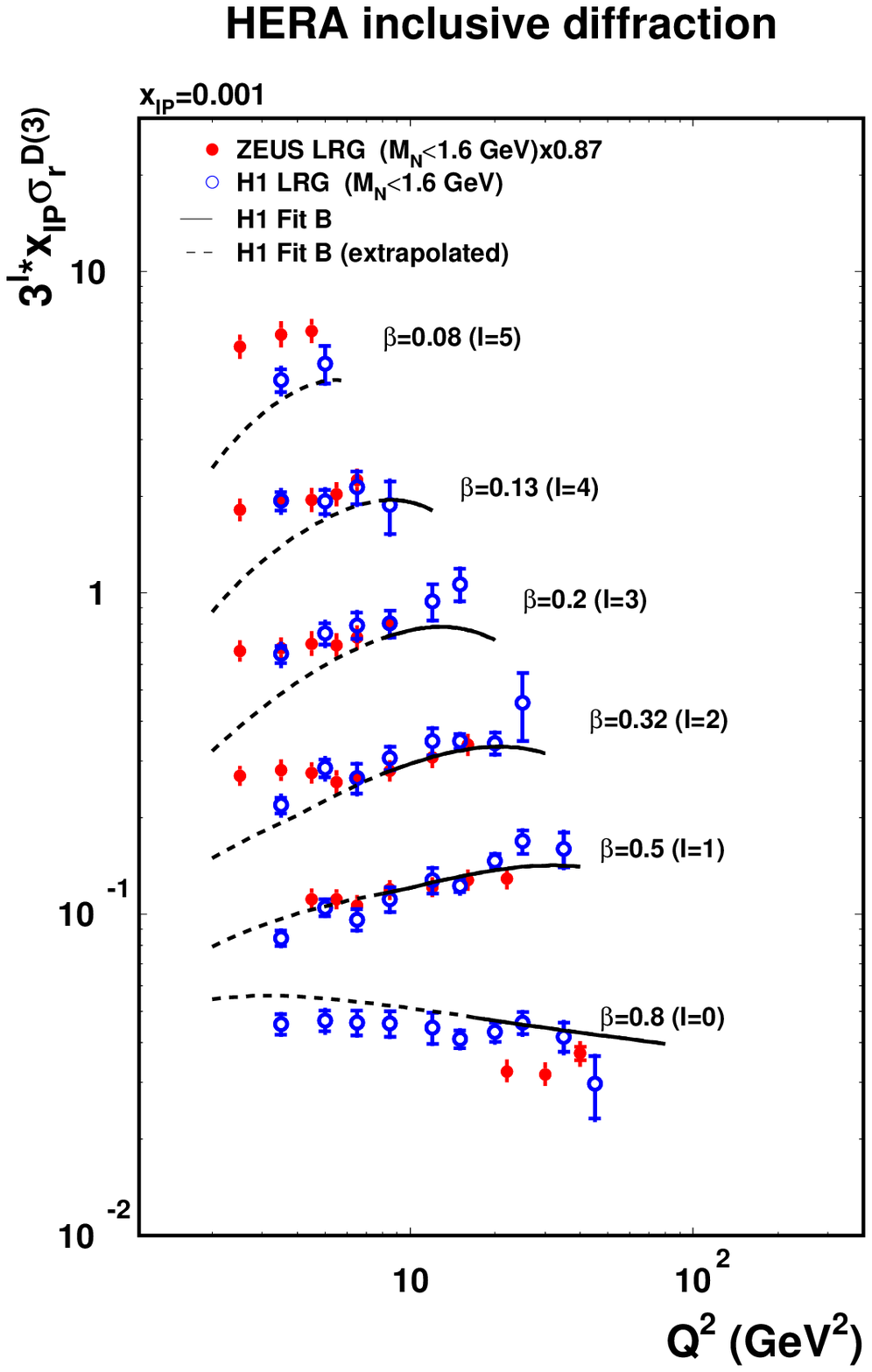,width=0.4\textwidth}}
      \put(-25,-5){\epsfig{file=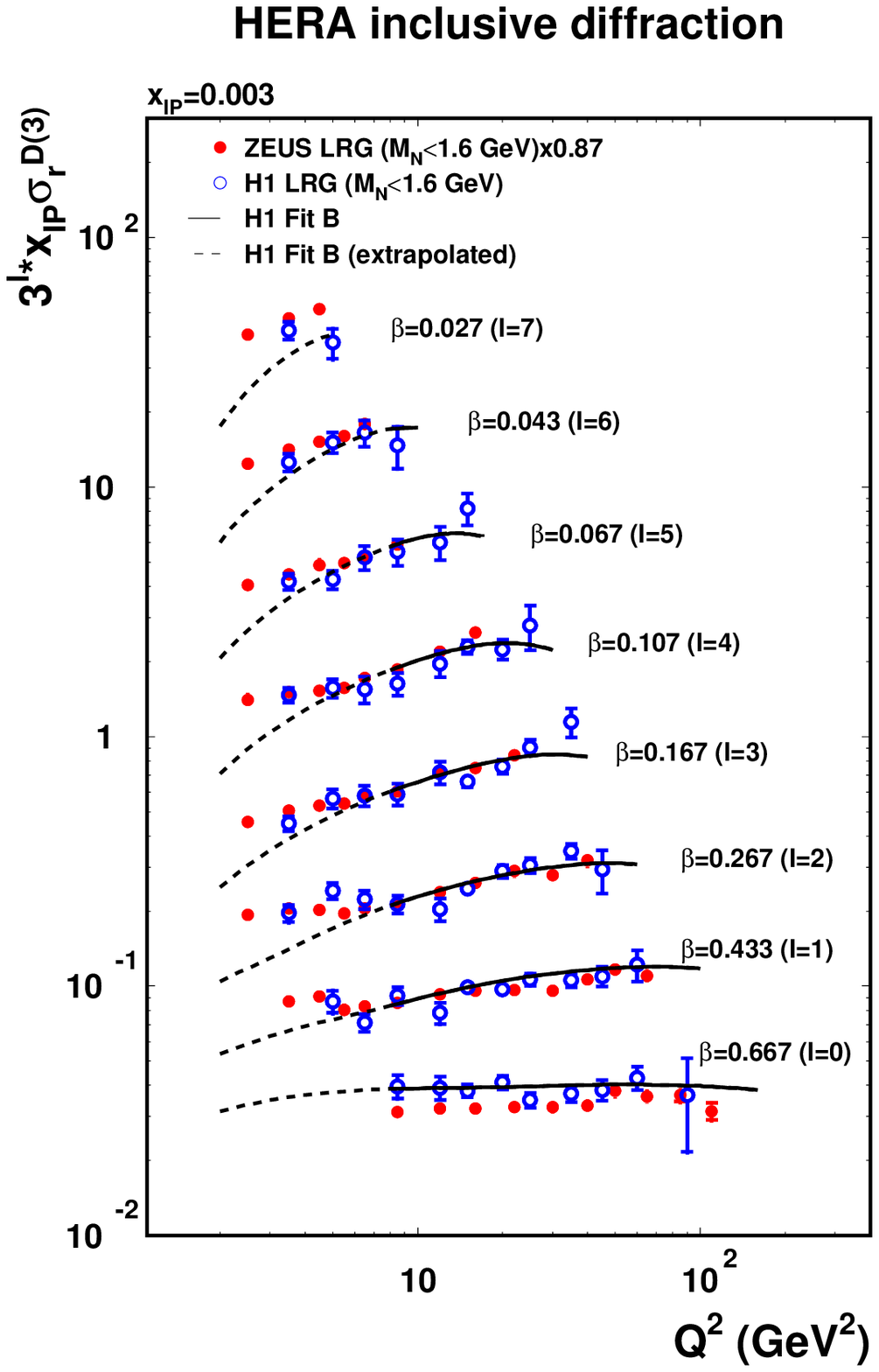,width=0.4\textwidth}}
      \put(-25,105){\epsfig{file=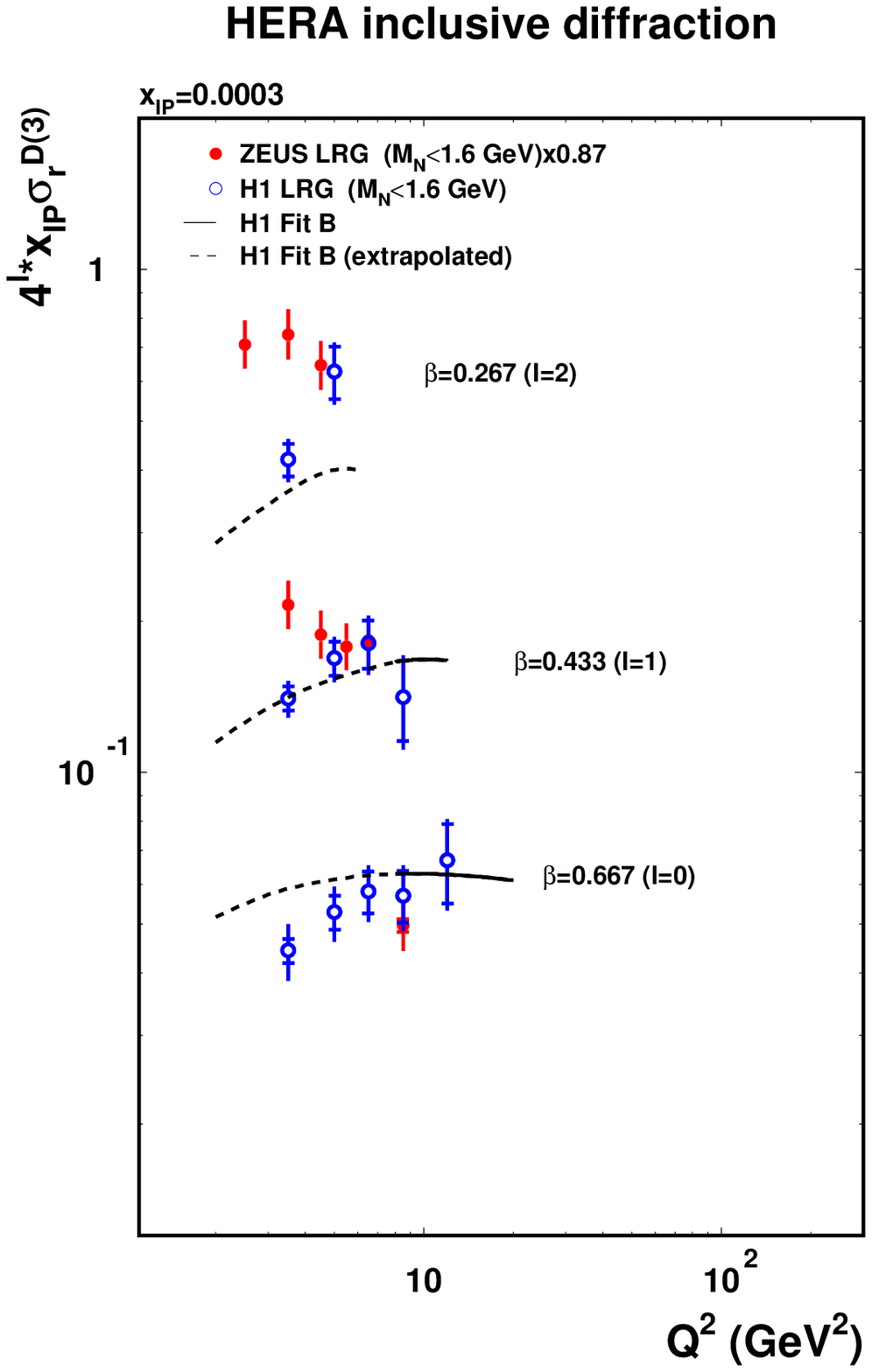,width=0.4\textwidth}}
    \end{picture}
  \end{center}
  \caption[]{Comparison between the H1 and ZEUS LRG measurements 
after correcting both data sets to $M_N  < 1.6 \ {\rm GeV}$ 
and applying a further
scale factor of 0.87 (corresponding to the average normalization difference)
to the ZEUS data. The measurements are compared with the results
of the QCD fit prediction (see text). Further H1 data at $x_{\pom} = 0.03$
are not shown.}
\label{fig:comp_lrg}
\end{figure}

A summary of present measurements using LRG event selection is
shown in Fig. \ref{fig:comp_lrg}.
The ZEUS LRG data are extracted at the H1 $\beta$ and $x_{\pom}$ 
values, but at different $Q^2$ values.
In order to match the $M_N~<~1.6$\,GeV range of the H1 data, 
a global factor of $0.91 \pm 0.07$, estimated with {\sc Pythia},  
is applied to the ZEUS LRG data in place of the correction
to an elastic proton cross section.
After this procedure, the ZEUS data remain higher than those of H1 
by $13\%$ on average, as discussed above. 
The results of the QCD fit  to 
H1 LRG data~\cite{f2dall} is 
also shown (see next section). 

%
%
\section{  Diffraction and the resolved Pomeron model}

\subsection{  Theoretical considerations}

Several theoretical formulations have been proposed to describe the  diffractive exchange.
The purpose is  to describe the {\it blob} 
displayed in Fig. \ref{fig1} 
in a quantitative 
way, leading to a proper description of data shown in Fig. \ref{figdata}.

Among the most popular models, the one based on a point-like structure of
the Pomeron   assumes that the exchanged object, the Pomeron, 
is a color-singlet quasi-particle whose structure is probed in
the reaction \cite{collins,is}. In this approach, 
diffractive parton distribution functions (diffractive PDFs)
are derived from the diffractive DIS cross sections 
in the same way as standard PDFs are extracted from DIS measurements.
It assumes also that a certain flux of Pomeron is emitted off the proton, depending on 
the variable $\xpom$, the fraction of the longitudinal momentum of
the proton lost during the interaction.
The partonic structure of the Pomeron is probed during 
the diffractive exchange \cite{collins,is}.  

In Fig. \ref{kin}, 
we illustrate this factorization property and
remind the notations for the kinematic variables used in this paper, as
the virtuality
$Q^2$ of the exchanged photon,
 the center-of-mass energy of the $\gamma^*p$ system $W$ and
 $M_X$ the mass of the diffractively produced hadronic system $X$.
It follows that the Bjorken variable $x_{Bj}$ verifies $x_{Bj} \simeq Q^2/W^2$ in the 
low $x_{Bj}$ kinematic domain of the H1 and ZEUS measurements ($x_{Bj}<0.01$). 
Also,  the Lorentz invariant variable $\beta$
defined in Fig. \ref{fig1} is equal to $x_{Bj}/\xpom$ and can be interpreted as
the fraction of longitudinal momentum of the struck parton in the (resolved) Pomeron.

Because the short-distance cross section ($\gamma^*-q$) 
of hard diffractive DIS is identical to inclusive DIS, the evolution 
of the diffractive parton distributions follows the same equations
as  ordinary 
parton distributions. Quantitatively,
QCD factorization is expected to hold for
$\ftwod$~\cite{collins,is} and
it may then be decomposed into diffractive parton distributions,
$f^D_i$, in a way similar to the inclusive $F_2$,
\begin{equation}
\frac{d\ftwod (x,Q^2,\xpom,t)}{d\xpom dt}=\sum_i \int_0^{\xpom} dz 
\frac{d f^D_i(z,\mu,\xpom,t)}{d\xpom dt} \hat{F}_{2,i}(\frac{x}{z},Q^2,\mu) \, ,
\label{eq:fact}
\end{equation}
where $\hat{F}_{2,i}$ is the universal structure function for DIS on
parton $i$, $\mu$ is the factorization scale at which $f^D_i$ are
probed and $z$ is the fraction of momentum of the proton carried by
the  parton $i$.  

The QCD evolution equation applies in the
same way as for the inclusive case.
Fig. \ref{fig8} is a simple experimental proof of this statement.  
For a fixed value of $\xpom$, the
evolution in $x$ and $Q^2$ is equivalent to the evolution in $\beta$
and $Q^2$.

If, following Ingelman and Schlein~\cite{is}, one further
assumes the validity of Regge factorization, $\ftwod$ may be decomposed
into a universal Pomeron flux and the structure function of the Pomeron,
\begin{equation}
\frac{d\ftwod (x,Q^2,\xpom,t)}{d\xpom dt}= f_{\pom/p}(\xpom,t)
\ftwopom (\beta,Q^2) \, ,
\label{eq:f2pom}
\end{equation}
where the normalization of either of the two components is arbitrary.
It implies that the $\xpom$ and $t$ dependence of the diffractive
cross section is universal, independent of $Q^2$ and $\beta$, and given
by \cite{f2dall,collins,is}
\begin{equation}
f_{\pom/p}(\xpom,t) \sim \left( \frac{1}{\xpom} \right)^{2\apom(0)-1}
e^{(b_0^{D}-2\aprime \ln \xpom)t} \, .
\label{eq:pomflux}
\end{equation}

In this approach, the mechanism for producing LRG is assumed to be
present at some scale and the evolution formalism allows to probe the
underlying partonic structure. The latter depends on the coupling of
quarks and gluons to the Pomeron.
It follows that the characteristics of diffraction 
are entirely contained in the input distributions at a given scale. It 
is therefore interesting to model these distributions. 

\subsection{  Diffractive parton densities}

\begin{figure}[htbp]
\begin{center}
\includegraphics[totalheight=9cm]{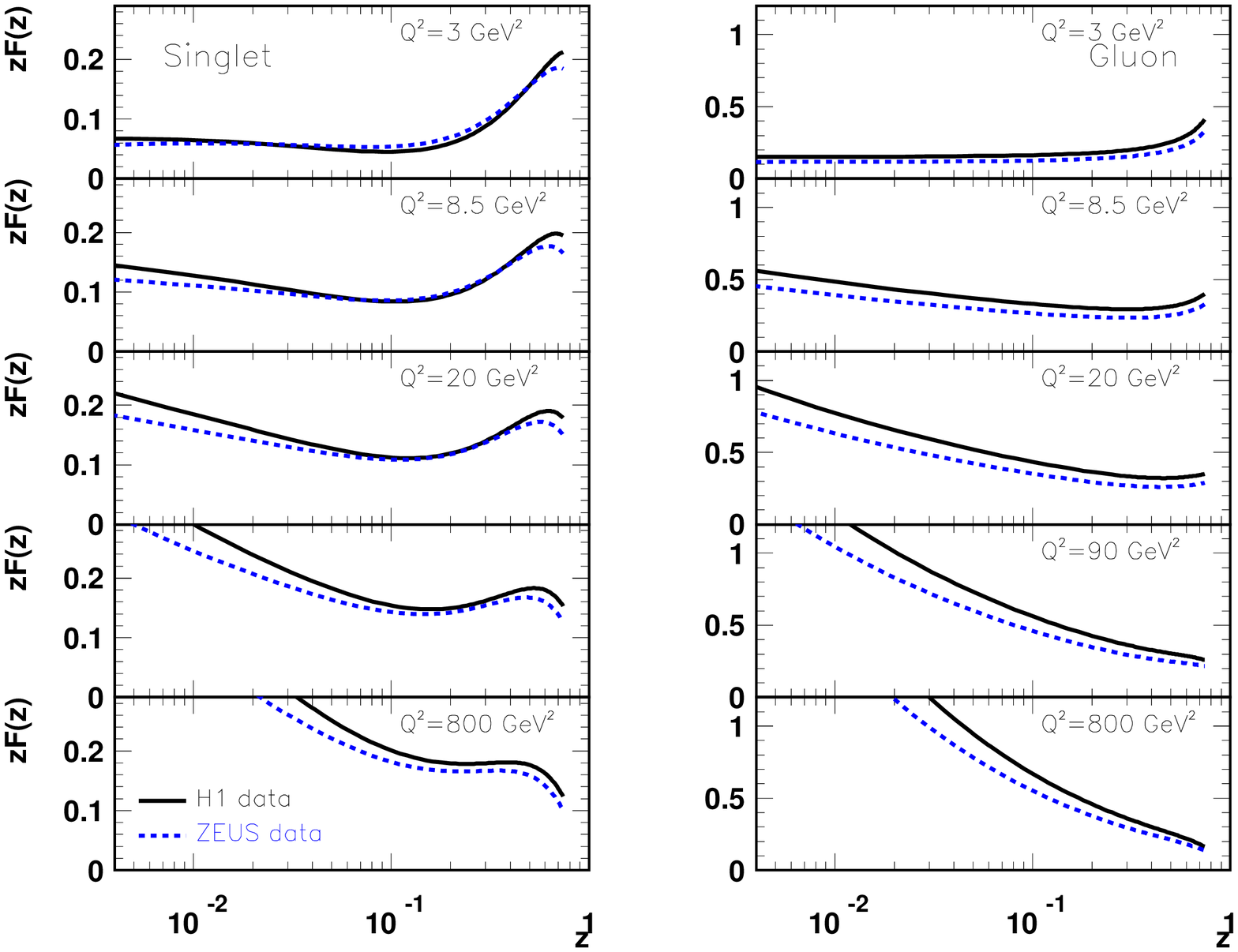}
\end{center}
\vspace{-0.5cm}
\caption{ Singlet and gluon distributions 
of the Pomeron as a function of $z$, the fractional momentum of the
Pomeron carried by the struck parton, derived from QCD fits on H1 and 
 ZEUS inclusive diffractive data (LRG)\cite{f2dall}. 
The parton densities are normalised to represent 
$\xpom$ times the true parton densities multiplied by the flux factor at
$\xpom = 0.003$ \cite{dpdfs_lolo}. 
A good agreement is observed between both diffractive PDFs, which indicates that
the underlying QCD dynamics derived in both experiments is similar.
}
\label{figfin}
\end{figure}

In Fig. \ref{figfin} we present the result for diffractive PDFs
(quark singlet and gluon densities),
obtained using the most recent inclusive
diffractive cross sections presented in Ref. \cite{f2dall}.
For each experiment (H1 and ZEUS),
we include measurements derived from Large Rapidity Gap (LRG) events 
in the QCD  analysis.

We follow  the procedure described in Ref. \cite{dpdfs_lolo},
with previous ZEUS data.
Note also that in all QCD fits, we let the global relative normalization
of the data set as a free parameter
(with respect to H1 LRG sample) \cite{dpdfs_lolo}. 
The typical uncertainties for the diffractive PDFs in Fig. \ref{figfin}
ranges from 5\% to 10\% for the singlet density and from 10\% to
25\% for the gluon distribution, with 25\% at large $z$
(which corresponds to large $\beta$
for quarks) \cite{dpdfs_lolo}. Similar results have been obtained by the
H1 collaboration \cite{f2dall}.

In order to analyze in more detail the large $z$ behavior of the gluon 
distribution $z{ {G}}(z,Q^2=Q_0^2)$ and give a quantitative estimate of the 
systematic error related to our parameterizations, we consider the possibility 
to change the gluon parameterization
by a multiplicative factor $(1-z)^{\nu}$ (see Ref. \cite{dpdfs_lolo}). 
If we include this
multiplicative factor $(1-z)^{\nu}$ in the QCD analysis, 
we derive a value of $\nu = 0.0 \pm 0.5$
(using the most recent data). Thus, we 
have to consider variations of $\nu$ in the interval $\pm 0.5$ 
in order to allow for the 
still large uncertainty of the gluon distribution 
(mainly at large $z$ values).  
The understanding of the
large $z$ behavior is of essential interest for any predictions at 
the Tevatron or LHC in central dijets production
(see below). In particular, a
proper determination of the uncertainty in this domain of momentum is
necessary and the method we propose in Ref. \cite{dpdfs_lolo} is a quantitative 
estimate,
that can be propagated easily to other measurements. 
\begin{figure}[htbp]
\begin{center}
\includegraphics[totalheight=7cm]{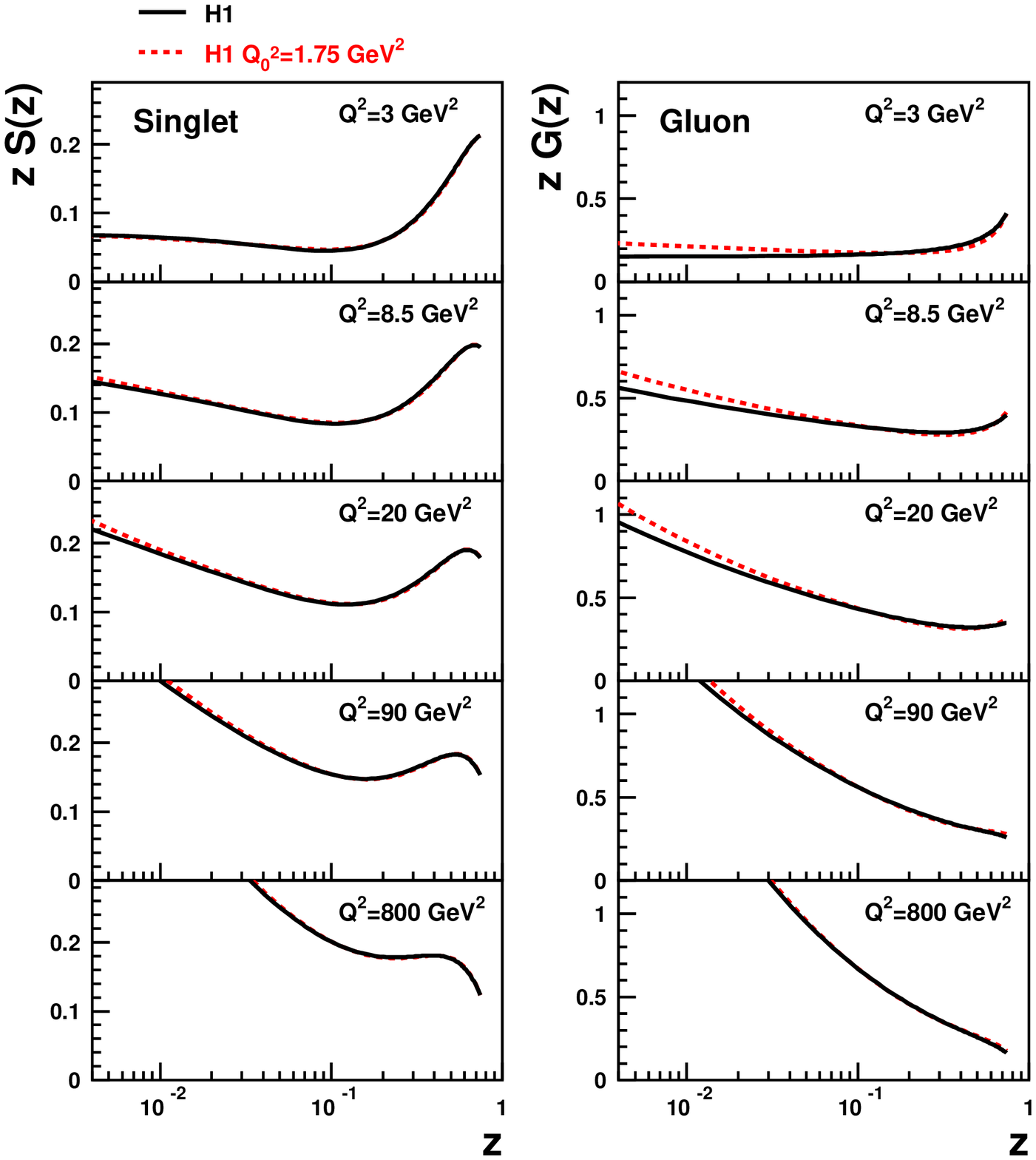}
\includegraphics[totalheight=7cm]{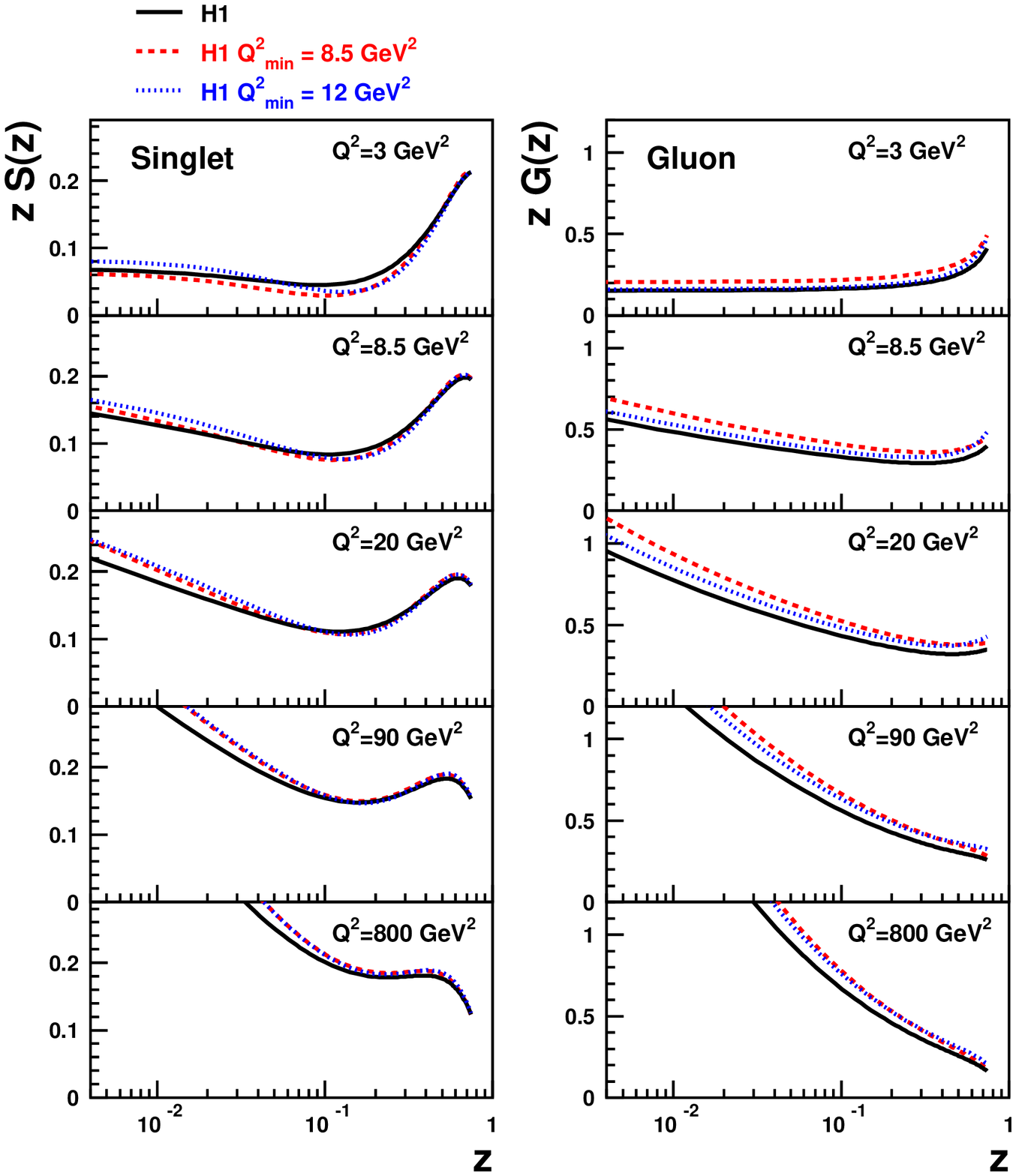}
\end{center}
\caption{ 
Left: Singlet and gluon distributions 
of the Pomeron as a function of $z$, the fractional momentum of the
Pomeron carried by the struck parton, derived from QCD fits on H1 data. 
Results are presented with $Q^2_0=3$~GeV$^2$ (full lines) and 
$Q^2_0=1.75$~GeV$^2$ (dashed lines). 
normalization follows the convention explained in Fig. \ref{figfin}.
Right:
Singlet and gluon distributions 
of the Pomeron as a function of $z$ derived from  QCD fits on H1 data. 
Results are presented with $Q^2_{min}=4.5$~GeV$^2$ (full lines),
$Q^2_{min}=8.5$~GeV$^2$ (dashed lines) and $Q^2_{min}=12$~GeV$^2$ (dotted lines). 
}
\label{fig0}
\end{figure}

Of course, 
several checks need to be done to analyze the stability of the QCD fits procedure
\cite{dpdfs_lolo}. We present two of them below:
\begin{itemize}
\item We have checked the dependence of the diffractive 
PDFs on variations of the 
starting scale $Q_0^2$ in Fig. \ref{fig0} (left).
Very small changes are observed while changing the starting scale form
3 to 1.75 GeV$^2$. 
\item We have checked 
the fit stability by changing the cut on $Q_{min}^2$, the lowest value
of $Q^2$ of data to be included in the fit. The results are given in Fig.
\ref{fig0} (right), where we show the results of the fits  
after applying a cut on $Q^2_{min}$ of 4.5, 8.5 and 12 GeV$^2$.
Differences are noticeable at small $\beta$ but well within the fit
uncertainties. No systematic behavior is observed within $Q^2_{min}$ variations.
\end{itemize}

\begin{figure}[t]
\begin{center}
 \includegraphics[totalheight=10cm]{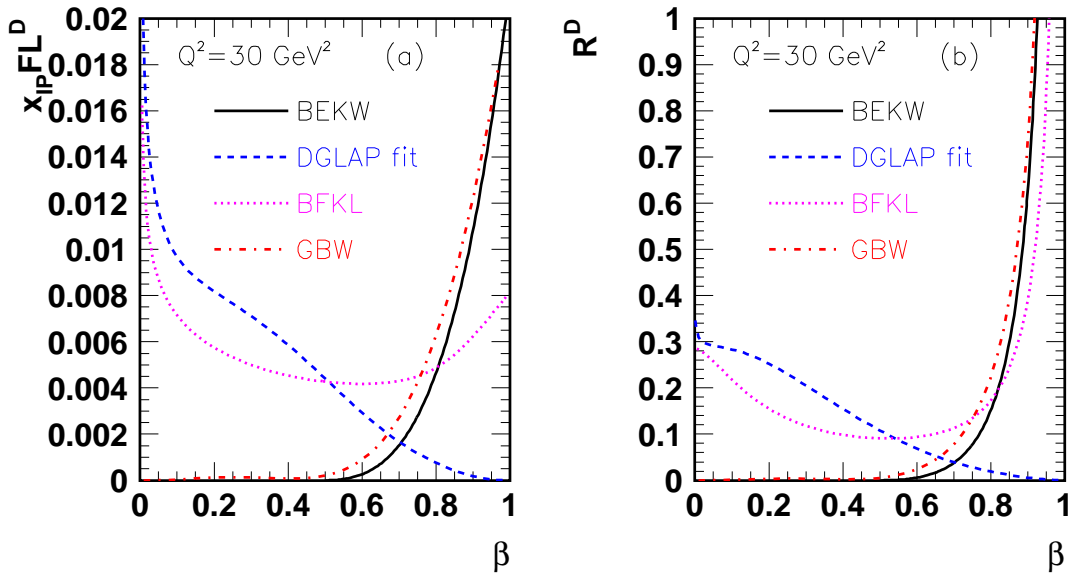}
\caption{ Predictions for $\xpom F_L^{D}$  and $R^D=\frac{F_L^D}{F_2^D-F_L^D}$  as a function of $\beta$
at $Q^2=30$~GeV$^2$ and $\xpom=10^{-3}$ \cite{dpdfs_lolo}.
The dashed line prediction refers to the diffractive PDFs analysis
discussed in this part. Note that
the longitudinal structure function $F_L^{D}$ is directly related to the
the reduced diffractive cross section:
$
\sigma_r^{D(3)} = F_2^{D(3)} -\frac{y^2}{1+(1-y)^2} F_L^{D(3)}
$.
Other curves
represent dipole model calculations (see next sections).
}
\label{Figdisc2}
\end{center}
\end{figure}

\begin{figure}[htbp]
\begin{center}
\includegraphics[width=8cm]{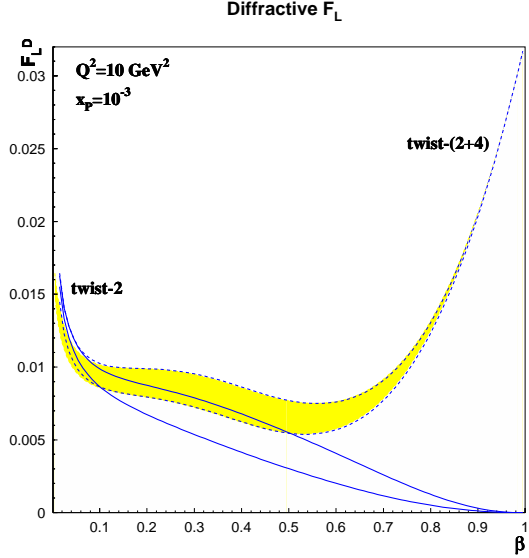}
\caption{Predictions for $F_L^{D}$ for $\xpom=10^{-3}$ and $Q^2=10$ GeV$^2$ 
from fits with twist--4, dominant only at large $\beta$, to the HERA data. 
The solid lines show predictions from pure (twist--2) QCD fits \cite{GolecBiernat:2007kv}.}
\label{fig:4}
\end{center}
\end{figure}

\begin{figure}[!]
\begin{center}
 \includegraphics[totalheight=8cm]{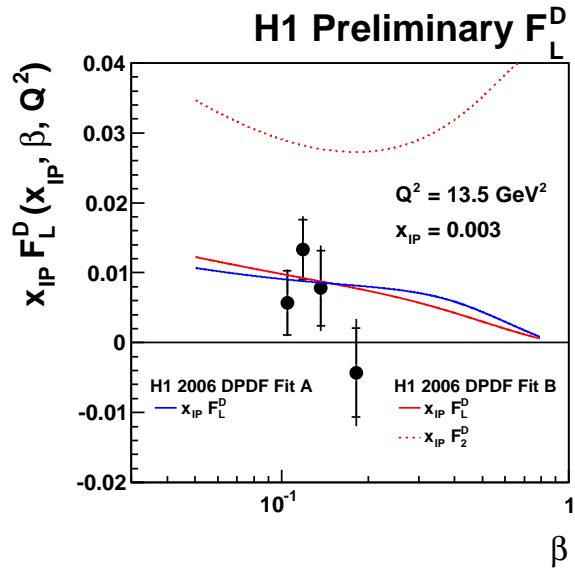}
\caption{ Measurements of the diffractive longitudinal structure function
$F_L^{D}$ multiplied by $\xpom$. The data are compared to QCD fits predictions (see text).
Also shown for comparison is the value of $F_2^{D}$ as a dashed line. }
\label{fldmeasurement}
\end{center}
\end{figure}

Then, an important conclusion  is the prediction for the
longitudinal diffractive structure function.
In Fig.~\ref{Figdisc2} \cite{dpdfs_lolo} and \ref{fig:4} 
\cite{GolecBiernat:2007kv}, we display this function with respect to
its dependence in $\beta$ 
 (Fig.~\ref{Figdisc2} (a)) and 
the ratio $R$ of the longitudinal to the transverse components 
of the diffractive structure function (Fig.~\ref{Figdisc2} (b)). 
A comment is in order about the large $\beta$ behavior.
$\xpom F_L^{D}$ is essentially zero at large $\beta$ from the 
pure QCD fits analysis. In fact, as illustrated in 
Fig.~\ref{Figdisc2}  and \ref{fig:4}, a non-zero contribution 
to the longitudinal structure function at large $\beta$
 corresponds to a twist--4, and is simply incorporated
in a dipole model formulation of diffraction (see next sections). 
Here, we give the qualitative feature of this effect 
on the predictions for $\xpom F_L^{D}$.
There is a 
significant difference between  predictions 
with or without this twist--4 component in the region of large
$\beta$. However, the difference is negligible
at low and medium $\beta$, where the
measurements are possible. 
Indeed, a 
first  measurements has been realized by the H1 collaboration  \cite{fldprel}, which is displayed 
in Fig. \ref{fldmeasurement}. It fits perfectly with the QCD fit prediction,
as well as with the dipole prediction in the kinematic range accessible experimentally
($\beta < 0.2$). This Fig. \ref{fldmeasurement} gives also directly the ratio
of  $F_L^{D}$ versus $F_2^{D}$.

\vfill
\clearpage

%
%
\section{  Diffraction at the Tevatron and prospects for LHC}

\subsection{Basics of diffraction at the Tevatron}

Once the gluon and quark densities in the Pomeron are known, it is easy to make
predictions for the Tevatron (or the LHC) if one assumes that the same mechanism is
the origin of diffraction in both cases. We assume the same structure of the Pomeron
at HERA and the Tevatron and we compute as an example the jet production
in single diffraction or double Pomeron exchange using the 
parton densities in the Pomeron measured at HERA. The interesting point is
to see if this simple argument works or not, or if the factorization
property between HERA and the Tevatron --- using the same parton distribution
functions --- holds or not \cite{cdfdiff,cdffact,royon}.
In other words, we need to know if it is possible to 
use the parton distributions in the Pomeron 
obtained at HERA to make predictions at the Tevatron, and
also further constrain the parton distribution functions in the Pomeron since
the reach in the diffractive kinematical plane at the Tevatron and HERA is
different. 

Theoretically, factorization is not expected to hold between the
Tevatron and HERA~\cite{collins} due to additional $pp$ or $p \bar{p}$ interactions. 
For instance, some soft gluon
exchanges between protons can occur at a longer time scale than the hard
interaction and destroy the rapidity gap or the proton does not remain intact
after interaction. The factorization break-up is confirmed by comparing the percentage of
diffractive events at HERA and the Tevatron (10\% at HERA and about 1\% of
single diffractive events at the Tevatron) showing already that factorization
does not hold. This introduces the concept of gap survival probability, the
probability that there is no soft additional interaction or that
the event remains diffractive. 

\begin{figure}
\begin{center}
\epsfig{file=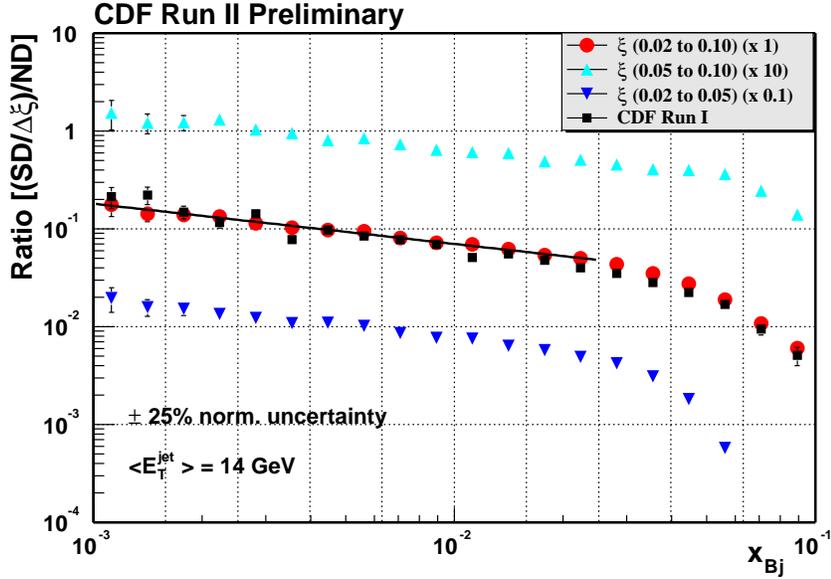,width=12cm}
\caption{Test of factorization within CDF data alone.
The percentage of diffractive events is presented
as a function of $x$ for different $\xi$ bins. The same $x$-dependence is observed within
systematic and statistical uncertainties
in all $\xi$ bins, supporting the fact that CDF data are consistent with 
factorization~\cite{cdfdiff}. }
\label{fig3}
\end{center}
\end{figure}

The first experimental test of factorization concerns CDF data
only. 
Fig.~\ref{fig3} shows the percentage of diffractive events
as a function of $x$ for different $\xi$ bins and shows the same $x$-dependence within
systematic and statistical uncertainties
in all $\xi$ bins supporting the fact that CDF data are consistent with 
factorization~\cite{cdfdiff}. The CDF collaboration also studied the $x$ dependence for
different $Q^2$ bins which leads to the same conclusions. 
 
A second step is to check whether factorization holds or not between Tevatron and
HERA data. The measurement of the diffractive structure function is possible
directly at the Tevatron. The CDF collaboration measured the ratio of dijet
events in single diffractive and non diffractive events, which is directly
proportional to the ratio of the diffractive to the standard proton structure
functions, where
\begin{eqnarray}
R(x) = \frac{Rate^{SD}_{jj} (x)}{Rate^{ND}_{jj} (x)} \sim
\frac{F^{SD}_{jj} (x)}{F^{ND}_{jj} (x)}.
\end{eqnarray} 
The comparison between the CDF measurement 
(black points, with systematics errors) and the
expectation from the diffractive QCD fits on HERA data in full line is shown in 
Fig.~\ref{fig4}~\cite{cdffact}. 

We notice a discrepancy of a factor 8 to 10 between the data and the predictions from
the QCD fit, showing that factorization does not hold. However, the difference
is compatible within systematic and statistical uncertainties
with a constant on a large part of the kinematical plane in
$\beta$, showing that the survival probability does not seem to be
$\beta$-dependent within experimental uncertainties. 

It would be interesting
to make these studies again in a wider kinematical domain both at the Tevatron and at
the LHC. The understanding of the survival probability and its dependence on the
kinematic variables is important to make precise predictions on inclusive diffraction
at the LHC.

\subsection{Discussion on the factorization breaking HERA/Tevatron}

In fact, from a fundamental point of view, 
it is natural that diffractive hard-scattering factorization does not apply to
hadron-hadron collisions.
Attempts to establish corresponding factorization theorems fail,
 because of interactions between spectator partons of the colliding
   hadrons.  The contribution of these interactions to the cross section
   does not decrease with the hard scale.  Since they are not associated
   with the hard-scattering subprocess, we no
   longer have factorization into a parton-level cross section and the
   parton densities of one of the colliding hadrons. These
interactions are generally soft, and we have at present to rely on
phenomenological models to quantify their effects. 

The yield of diffractive events in hadron-hadron collisions is then lowered
precisely because of these soft interactions between spectator partons
(often referred to as re-interactions or multiple scatterings).  
They can produce additional final-state particles which fill the would-be
rapidity gap (hence the often-used term rapidity gap survival).  When
such additional particles are produced, a very fast proton can no longer
appear in the final state because of energy conservation.  

Diffractive
factorization breaking is thus intimately related to multiple scattering
in hadron-hadron collisions.
We can also remark simply that
the collision partners, in hadron-hadron reactions, are both
composite systems of large transverse size, and it is not too
surprising that multiple interactions between their constituents can
be substantial.  

In contrast, the virtual photon in $\gamma^* p$
collisions has small transverse size, which disfavors multiple
interactions and enables diffractive factorization to hold.  According
to our discussion, we may expect that for
decreasing virtuality $Q^2$ the photon behaves more and more like a
hadron, and diffractive factorization may again be broken.

\subsection{Restoring factorization at the Tevatron}

\begin{figure}
\begin{center}
\epsfig{file=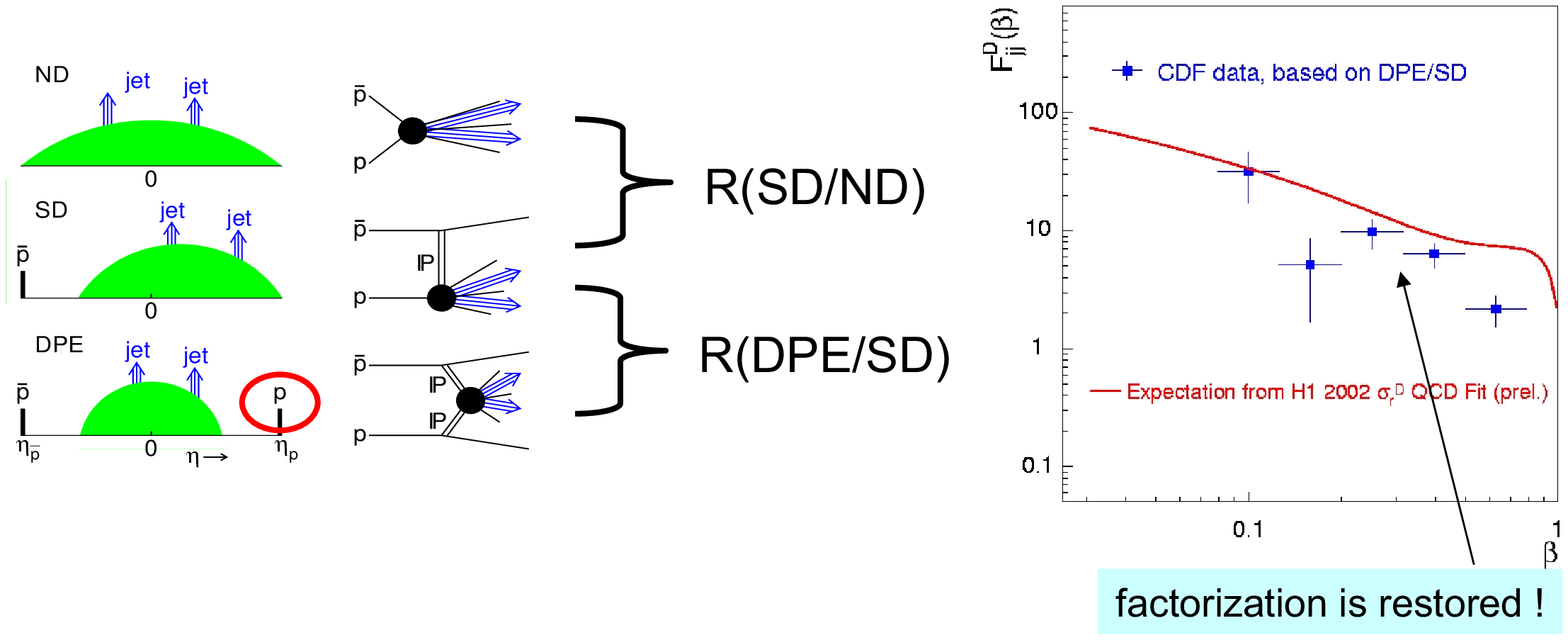,width=6cm,angle=270}
\caption{Restoration of factorization for the ratio of double Pomeron exchange
to single diffractive events (CDF Coll.).
Whereas
factorization was not true for the ratio of single diffraction to non diffractive events,
factorization holds for the ratio of double Pomeron exchange to single
diffraction. The price to pay for one gap is the same as the
price to pay for two gaps. 
}
\label{fig5}
\end{center}
\end{figure}

The other interesting measurement which can be also performed at the Tevatron is
the test of factorization between single diffraction and double Pomeron
exchange. The results from the CDF collaboration are shown in 
Fig.~\ref{fig5}~\cite{cdffact}.
The left plot shows the definition of the two ratios while the right figure
shows the comparison between the ratio of double Pomeron exchange to single
diffraction and the QCD predictions using HERA data in full line. 

Whereas
factorization was not true for the ratio of single diffraction to non diffractive events,
factorization holds for the ratio of double Pomeron exchange to single
diffraction. In other words, the price to pay for one gap is the same as the
price to pay for two gaps. The survival probability, i.e. the probability not to emit an additional
soft gluon after the hard interaction needs to be applied only once to require
the existence of a diffractive event, but should not be applied again for double
Pomeron exchange.

\ffig{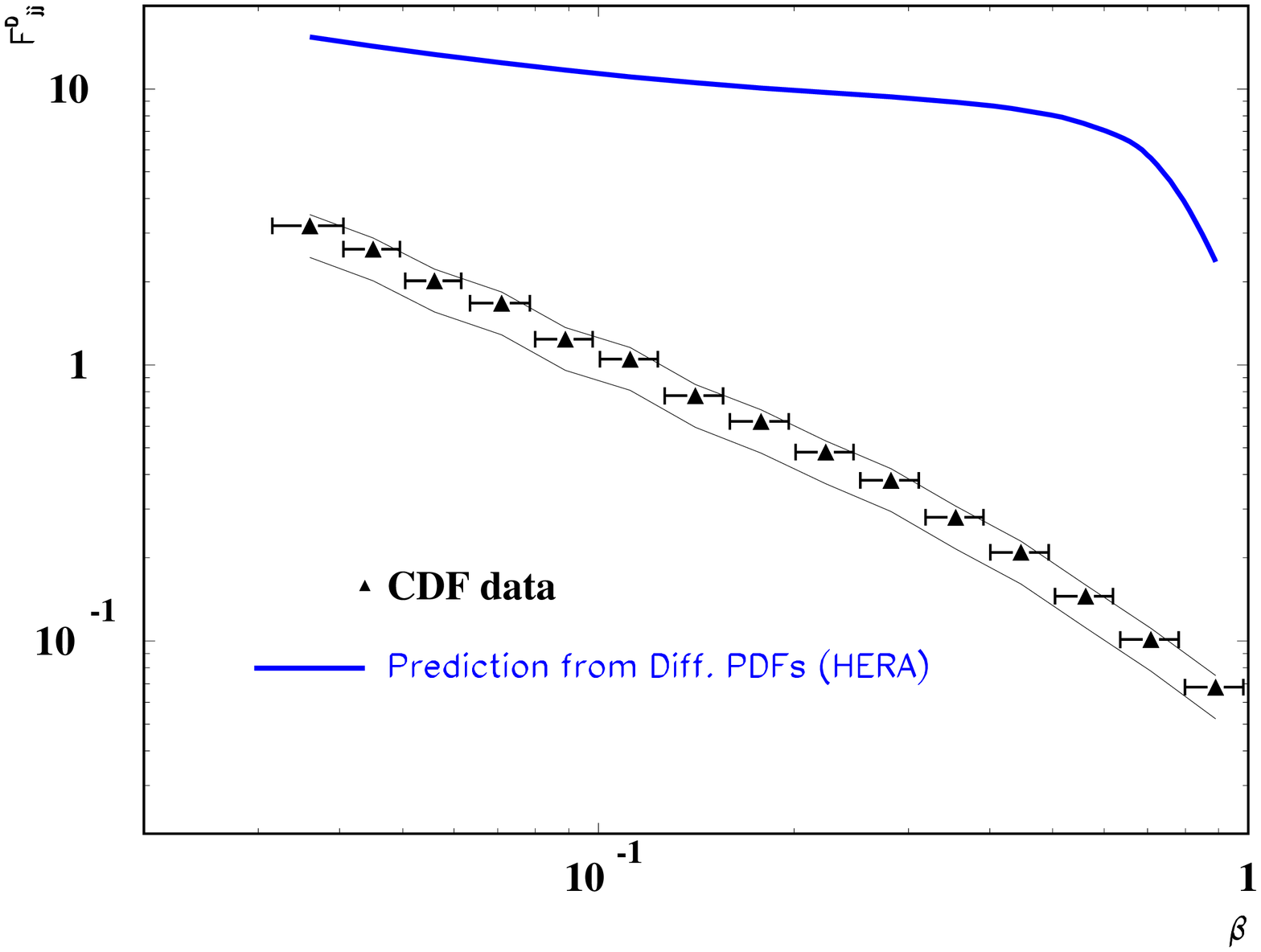}{80 mm}
{Comparison between the CDF measurements 
($Q^2=75$ GeV$^2$, $0.035<\xi<0.095$ and $|t|<1$ GeV$^2$) of diffractive structure
function (black points) with the expectation of the HERA (using first H1
diffractive data) 
diffractive PDFs \cite{dpdfs_lolo}. 
 The large discrepancy both 
in shape and normalization between HERA 
predictions and CDF data illustrates the breaking of factorization 
at the Tevatron.
Using the most recent measurements in QCD fits (and diffractive PDFs 
extraction) does
not change this conclusion.
 }
{fig4}

To summarize, factorization does not hold between HERA and Tevatron as expected
because of the long term additional soft exchanges with respect to the the hard
interaction. 
However, experimentally, factorization holds with CDF data themselves and also
between single diffraction and double Pomeron exchange which means that the soft
exchanges do not depend on the hard scattering, which is somehow natural.

\subsection{Interest of exclusive events}

\begin{figure}
\begin{center}
\epsfig{file=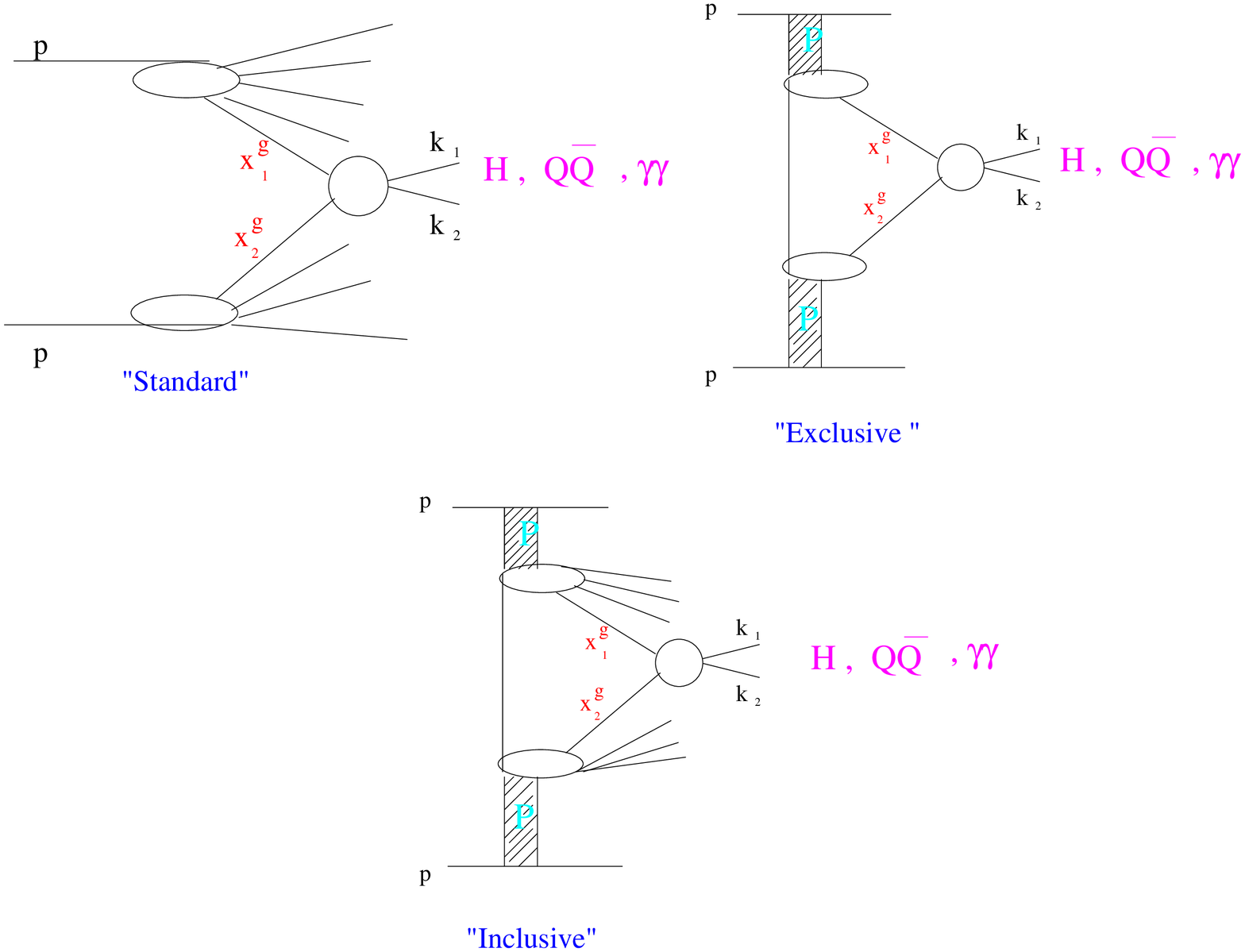,width=12cm}
\caption{Scheme of non diffractive, inclusive double Pomeron exchange,
exclusive diffractive events at the Tevatron or the LHC.}
\label{fig7}
\end{center}
\end{figure}

Once established some basics of the diffraction at the Tevatron, 
a fundamental topic concerns the analysis of exclusive events.
A schematic view of non diffractive, inclusive double Pomeron exchange,
exclusive diffractive events at the Tevatron or the LHC is displayed in
Fig.~\ref{fig7}.
The upper left plot shows the standard non diffractive events
where the Higgs boson, the dijet or diphotons are produced directly by a
coupling to the proton and shows proton remnants. The bottom plot displays 
the standard diffractive double
Pomeron exchange where the protons remain intact after interaction and the total
available energy is used to produce the heavy object (Higgs boson, dijets,
diphotons...) and the Pomeron remnants. We have so far only discussed
this kind of events and their diffractive production using the
parton densities measured at HERA. 

There may be a third class of processes
displayed in the upper right figure, namely the exclusive diffractive
production. In this kind of events, the full energy is used to produce the heavy
object (Higgs boson, dijets, diphotons...) and no energy is lost in Pomeron
remnants. There is an important kinematical consequence. The mass of the
produced object can be computed using roman pot detectors and tagged protons

\begin{eqnarray}
M = \sqrt{\xi_1 \xi_2 S}.
\end{eqnarray} 
We see immediately the advantage of these processes. We can benefit from the
good roman pot resolution on $\xi_1$ and $\xi_2$ to get a good resolution on mass. It is then
possible to measure the mass and the kinematical properties of the produced
object and use this information to increase the signal over background ratio by reducing the
mass window of measurement. It is thus important to know if this kind of events
exist or not. 

In the following, we give some details of the search for exclusive events in the
different channels which are performed by the CDF and D0 collaborations at the Tevatron.
Prospects for the LHC are then outlined.

\subsection{Search for exclusive events in $\chi_c$ production}

For example,
one way to look for exclusive events at the Tevatron is to search 
for the diffractive exclusive production of 
light particles like the $\chi$ mesons. This would give rise to high enough 
cross sections -- contrary to the diffractive exclusive production of heavy
mass objects such as Higgs bosons --- to check the dynamical 
mechanisms and the existence of exclusive events. 
Exclusive production of $\chi_{c}$ has been studied by the 
CDF collaboration~\cite{royon} with an upper limit for the cross section 
of $\sigma_{exc}(p\bar{p} \rightarrow p+J/\psi
 + \gamma+\bar{p}) \sim 49 \pm 18 (stat)
\pm 39 (sys)\  $ pb, where the $\chi_c$ decays into $J/\Psi$ and $\gamma$, the
$J/\Psi$ decaying itself into two muons. The experimental signature is thus two
muons in the final state and an isolated photon, which is a very clear signal.

Unfortunately, the cosmics contamination is difficult to compute and this is
why the CDF collaboration only quotes an upper limit on the $\chi_c$ production
cross section.
To know if the production is really
exclusive, it is important to study the tail of inclusive diffraction which is a
direct contamination of the exclusive signal. The tail of inclusive diffraction
corresponds to events which show very little energy in the forward direction, or in
other words where the Pomeron remnants carry very little energy. This is why these
events can be called quasi-exclusive.

In Ref.~\cite{chic}, it is shown that the contamination of
inclusive events into the signal region depends strongly on the
assumptions on the gluon distribution in the Pomeron at high $\beta$, which
is poorly known as we mentioned in a previous section. 
Therefore, this channel is  not
conclusive concerning the existence of exclusive events.
In the same spirit,
the CDF collaboration also looked for the exclusive production of dilepton and
diphoton~\cite{cdfgamma}.

\subsection{Search for exclusive events using the dijet 
mass fraction}

Another very important aspect of diffraction at the Tevatron 
is related to the diffractive production of dijet events in double Pomeron exchange
(see Fig. \ref{fig7}).
The CDF collaboration measured the so-called dijet mass fraction
in dijet events --- the ratio of the mass carried by the two jets produced in the event 
divided by the
total diffractive mass --- when the antiproton is tagged in the roman pot
detectors and when there is a rapidity gap on the proton side to ensure that the
event corresponds to a double Pomeron exchange. 

\begin{figure}
\begin{center}
\epsfig{figure=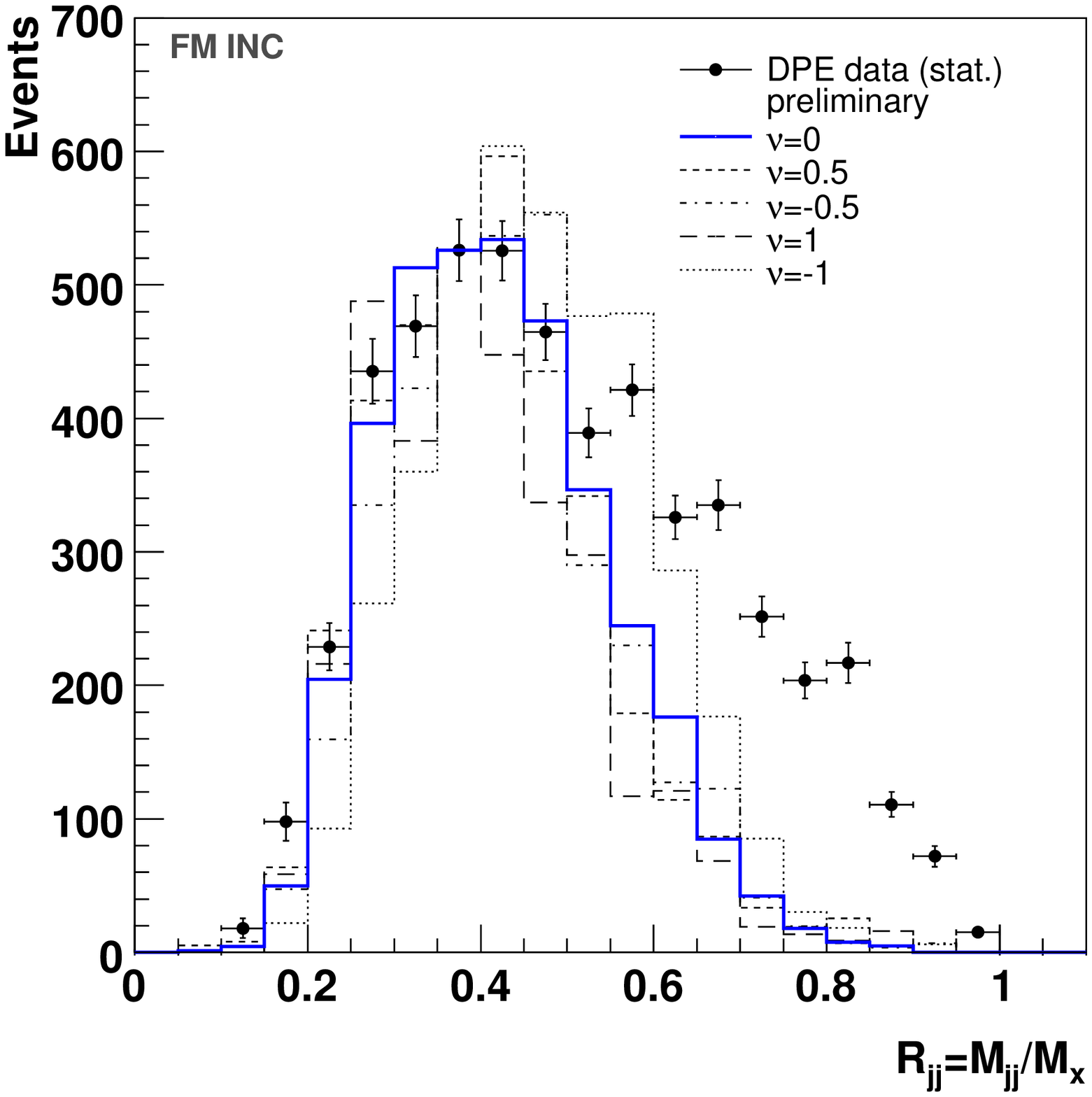,height=2.5in}  
\epsfig{figure=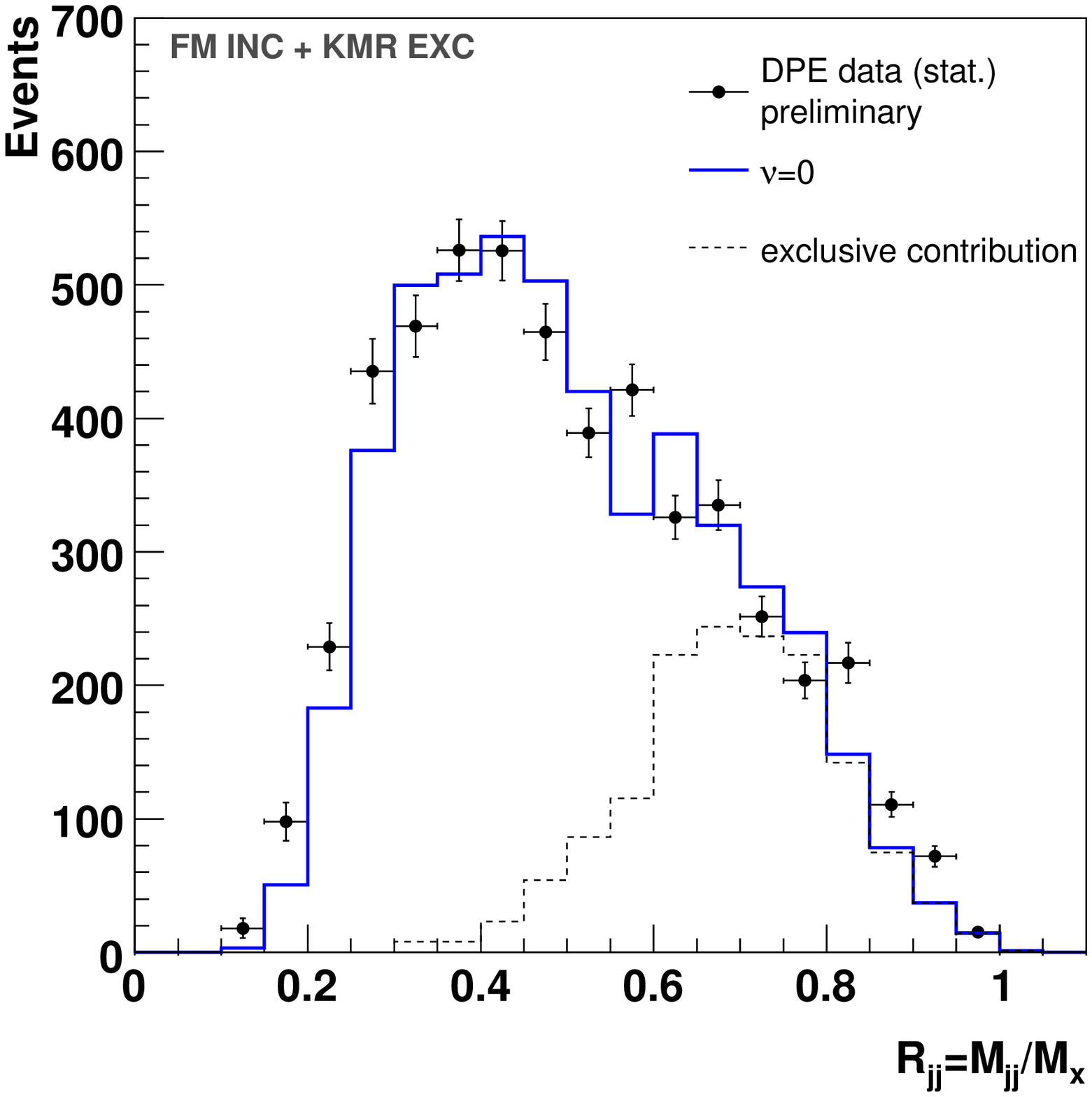,height=2.5in} 
\caption{Left: Dijet mass fraction measured by the CDF collaboration compared to the
prediction from inclusive diffraction based on the parton densities
in the Pomeron measured at HERA. The gluon
density in the Pomeron at high $\beta$ was modified by varying the parameter
$\nu$.
Right: Dijet mass fraction measured by the CDF collaboration compared to the
prediction adding the contributions from inclusive and exclusive diffraction.}
\label{compare2}
\end{center}
\end{figure}

The CDF collaboration has measured this
quantity for different jet $p_T$ cuts~\cite{cdfrjj}. 
In Fig. \ref{compare2}, we compare this measurement
to the expectation coming from the structure of the Pomeron coming from HERA.
For this sake, one takes the gluon and quark densities in the Pomeron measured at HERA
as described in Ref.~\cite{dpdfs_lolo} and the factorization breaking between
HERA and the Tevatron is assumed to come only through the gap survival probability
(0.1 at the Tevatron). 
The comparison between the CDF data for a jet $p_T$ cut of 10 GeV as an
example and the predictions from inclusive diffraction is given in 
Fig.~\ref{compare2}, left. 

We also display
in the same figure the effects of changing the gluon density at high $\beta$ (by
changing the value of the $\nu$ parameter) and we note that inclusive
diffraction is not able to describe the CDF data at high dijet mass fraction,
even after increasing the gluon density in the Pomeron at high $\beta$ (multiplying
it by $1/(1-\beta)$),
where exclusive events are expected to appear~\cite{dpdfs_lolo,oldab}. 

The conclusion
remains unchanged when jets with $p_T>25$ GeV are considered~\cite{dpdfs_lolo,oldab}.
Adding exclusive events to the distribution of the dijet mass fraction leads to
a good description of data~\cite{dpdfs_lolo,oldab} as shown in Fig.~\ref{compare2}, right, 
where we superimpose the
predictions from inclusive and exclusive diffraction. 

This study does not prove explicitly
that exclusive events exist but shows that some additional component with respect to
inclusive diffraction is needed to explain CDF data. Adding exclusive
diffraction allows to explain the CDF measurement. 
To be sure of the existence
of exclusive events, the observation will have to be done in different channels
and the different cross sections to be compared with theoretical expectations
\footnote{
In Ref.~\cite{oldab}, the CDF data were also compared to the soft color
interaction models. While the need for exclusive events is less
obvious for this model, especially at high jet $p_T$, the jet rapidity
distribution measured by the CDF collaboration is badly reproduced. This is due
to the fact that, in the SCI model, there is a large difference between
requesting an intact proton in the final state and a rapidity gap.}.
 
\begin{figure}[htbp]
\begin{center}
\epsfig{file=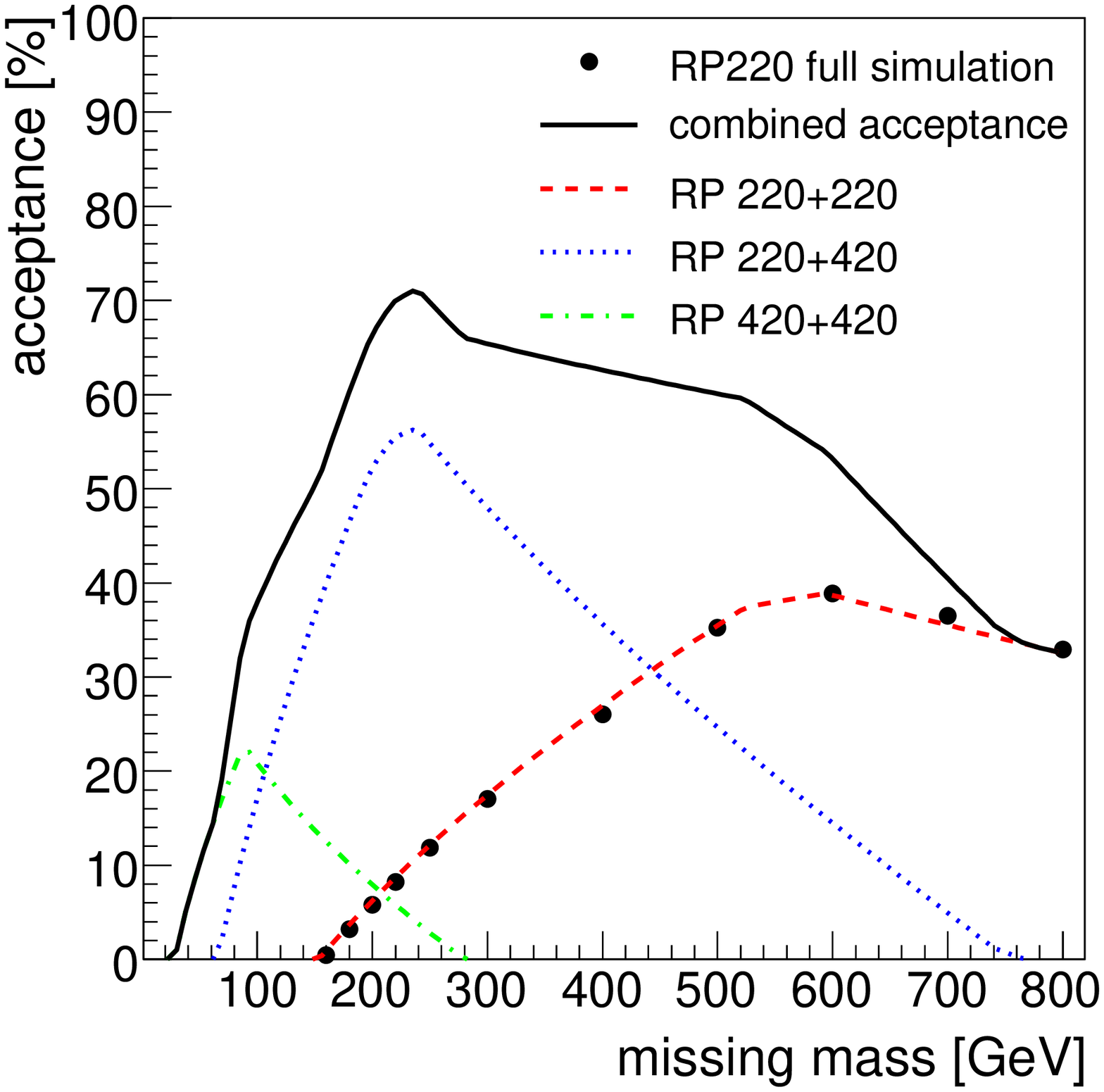,width=9cm}
\caption{Roman pot detector acceptance as a function of missing mass
assuming a 10$\sigma$ operating positions, a dead edge for the detector of 50
$\mu m$ and a thin window of 200 $\mu m$.}
\label{accept1}
\end{center}
\end{figure}

\subsection{Prospects for LHC}

The search for exclusive events at the LHC can be performed in the same channels
as the ones used at the Tevatron. 
Let us recall that a strong motivation for this idea is that heavy objects,
like Higgs boson, could be produced in double pomeron exchange at the LHC~\cite{atlas}.
In addition, some other possibilities
benefiting from the high luminosity of the LHC appear. One of the cleanest ways
to show the existence of exclusive events would be to measure the dilepton and
diphoton cross section ratios as a function of the 
dilepton/diphoton mass~\cite{atlas}. If
exclusive events exist, this distribution should show a bump towards high values
of the dilepton/diphoton mass since it is possible to produce exclusively
diphotons but not dileptons at leading order as we mentioned in the previous
paragraph. 

The motivation to install forward detectors at in ATLAS and CMS is then quite
clear. In addition, 
it extends nicely the project of measuring the total cross sections
in ATLAS and TOTEM by measuring hard diffraction at high
luminosity at the LHC. 
Of course, this is a very challenging technical project.

Without entering into details, a few technical issues can
be discussed simply. Two locations for the forward detectors are considered at
220 and 420m respectively to ensure a good coverage in $\xi$ or in mass of the
diffractively produced object  \cite{atlas}. Installing
forward detectors at 420m is quite challenging since the detectors will be
located in the cold region of the LHC and the cryostat has to be modified to
accommodate the detectors. In addition, the space available is quite small and
some special mechanism called movable beam pipe are used to move the detectors
close to the beam when the beam is stable enough. The situation at 220m is
easier since it is located in the warm region of the LHC and both roman pot
and movable beam pipe technics can be used. The AFP (ATLAS Forward Physics)
project is under discussion in the ATLAS collaboration and includes both 220
and 420m detectors on both sides of the main ATLAS detector \cite{atlas}.

To conclude on the diffraction at the LHC,
the missing mass acceptance is given in Fig.~\ref{accept1}. 
The missing mass acceptance using only
the 220m pots starts at 135 GeV, but increases slowly as a function of missing
mass. It is clear that one needs both detectors at 220 and 420m to obtain a good
acceptance on a wide range of masses since most events are asymmetric (one tag
at 220m and another one at 420m). The precision on mass reconstruction using
either two tags at 220m or one tag at 220m and another one at 420m is of the
order of 2-4 \% on the full mass range, whereas it goes
down to 1\% for symmetric 420m tags \cite{atlas}.

\vfill
\clearpage

%
%
\section{  Diffraction and the dipole model}

\subsection{  Simple elements of theory}

\begin{figure}[t]
   \vspace*{-1cm}
    \centerline{
     \epsfig{figure=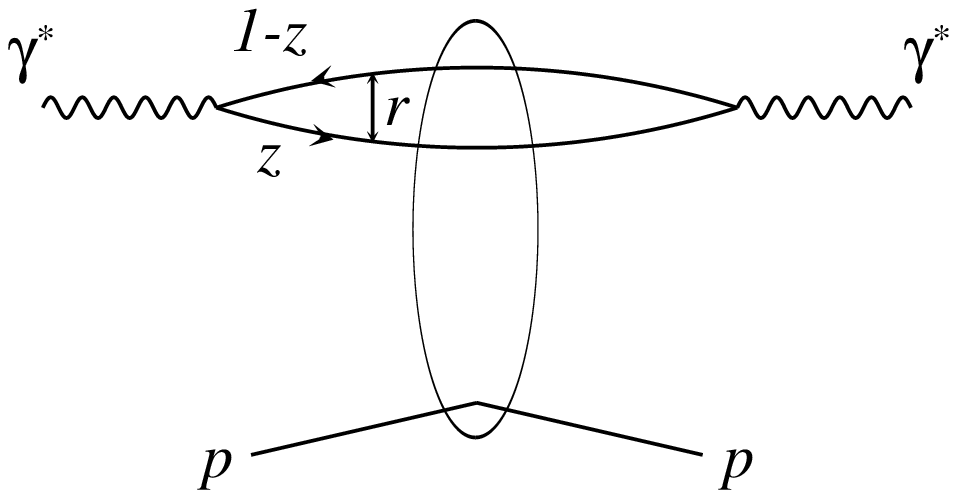,width=8cm}
               }
    \vspace*{-0.5cm}
\caption{Picture for the total
cross section ($\gamma^{*
}p\rightarrow\gamma^{*}p$) in the dipole model.
}
\label{f2dipoletot}
\end{figure}

The physical picture of hard diffraction at HERA is interesting in the 
proton rest frame and reminiscent of the aligned jet model. In the 
proton rest frame, at small $x_{Bj}$, the virtual photon splits into a 
$q\bar{q}$ pair long before it hits the proton
\cite{dipole_1,dipole_2,dipole_2bis,dipole_2golec} (see Fig. \ref{f2dipoletot}). 
The $q\bar{q}$ 
wave-function of the virtual photon suppresses configurations in which 
one of the quarks carries almost all momentum. In fact, these 
configurations are the ones 
that give rise to a large diffractive cross section, just 
because the wave-function suppression is compensated by the large cross 
section for the scattering of a $q\bar{q}$ pair of hadronic transverse 
size off the proton. The harder of the two quarks is essentially a 
spectator to diffractive scattering. 

The scattering of the softer quark 
off the proton is non-perturbative and cannot be described by exchange 
of a finite number of gluons. Hence there is an unsuppressed probability 
that the softer quark leaves the proton intact. 
This explains simply the idea behind the leading twist 
nature of hard diffraction. The details of the scattering of the softer 
quark off the proton are encoded in the diffractive quark 
distribution. In a similar way, the $q\bar{q}g$ configuration in the 
virtual photon, in which the $q\bar{q}$ pair carries almost all momentum,  
gives rise to the diffractive gluon distribution.

Also, in the simplest case, the  colorless exchange
responsible for the rapidity gap is modeled by the exchange of two gluons
(projected onto the color singlet state)
coupled to the proton with some form factor or to a
heavy onium which serves as a model of the proton
\cite{dipole_2,dipole_2bis,dipole_2golec}.
We focus the following discussion on 
 dipole approaches of diffractive interactions,
 that follow exactly these ideas. 

Then, we can model the reaction in three different phases, as displayed in Fig. \ref{figbekw} -top-:
\begin{itemize}
\item[(1)] the
transition of the virtual photon to the $q\bar{q}$ pair (the color
dipole) at a large distance
$
l \sim \frac{1}{m_N x}
$ of about 10-100 fm for HERA kinematics,
upstream the target, 
\item[(2)] the interaction of the
color dipole with the target nucleon, and 
\item[(3)] the projection of
the scattered $q\bar{q}$ onto the diffractive system $X$. 
\end{itemize}

\subsection{  Confrontation of HERA measurements to the dipole approach}

\begin{figure}[t]
\begin{center}
  \hspace*{1.5cm}
         \psfig{figure=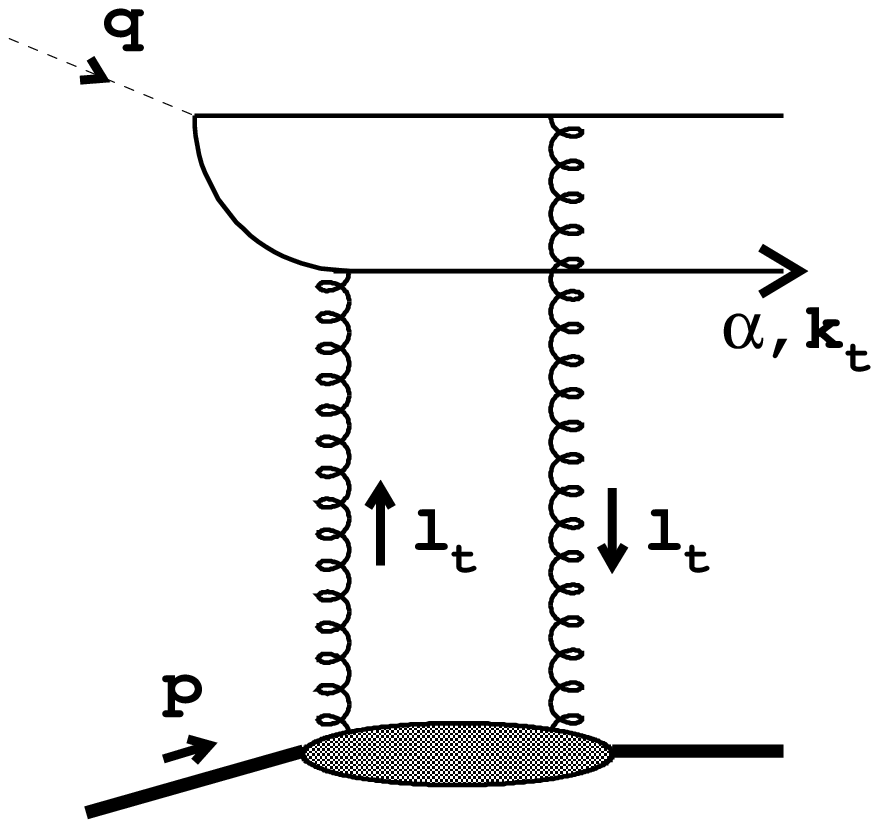,width=6.5cm}
         \psfig{figure=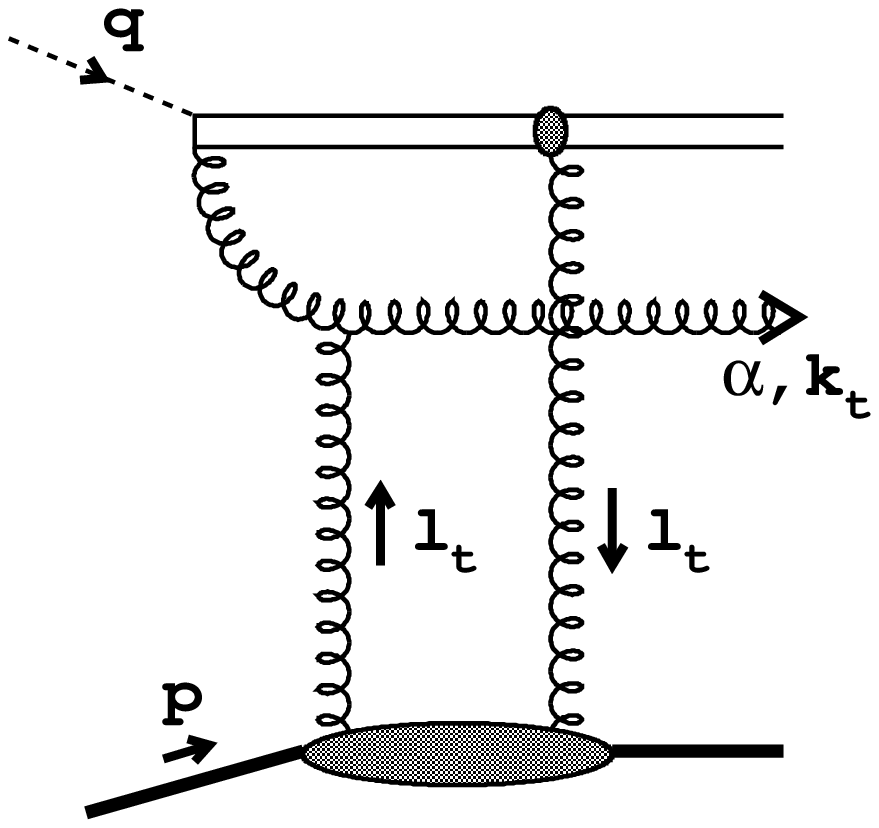,width=6.5cm}
         \includegraphics[width=12cm,height=6.5cm]{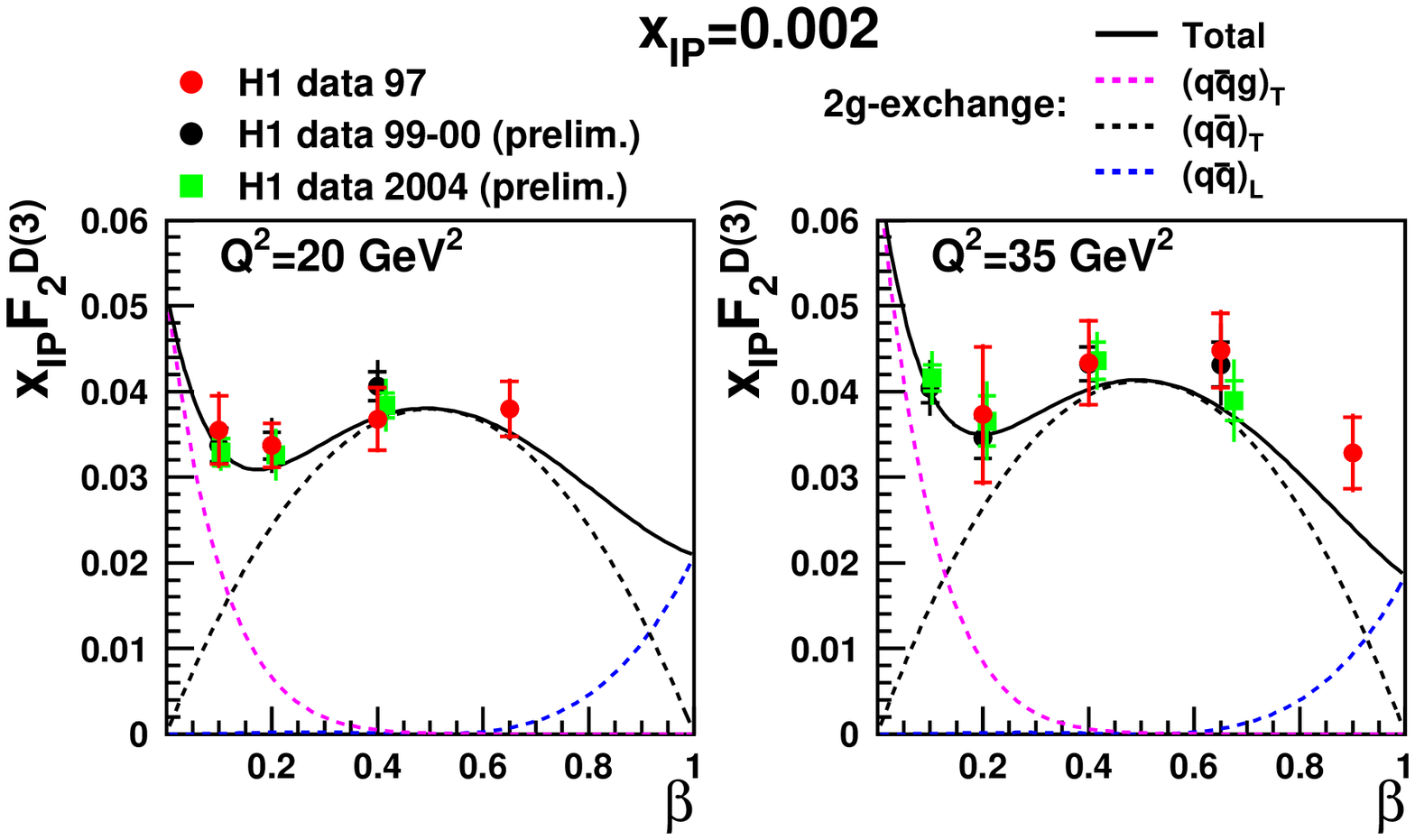}
\caption{ Top: The $q\bar{q}$ and $q\bar{q}g$ components of the diffractive system.
Bottom:
The diffractive structure function
$\xpom F_2^{D(3)}$ is presented  as a function of $\beta$
for two values 
of $Q^2$. The different components of the two-gluon
exchange model are displayed (see text). They add up to give a good description of the data. The structure function 
  $ \xpom F_2^{D(3)}$
is obtained directly from the
measured  diffractive cross section using the relation : 
$
\frac{d^3 \sigma^{ep\rightarrow eXp}}{d\xpom\ dx\ 
dQ^2} \simeq \frac{4\pi\alpha_{em}^2}{xQ^4}
({1-y+\frac{y^2}{2}}) F_2^{D(3)}(\xpom,x,Q^2)
$, where $y$ represents the inelasticity of the reaction.}
\label{figbekw}
\end{center}
\end{figure}

Following the arguments above,
the inclusive diffractive
cross section is  described with
three main contributions in dipole approaches. The first one describes the diffractive 
production of a $q \bar{q}$ pair from
a transversely polarized photon, the second one the production of 
a diffractive $q \bar{q} g$ system, and the third one the production of a
$q \bar{q}$ component from a longitudinally polarized photon (see Fig. \ref{figbekw} -top-).
In Fig. \ref{figbekw} -bottom-, we show that this approach,
also called two-gluon exchange model gives a good description
of the diffractive cross section measurements \cite{dipole_2,dipole_2bis,dipole_2golec}. 

One of the great advantage of the dipole model is that it provides a natural explanation
of the rapidity gap formation.
Another  great advantage of the dipole formulation
is that it provides a natural explanation of the 
experimental observation  that
$\sigma^{diff}/\sigma^{tot}\simeq const$ as a function of energy $W$
(see Fig. \ref{fig8}) \cite{dipole_2golec}.
Indeed, the dipole 
picture is valid in the frame in which the $q\bar{q}$ pair (dipole) carries most of
the available rapidity $Y\sim \ln(1/x)$ of the system.
The gluon radiation from the parent dipole can then
be interpreted (in the large $N_c$ limit)
as a collection of dipoles of different transverse sizes which interact with the proton. 
If the proton stays intact,  diffractive events with large rapidity gap are formed. 
In such case, the diffractive system is given by the color dipoles and the 
diffractive exchange
can be modeled by color singlet gluons exchange (two-gluon exchange) between
the dipole and the proton (see Fig. \ref{f2dipoletot} and \ref{figbekw}-top-). 
When only the parent $q\bar{q}$ dipole forms a diffractive system,
the diffractive cross section at $t=0$ reads 
\begin{equation}
\label{eq:5}
\frac{d\,\sigma^{diff}}{dt}_{\mid\, t=0}
\,=\,
\frac{1}{16\,\pi}\,
\int d^2 r\, dz\,
|\Psi^\gamma(r,z,Q^2)|^2\ \hat\sigma^2(x,r),
\end{equation}
where   $\Psi^\gamma$ is the well known light-cone wave function of the virtual photon,
$r$ is the dipole transverse size and $z$ is a fraction of the photon
momentum carried by the quark. 
Applying the $q\bar{q}$ dipole picture to the total inclusive cross section,
$\sigma^{tot}$, 
the following relation holds in the small-$x$ limit 
\begin{equation}
\label{eq:6}
\sigma^{tot}\,=\,
\int d^2 r\, dz\,
|\Psi^\gamma(r,z,Q^2)|^2\ \hat\sigma(x,r),
\end{equation}
with the same dipole cross $\hat\sigma(x,r)$ as in Eq. (\ref{eq:5}).
This Eq. (\ref{eq:6}) is pictured in Fig. \ref{f2dipoletot}.

\subsection{  Saturation, concepts and practice}

In Eq. \ref{eq:5} and \ref{eq:6},
the parameterization of $\hat\sigma(x,r)$ must be realized with
caution \cite{dipole_2golec,dipole_2bis}.
There are several features to consider.
First, the  density of gluons at given $x$ increases with
increasing $Q^2$, as described in perturbative QCD.  According to  QCD evolution it
also increases at given $Q^2$ when $x$ becomes smaller, so that the
gluons become more and more densely packed.  At some point, they 
start to overlap and thus re-interact and screen each other.  Then, we
enter a regime where the density of partons saturates and where the
linear QCD evolution equations cease to be valid.  
To quantify these
effects, a saturation scale $Q^2_s$ can be introduced, which also depends on $x$,
such that for $Q^2 \sim Q^2_s(x)$ these effects of saturation become important.

\begin{figure}[t]
\begin{center}
\includegraphics[width=0.55\textwidth]{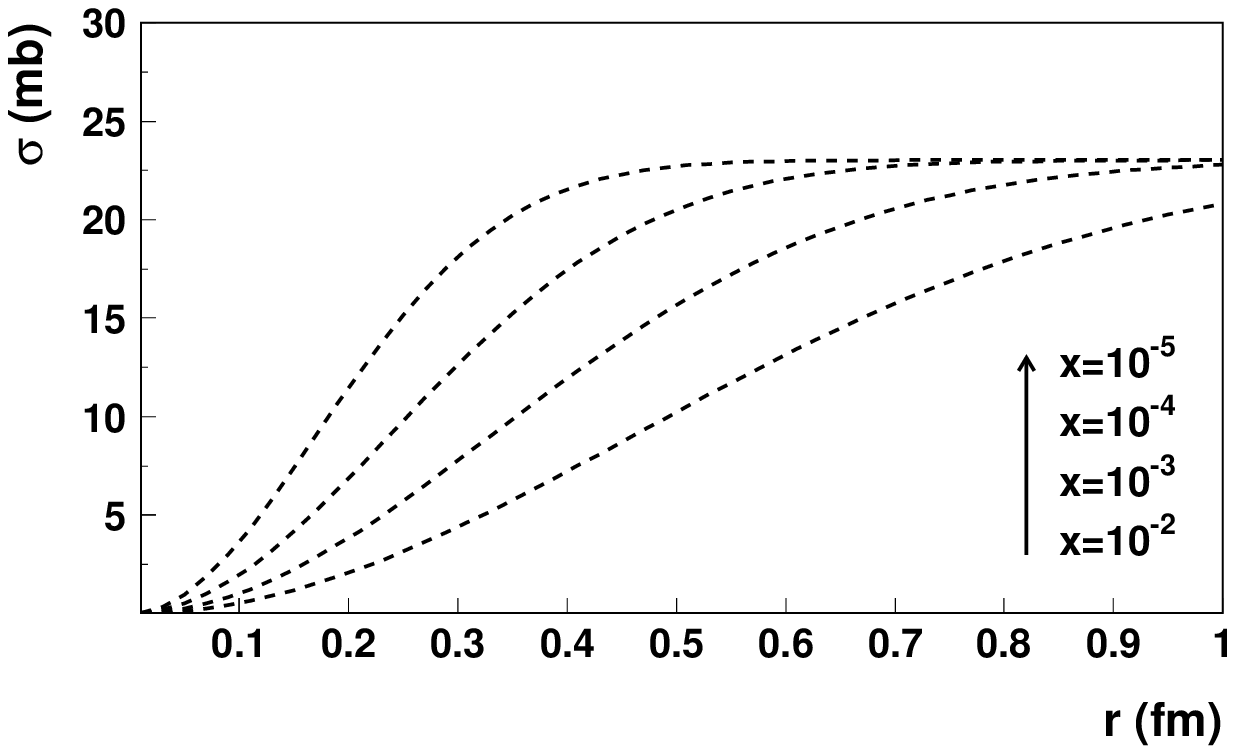}
\caption{\label{gbw-xsect}  The dipole cross section $\sigma_{q\bar{q}}$ in 
the saturation model as a function of dipole size $r$ for
  different $x$ (see text).}
\end{center}
\end{figure}

In practice,
essential features of the saturation phenomenon are 
verified in the following parameterization for the
dipole cross section first
proposed in  Ref. \cite{dipole_2golec}
\begin{equation}
\label{eq:7}
\hat{\sigma}(x,r)\,=\,\sigma_0\, \{1-\exp(-r^2 Q^2_s(x))\}\,,
\end{equation}
where $Q_s(x)=Q_0\,(x/x_0)^{-\lambda}$ is the saturation scale.
In Fig.~\ref{gbw-xsect}, we  display the dipole cross section
dependence of Eq. (\ref{eq:7}) as a function of $r$ at
given $x$ in this model.  At small dipole size 
$r \sim 1/Q$ (large $Q^2$), the cross section rises
following the relation $\hat{\sigma}(x,r) \propto r^2 x g(x)$. 
At some value $R_s(x)$ of $r$, the dipole cross section is so large
that this relation ceases to be valid, and $\hat{\sigma}(x,r)$ starts to
deviate from the quadratic behavior in $r$.
Therefore, $R_s(x) = 1/Q_s(x)$ represents a typical saturation scale.
As $r$ continues to increase, $\hat{\sigma}(x,r)$ eventually saturates 
at a value typical of a meson-proton cross section.
For
smaller values of $x$, the initial growth of $\sigma_{q\bar{q}}$ with $r$ 
is stronger because the gluon distribution is larger.  The target is thus
more opaque and  saturation sets in at lower $r$.

\begin{figure}[ht]
\begin{center}
\epsfig{file=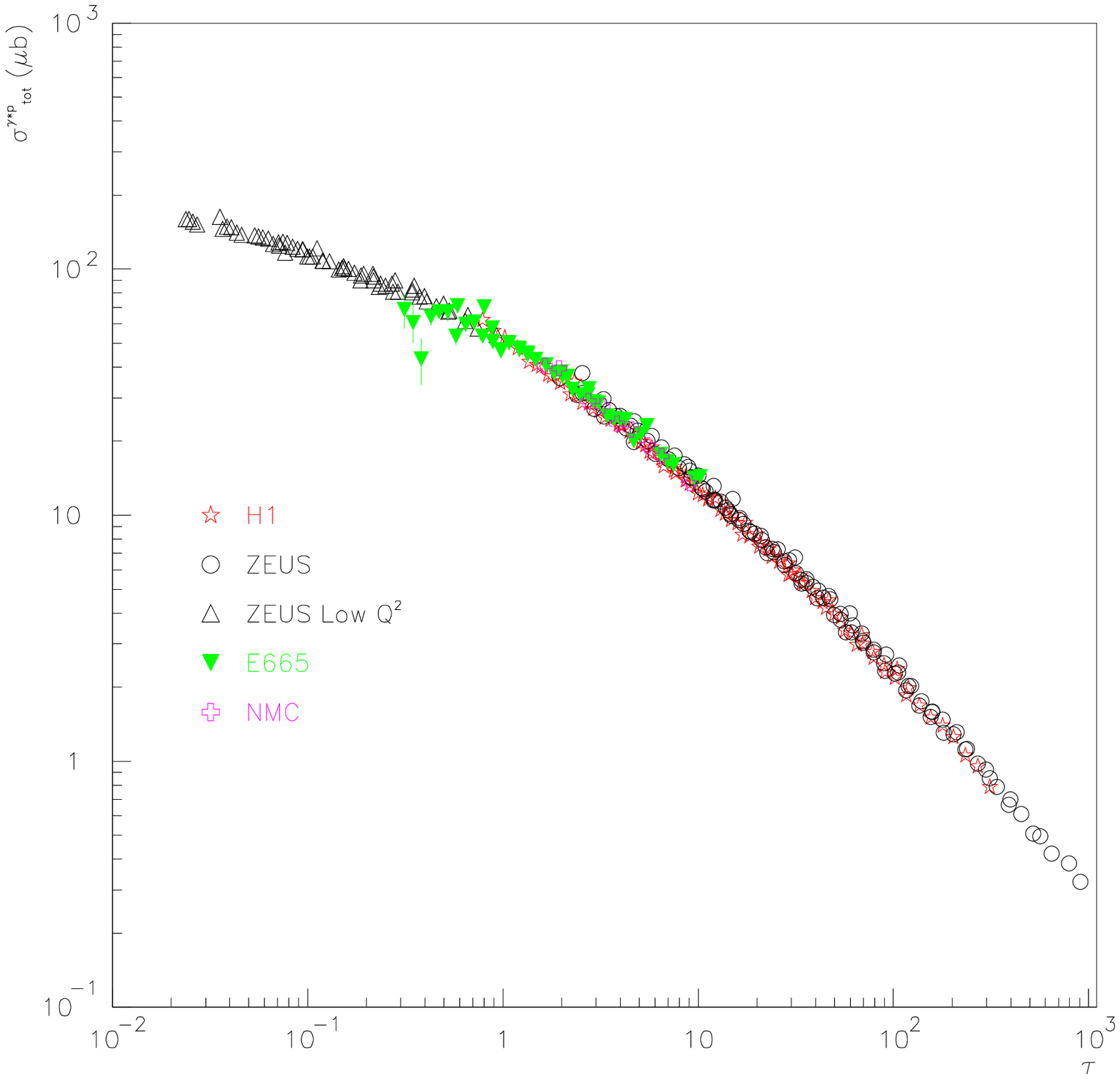,width=10cm}
\caption{The total cross section $\sigma^{\gamma^*p\rightarrow X}_{tot}$ as a 
function of $\tau=Q^2/Q^2_s$ for $x\!<\!0.01$. 
The saturation regime is  reached when $Q_s\!\sim\!Q$. 
For inclusive events in deep inelastic scattering, this feature manifests itself 
(as displayed)
via the geometric scaling property: instead of being a function of 
$Q^2/Q_0^2$ and $x$ separately, the total cross-section is only a function of 
$\tau\!=\!Q^2/Q_s^2(x),$ up to large values of $\tau$.
}
\end{center}
\label{F1gs}
\end{figure}

Parameters of the
dipole cross section of Eq. (\ref{eq:7}) are 
obtained from the analysis of inclusive data, 
and then can be used to predict diffractive cross section in DIS,
and even more processes as we discuss in the next section.
An important aspect of 
 Eq. (\ref{eq:7}), in which $r$ and $x$ are combined
into one dimensionless variable $r Q_s(x)$, is  what is called
geometric scaling, a new scaling property in  inclusive DIS  at small $x$.
In Ref. \cite{dipole_2golec}, it has been shown to be valid for the
total cross section (see Fig. \ref{F1gs}).

It happens that
 diffraction in DIS is an ideal process to study parton saturation  since 
 this process is
especially sensitive to the large dipole contribution, $r>1/Q_s(x)$ \cite{dipole_2golec}.
Unlike inclusive DIS, the region below, $r<1/Q_s(x)$, is suppressed by an additional power of $1/Q^2$.
This makes obviously diffractive interactions very important 
for tracting saturation effects.
As already mentioned,
the dipole cross section with saturation (see Eq. (\ref{eq:7})) leads in a natural way
to the constant ratio  (up to logarithms)
\begin{equation}
\label{eq:8}
\frac{\sigma^{diff}}{\sigma^{tot}} \sim \frac{1}{\ln(Q^2/Q^2_s(x))}\,.
\end{equation}

We can present very simply the main elements of the calculation
that bring this result. 
Indeed, the photon wave function, in Eq. (\ref{eq:5}), favors
small dipoles  (small $r \sim 1/Q$), which gives
$$
\frac{d\,\sigma^{diff}}{dt}_{\mid\, t=0}
=
\frac{1}{16\,\pi}\,
\int d^2 r\, dz\,
|\Psi^\gamma(r,z,Q^2)|^2\ \hat\sigma^2(x,r)
\sim \frac{1}{Q^2} \int_{1/Q^2}^{\infty} \frac{dr^2}{r^4}  \hat\sigma^2(x,r).
$$

On the other hand, the dipole cross section favors relatively large
dipoles, with $\hat\sigma(x,r) \sim r^2$.
However, as discussed above in the building of  Eq. (\ref{eq:7}),
at sufficiently high energy, saturation cuts off the large
dipoles already on the semi-hard scale $1/Q_s$. This leads to
\begin{equation}
\frac{d\,\sigma^{diff}}{dt}_{\mid\, t=0}
\sim \frac{1}{Q^2} \int_{1/Q^2}^{1/Q_s^2} \frac{dr^2}{r^4}  (r^2 Q_s^2)^2 \sim \frac{Q_s^2(x)}{Q^2}
\propto x^{-\lambda}
\label{edmond}
\end{equation}
and it follows immediately that $\frac{\sigma^{diff}}{\sigma^{tot}}$ is a constant of $x$
at fixed values of $Q^2$.
This result is illustrated experimentally  at the beginning of
this review, in Fig. \ref{fig8}.

\begin{figure}[htbp]
   \centering
   \epsfig{file=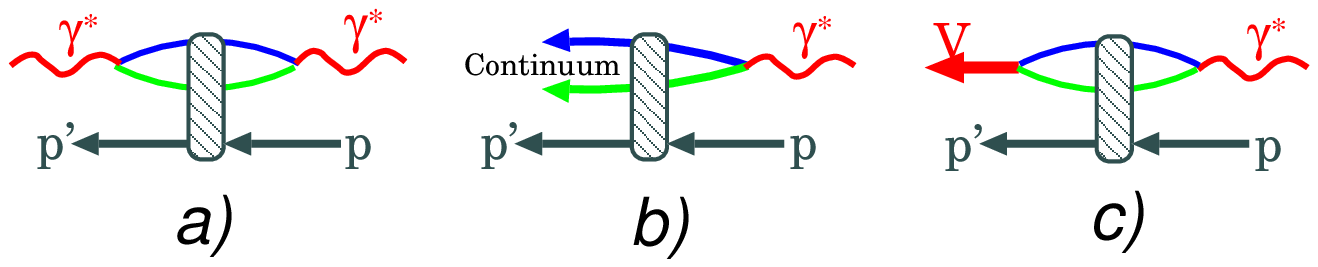,width=150mm}
   \caption{    The unified picture of Compton scattering,
diffraction excitation of the photon into hadronic continuum
states and into the diffractive vector meson
\label{figunified}}
\end{figure}

\begin{figure}[htb]
\begin{center}
\epsfig{file=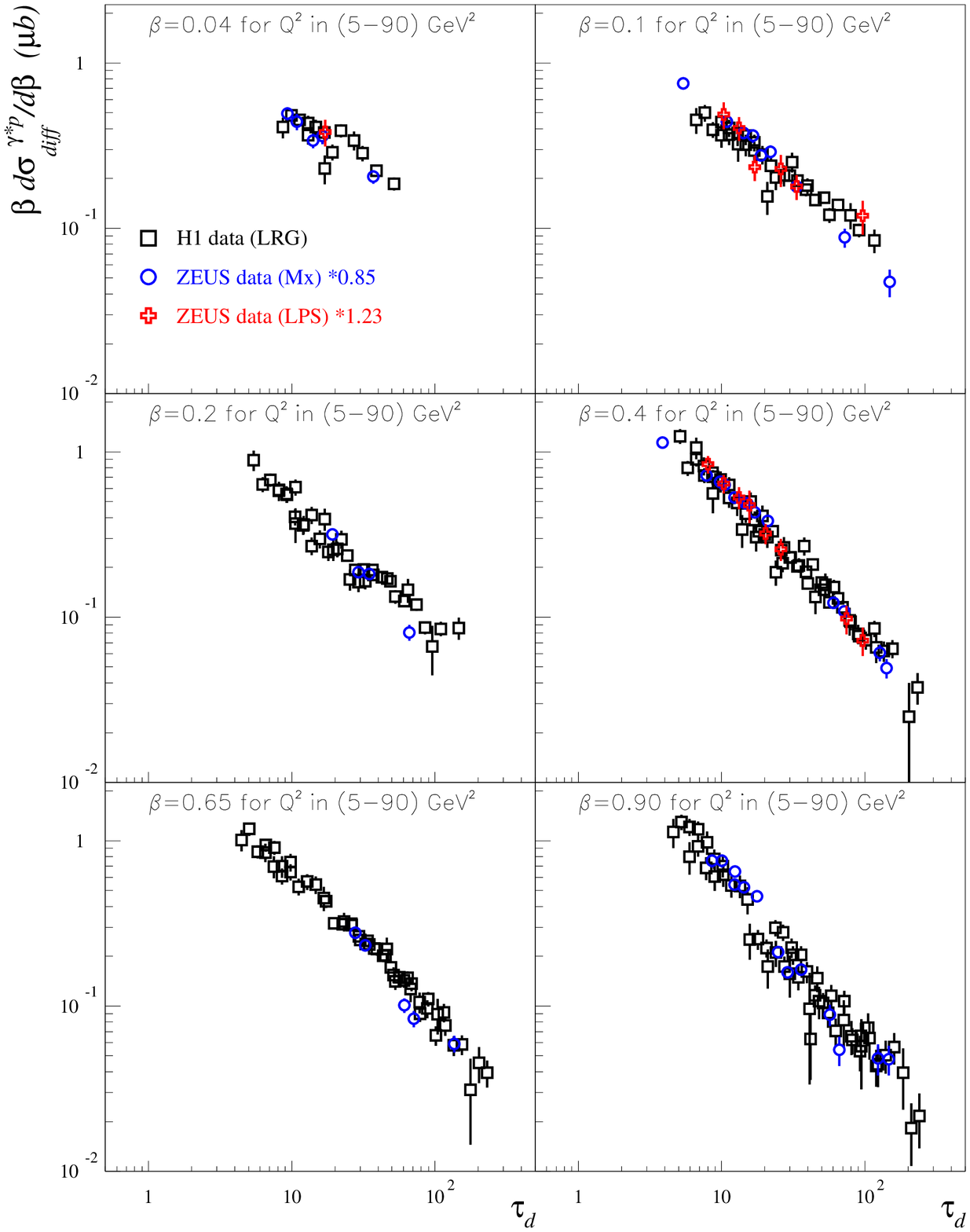,width=13cm}
\caption{\label{figlsmarquet}The diffractive cross section
$\beta\ d\sigma^{\gamma^*p\rightarrow Xp}_{diff}/d\beta$ 
from H1 and ZEUS 
measurements, as a function of $\tau_d= Q^2 /Q^2_s(\xpom)$ in bins of $\beta$ for $Q^2$ values in 
the range $[5;90]\ \mbox{GeV}^2$ and for $\xpom\!<\!0.01$ \cite{marquet}.
The geometric scaling is confirmed with a good precision. }
\end{center}
\end{figure}

With Eq. \ref{eq:7},
we have also introduced  above an interesting consequence 
of the dipole model
for the total cross section,
the geometric scaling property.
Namely,
the total cross section does
not depend on $x$ and $Q^2$ independently but 
can be expressed as a function of  a single variable $\tau= Q^2 /Q^2_s(x)$ \cite{dipole_2golec}.
This property has also been shown recently to be verified 
under minimal assumptions for all
diffractive processes  \cite{marquet} (see Fig. \ref{figlsmarquet}).
The experimental confirmation of this relation
 is  an interesting
piece of evidence that saturation effects are (already) visible in the
inclusive diffractive DIS data.
Extensions of these ideas at non-zero $t$ values, 
rooted on fundamental grounds, have
also been recently derived \cite{dipole_2bis}. This provides 
essential perspectives to understand the transverse degrees
of freedom which are discussed in the next sections.

\subsection{  Towards a common description of all diffractive processes }

Let us mention that
one of the great interest of the dipole model in its two-gluon exchange formulation
is that it provides a unified description of 
different processes measured in $\gamma^* p$ collisions at HERA:
inclusive $\gamma^* p \rightarrow X$, diffractive $\gamma^* p \rightarrow X  \ p'$
and (diffractive) exclusive vector mesons (VM) production 
$\gamma^* p \rightarrow VM \ p'$
(see Fig. \ref{figunified}). 
In the last case, the step (3) described 
in Fig. \ref{f2dipoletot} consists
in the recombination of
the scattered pair $q\bar{q}$ onto a real VM (as $J/\Psi$, $\rho^0$, $\phi$,...)
\cite{rho,phi,jpsi,upsilon} or onto 
a real photon for the reaction $\gamma^* p \rightarrow \gamma \ p'$.
This last process is called deeply virtual Compton scattering (DVCS)
\cite{dvcsh1,dvcszeus}.
Also, we understand immediately the fundamental interest of
exclusive VM production to clarify the generic mechanism of diffractive DIS.
Indeed, the scales involved in the VM process
can act as triggers to isolate when the virtual photon
fluctuates mainly into small-size (small $r$) $q\bar{q}$ pair configurations,
or mainly into large-size configurations.
For example, a small-size $q\bar{q}$ dipole is most likely to be
produced if the virtual photon is polarized longitudinally or if the
dipole is built with heavy quarks. Therefore, we can already 
state that exclusive $J/\Psi$ production is a good candidate for
a hard diffractive process fully calculable in perturbative QCD.
We discuss completely these  ideas in the next section.

\vfill
\clearpage

%
%
\section{  Exclusive particle production at HERA}

\subsection{ Triggering the generic mechanism of diffractive production}

\begin{figure}[htbp]
\centerline{\includegraphics[width=0.5\columnwidth]{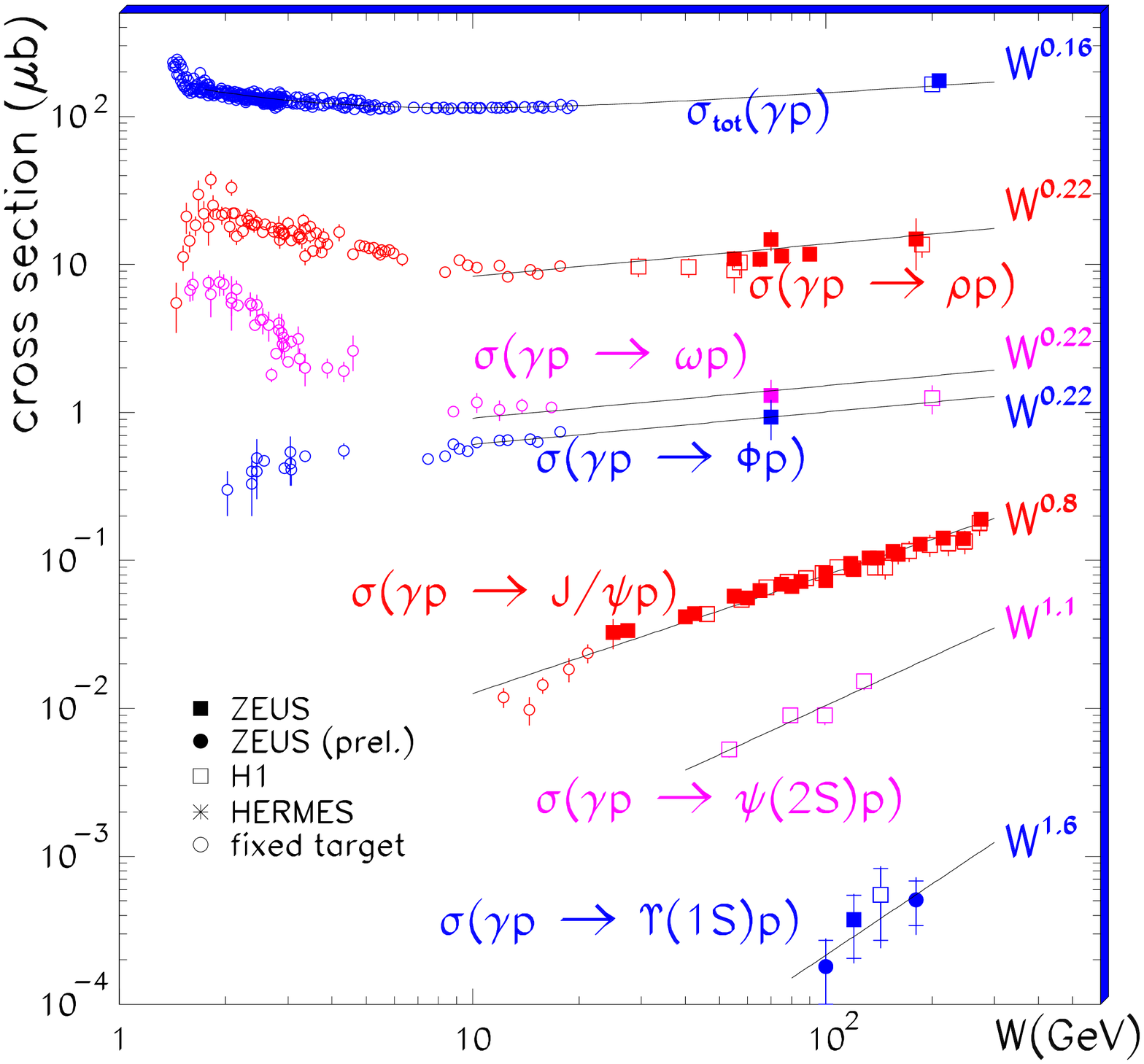}}
\caption{ 
$W$ dependence of the exclusive vector meson cross section in
photoproduction, $\sigma(\gamma p \to V p)$. The total photoproduction
cross section is also shown. The lines are the fit result of the
form $ W^\delta$ to the high energy part of the data.}
\label{fig:sigvm}
\end{figure}

There is a long experimental and theoretical history to the study of
vector meson production, revived  with the
advent of HERA. On the experimental side, the important result is that the cross
sections for exclusive vector meson production rise strongly with
energy (if a hard scale is
present) when compared to fixed target experiments. 
A compilation of experimental 
measurements are shown in
Fig. \ref{fig:sigvm} \cite{rho,phi,jpsi,upsilon}. 
We observe some statements mentioned
briefly at the very end the previous section. 
For example, for  $J/\psi$ exclusive production,
the $W$ dependence of the cross section is typical 
of a hard process. Indeed,
the mass of the $J/\psi$ plays the role of the large scale,
which mainly triggers
small-size (small $r$) $q\bar{q}$ pair configurations
of the initial virtual photon, which then build the hard process.
If we follow the 
discussion of the previous section,
we can also easily write the above argument at a quantitative level.
Indeed, VM cross sections in the (hard) perturbative regime,
$\gamma^* p \rightarrow VM \ p'$, depend on the square of
the gluon density in the proton.  
A first approximation of the cross section can then be written as
\begin{equation}
  \left| \frac{d\sigma} {dt}\right|_{t=0}
     (\gamma^*N\rightarrow VN) 
   = {4\pi^3\Gamma_V m_V \alpha_s^2(Q) 
  \frac{ \eta_V^2\left(xg(x, Q^2)\right)^2 }{ 
    3\alpha_{\rm em} Q^6}} \ , 
    \label{jpsieq}
\end{equation}
where the dependence on the meson structure 
is in the parameter
\begin{equation}
\eta_V = \frac{1}{2} \int \frac{dz}{z(1-z)} \phi^V(z)
     \left( \int dz \phi^V(z) \right)^{-1} 
\end{equation}
and $\phi^V(z)$ is the leading-twist light-cone 
wave function. 

Fig. \ref{fig:sigvm} presents also interesting features that
we can  comment at this level of the discussion.
It shows the transition from soft to hard
processes, using the mass of the VM as a trigger.
From
the lightest one, $\rho^0$, up to the
$\Upsilon$, Fig.~\ref{fig:sigvm} shows
$\sigma(\gamma p \to V p)$ as a function of $W$. 
For comparison, the total photoproduction cross
section, $\sigma_{tot}(\gamma p)$, is also shown. The data at high $W$
can be parameterized as $W^\delta$, and the value of $\delta$ is
displayed in Fig. \ref{fig:sigvm} for each reaction. One sees clearly the
transition from a shallow $W$ dependence for low mass VM (soft) to a steeper
one as the mass of the VM increases (hard)
\cite{rho,phi,jpsi,upsilon}.

\begin{figure}[htbp]
\includegraphics[width=0.5\textwidth]{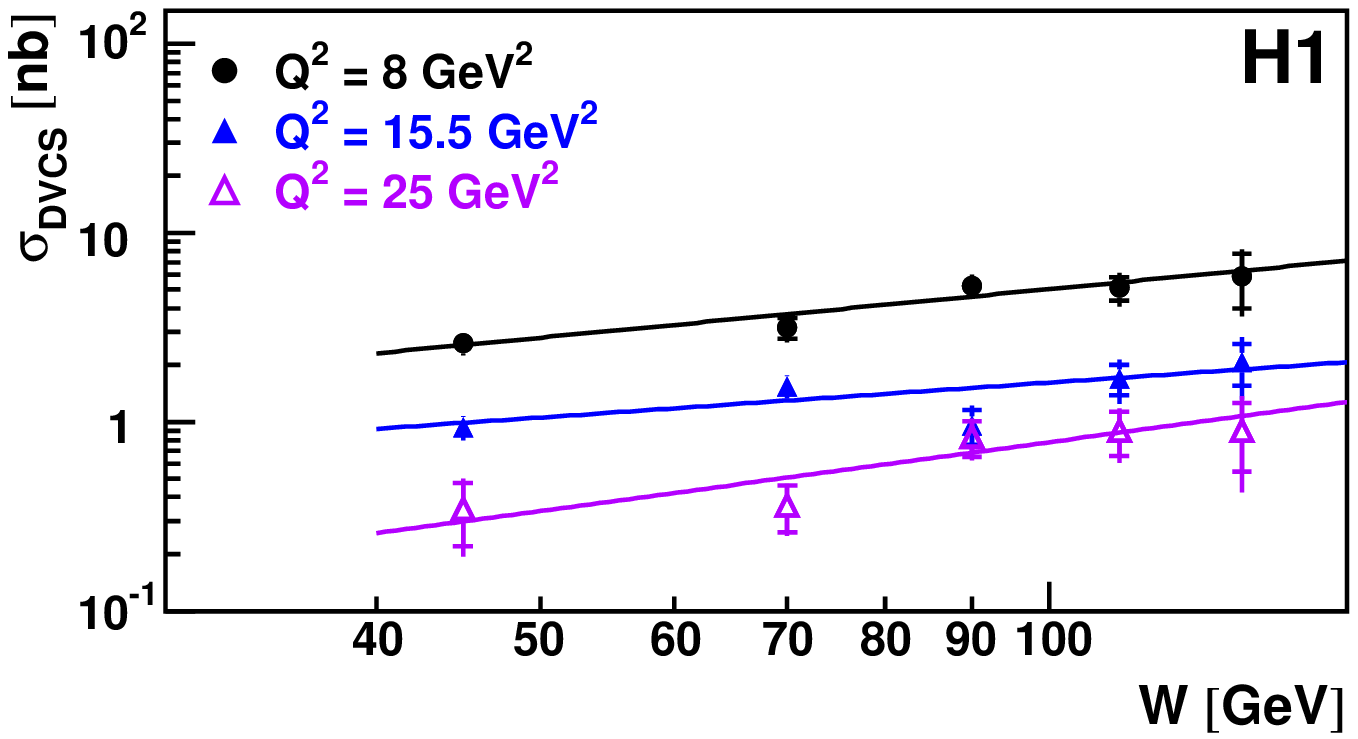}
\includegraphics[width=0.45\textwidth]{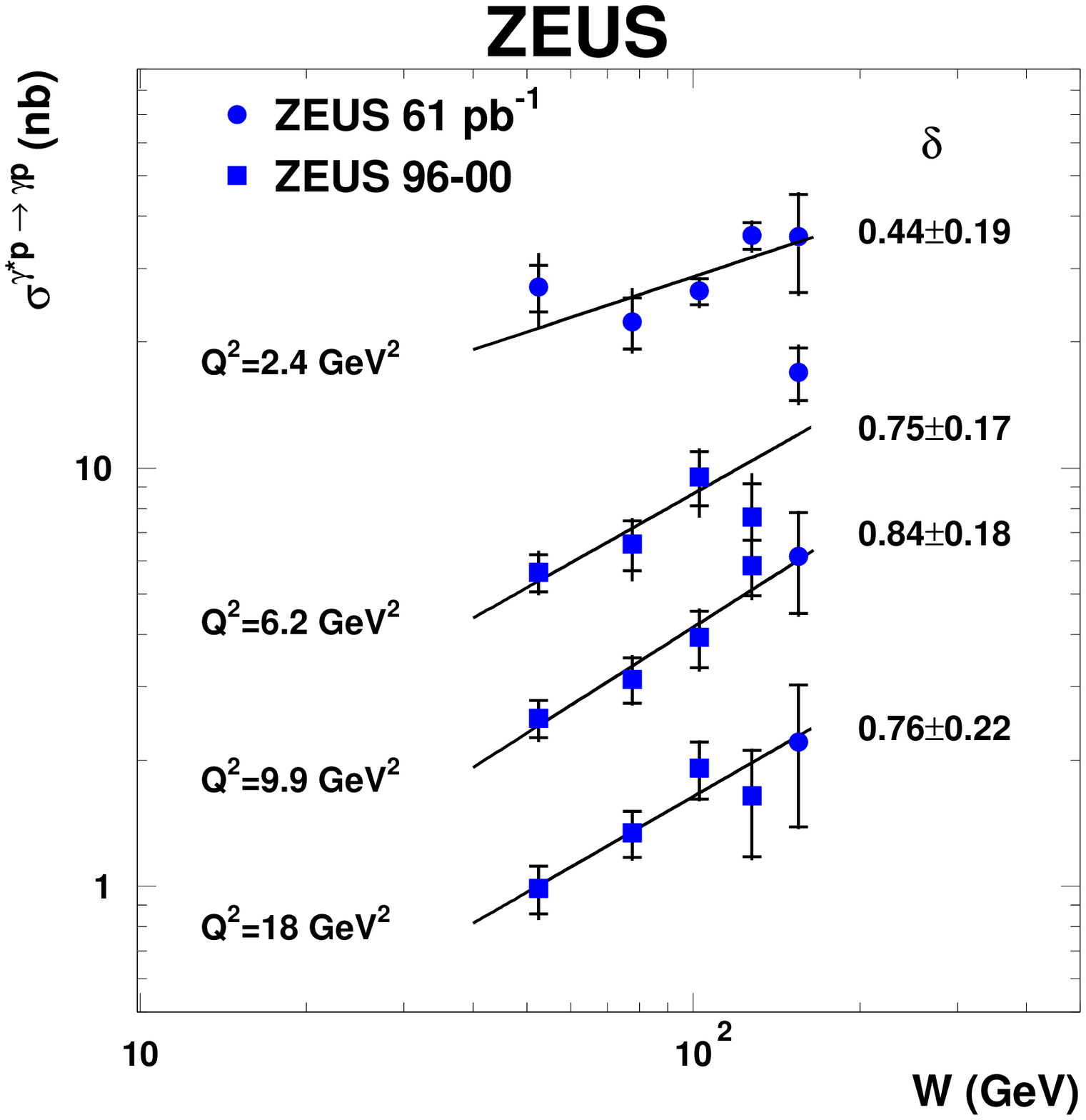}
\caption{
The DVCS cross section, $\sigma^{\gamma ^*p\rightarrow \gamma p}$, as a function of 
$~W$ for  different $Q^2$ values. 
The solid lines are the results of a fit of the
form $\sigma^{\gamma ^*p\rightarrow \gamma p} \propto W^\delta$. 
The values of $\delta$ and their statistical uncertainties are given in the figure.  
}
\label{fig:wfit}
\end{figure}

An interesting phenomenon is observed for the DVCS cross section
(see Fig. \ref{fig:wfit}), which
presents
the same hard $W$ dependence as for the $J/\psi$ \cite{dvcsh1,dvcszeus},
with a (zero mass) photon in the final state.
It does not seem to follow the logic of the above argument and
we come back later of this point.
Obviously, Fig.~\ref{fig:sigvm} displays only one aspect of 
the problem, using the mass of the VM as the scale trigger for
the soft-hard diffractive process.
It is  clear that
the scale $Q^2$ is also particularly
well suited, always for the 
exclusive electroproduction of light vector mesons \cite{rho,phi,jpsi,upsilon}
and DVCS \cite{dvcsh1,dvcszeus}. 
The soft-hard transition 
can be observed experimentally in different ways when varying $Q^2$:
\begin{itemize}
\item[(1)]
In the change of the logarithmic derivative $\delta$ of the
process cross section $\sigma$ with respect to the $\gamma^* p$ center-of-mass
energy $W$ ($\sigma \sim W^\delta$). We expect a variation  
from a value of about 0.2 in the soft regime (low $Q^2$ values)
to 0.8 in the hard one (large $Q^2$ values).
\item[(2)]
In the decrease of the exponential slope $b$ of
the differential cross section with respect to the
squared-four-momentum transfer $t$ ($d\sigma/dt \sim e^{-b|t|}$), 
from a value of about 
10~GeV$^{-2}$ to an asymptotic value of about 5~GeV$^{-2}$ when the
virtuality $Q^2$ of the photon increases.
\end{itemize}

\begin{figure}[htbp]
\centerline{\includegraphics[width=0.5\columnwidth]{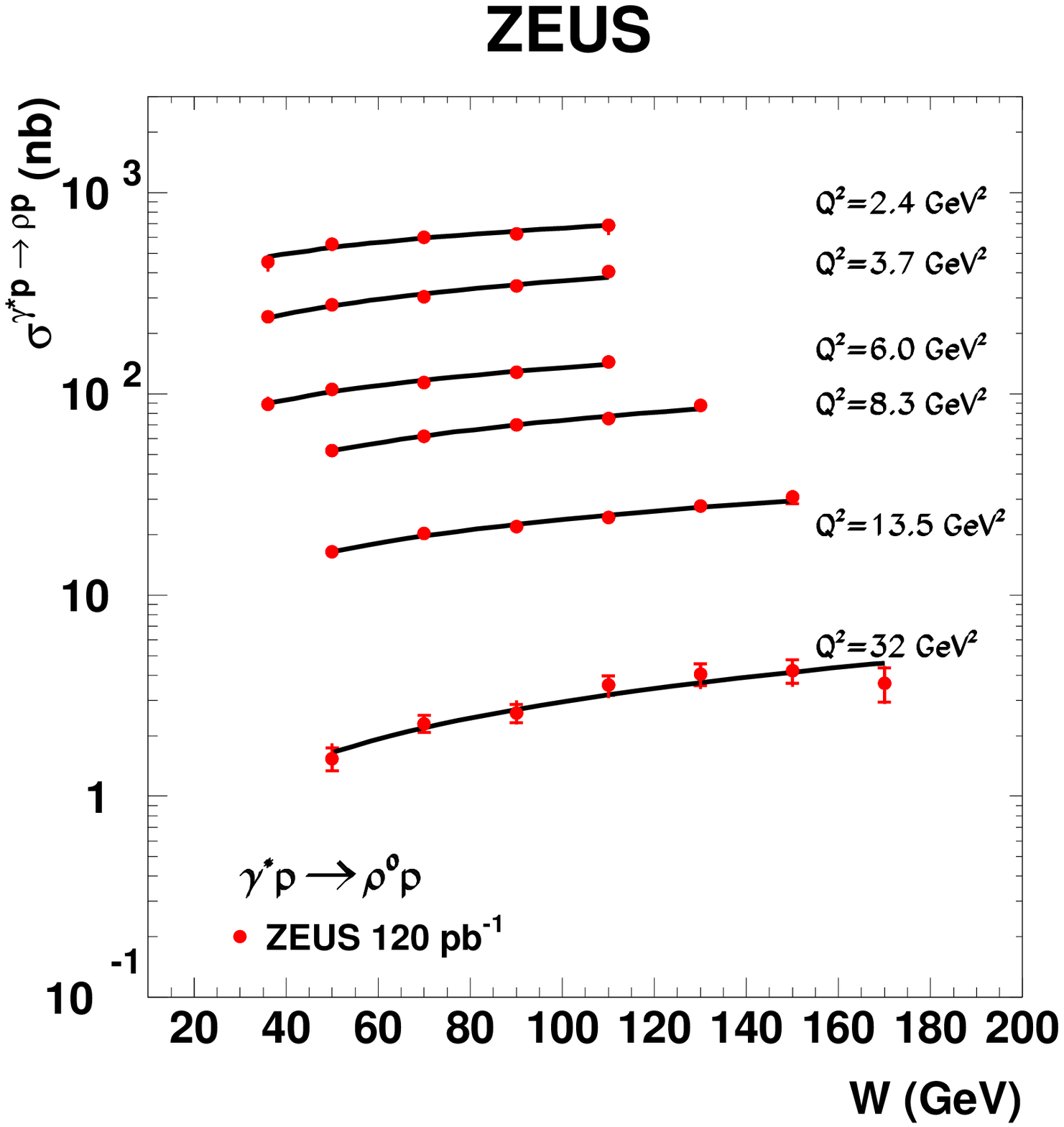}}
\caption{ 
$W$ dependence of the cross section for exclusive $\rho^0$
electroproduction, for different $Q^2$ values, as indicated in the
figure.  The lines are the fit results of the form $ W^\delta$
to data.}
\label{fig:w}
\end{figure}

We illustrate this procedure on recent data on 
$\rho^0$ production \cite{rho}.
The cross section $\sigma (\gamma^* p \to \rho^0 p)$ is
presented in Fig.~\ref{fig:w} as a function of $W$, for different
values of $Q^2$. The cross section rises with $W$ in all $Q^2$ bins.
The same conclusion holds for DVCS, as already discussed and shown in Fig. \ref{fig:wfit}
\cite{dvcsh1,dvcszeus}.

\begin{figure}[htbp]
\centerline{\includegraphics[width=0.65\columnwidth]{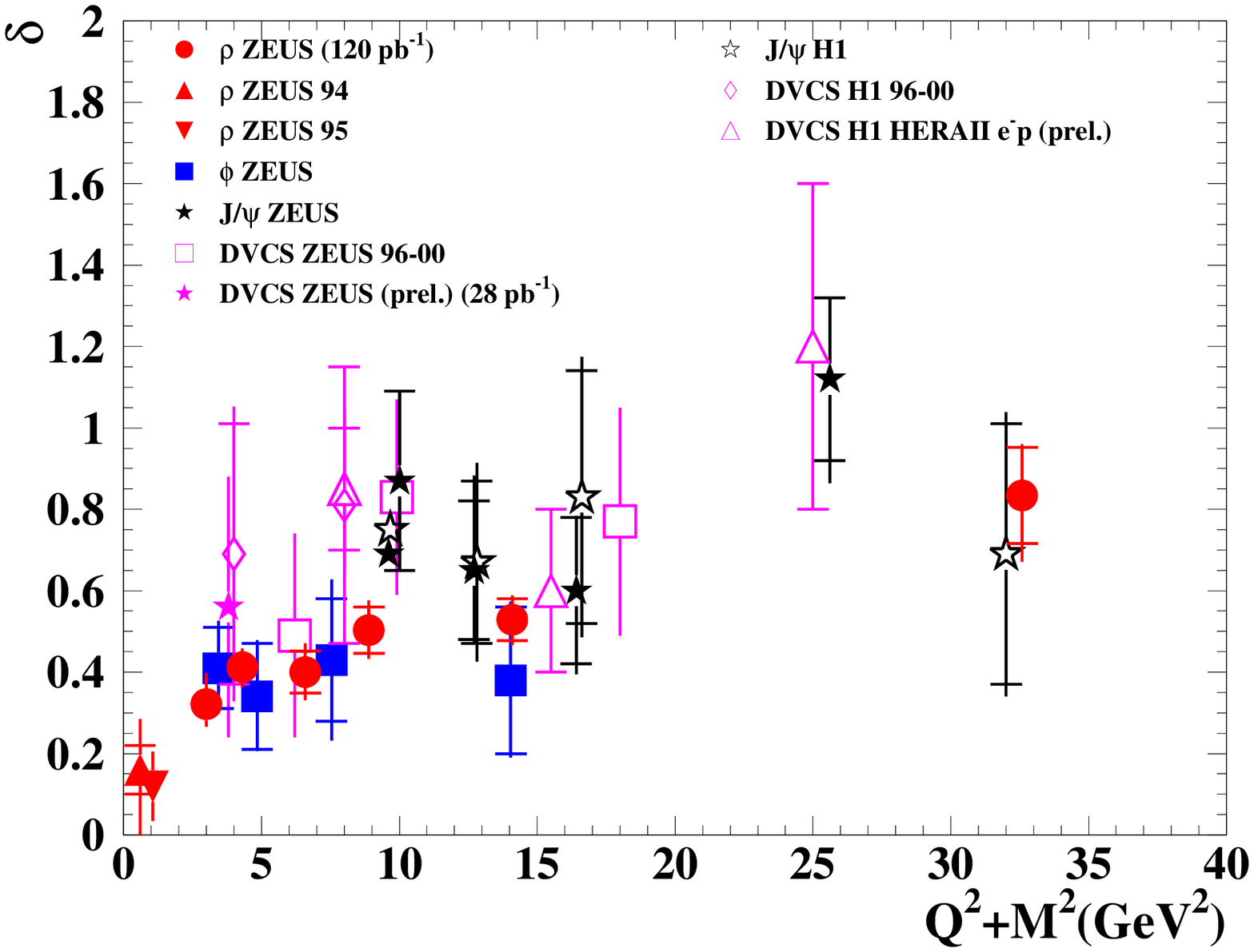}}
\caption{ 
A compilation of values of $\delta$ from fits of the form
$W^\delta$ for exclusive VM electroproduction, as a function
of $Q^2+M^2$. It includes also the DVCS results.}
\label{fig:del07}
\end{figure}

A compilation of 
values of $\delta$ from DVCS and VM measurements are
presented in Fig~\ref{fig:del07}.  
Results are plotted as a function of
$Q^2+M^2$, where $M$ is the mass of the vector meson
(equal to zero in case of DVCS).  We observe a
universal behavior, showing an increase of $\delta$ as the scale
becomes larger. The value of $\delta$ at low scale is the one expected
from the soft Pomeron intercept, while the one at large
scale is in accordance with twice the logarithmic derivative of the
gluon density with respect to $W$.

\subsection{ Deeply virtual Compton scattering}

\begin{figure}[htb]
 \begin{center}
  \epsfig{figure=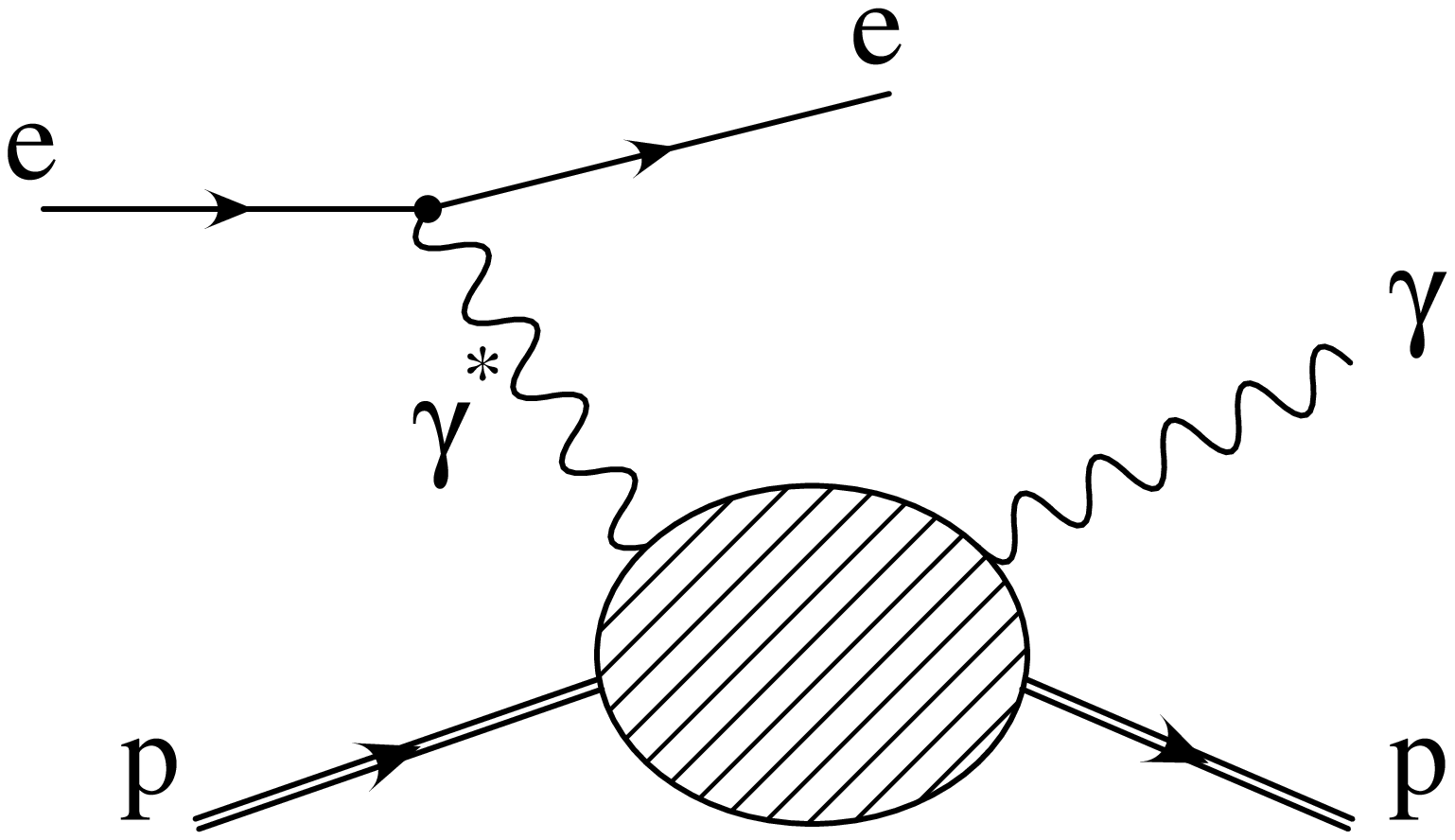,height=0.22\textwidth}
  $\qquad$
  \epsfig{figure=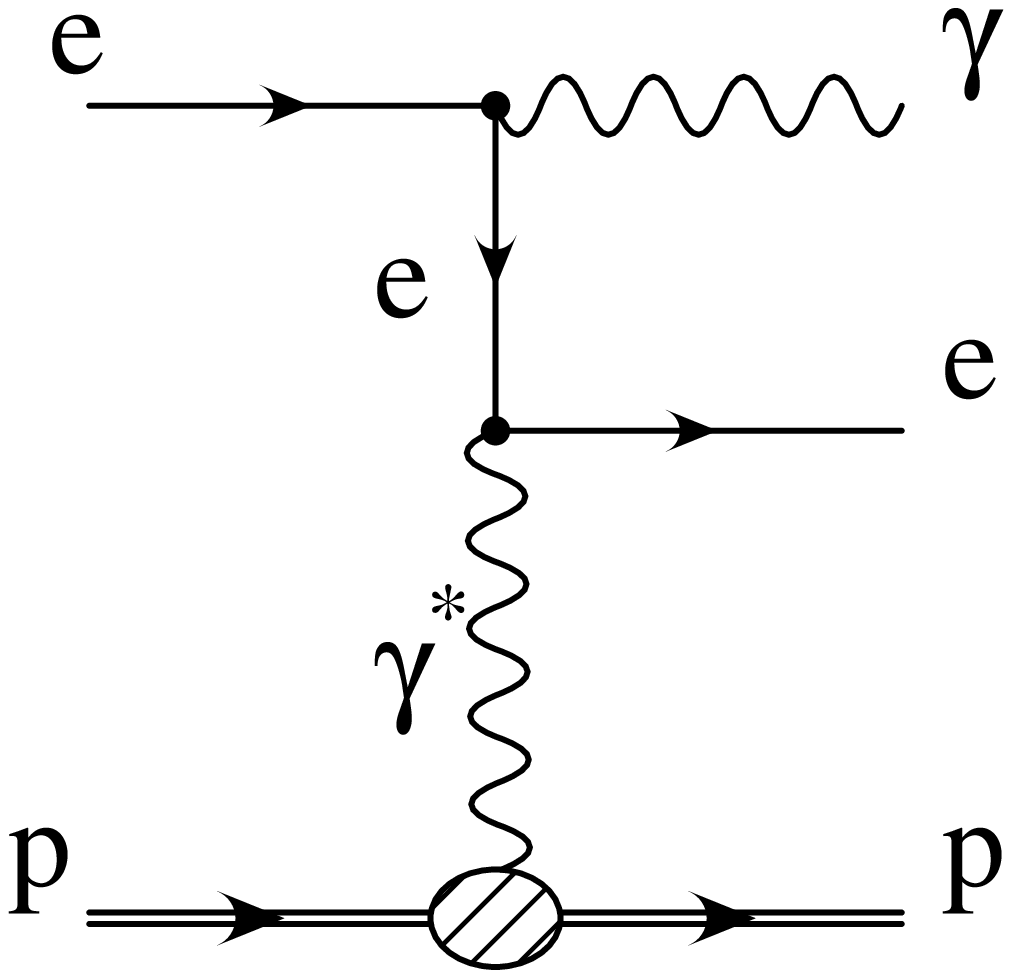,height=0.22\textwidth}
  $\qquad$
  \epsfig{figure=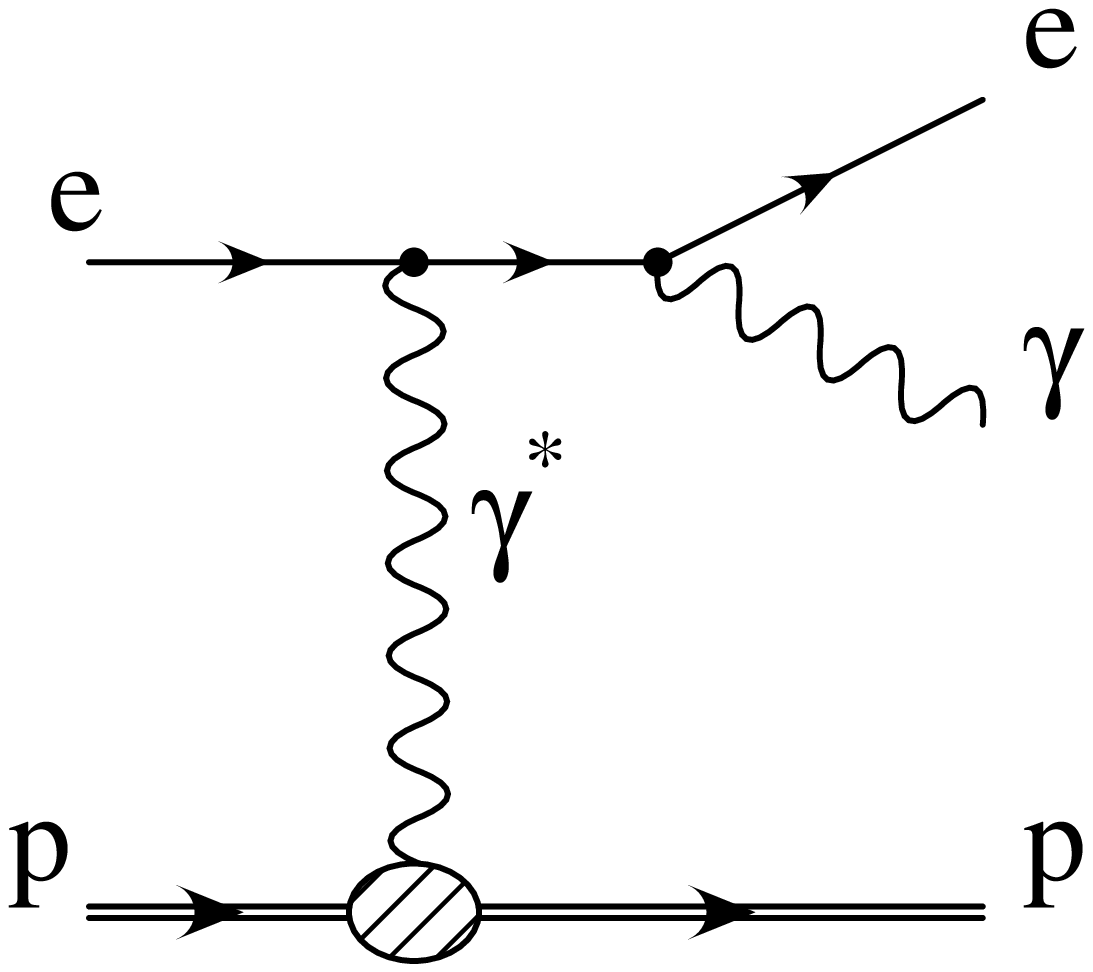,height=0.22\textwidth}
  \\
  \begin{picture}(0,0)(100,0)
  \end{picture}
  \caption{ Diagrams illustrating the DVCS (left) and the Bethe-Heitler 
    (middle and right) processes.
The reaction studied receives contributions from both the DVCS 
process, whose origin lies in the strong interaction, and the purely 
electromagnetic Bethe-Heitler (BH) process, 
where the photon is emitted from the positron. 
The BH cross section can be precisely calculated  in QED 
using elastic proton form factors.
    }
  \label{fig:bh}
 \end{center}
\end{figure}

Let us comment in more details the analysis of the DVCS signal, which
we discuss in a different context in further sections.
The DVCS process,  $ep \rightarrow ep \gamma$,
also receives a contribution from the purely 
electromagnetic Bethe-Heitler (BH) process, where the photon is emitted from the electron,
as displayed in Fig. \ref{fig:bh}.

\begin{figure}[htbp]
\vspace*{2cm}
\begin{center}
 \includegraphics[totalheight=6cm]{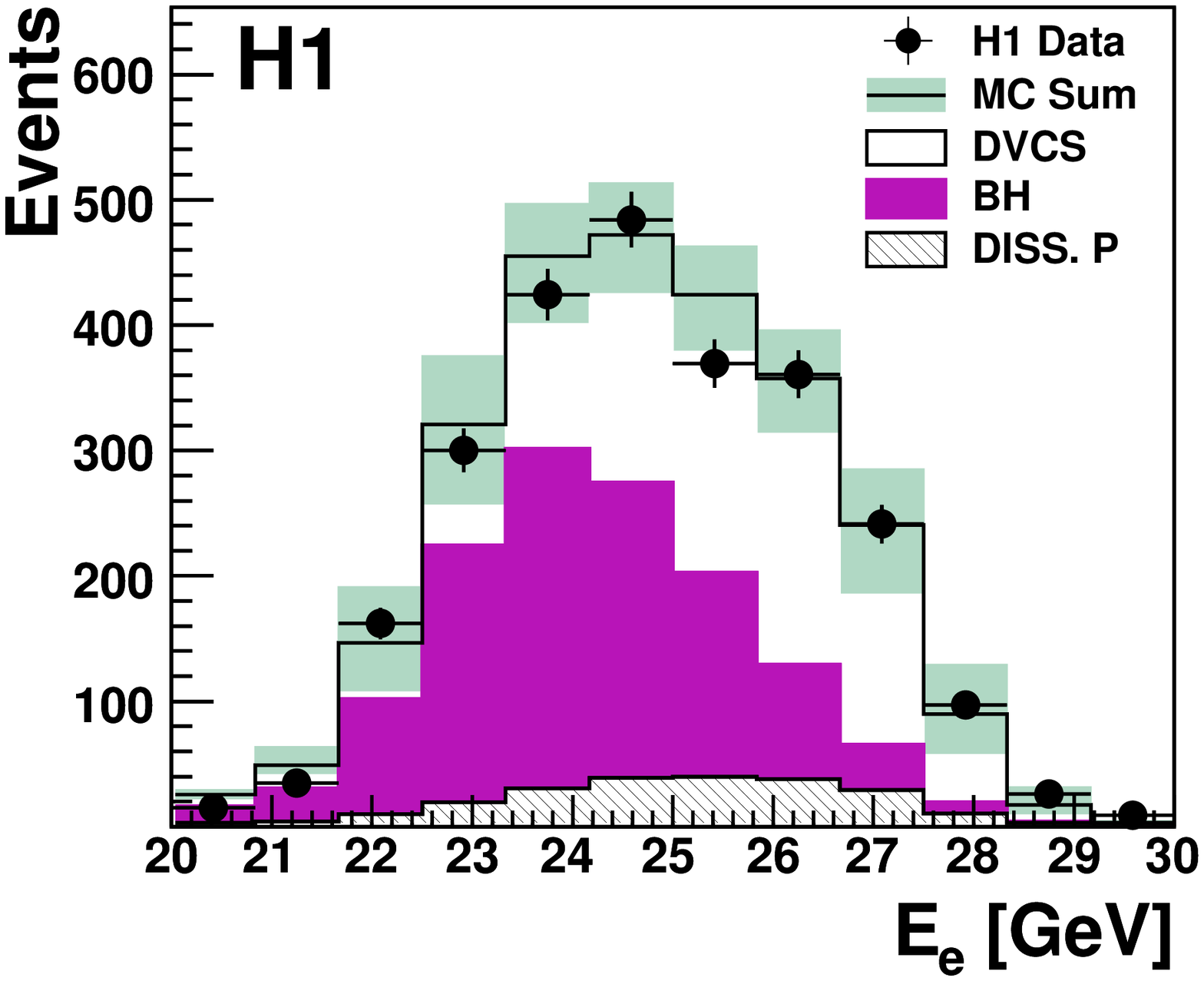}
 \includegraphics[totalheight=6cm]{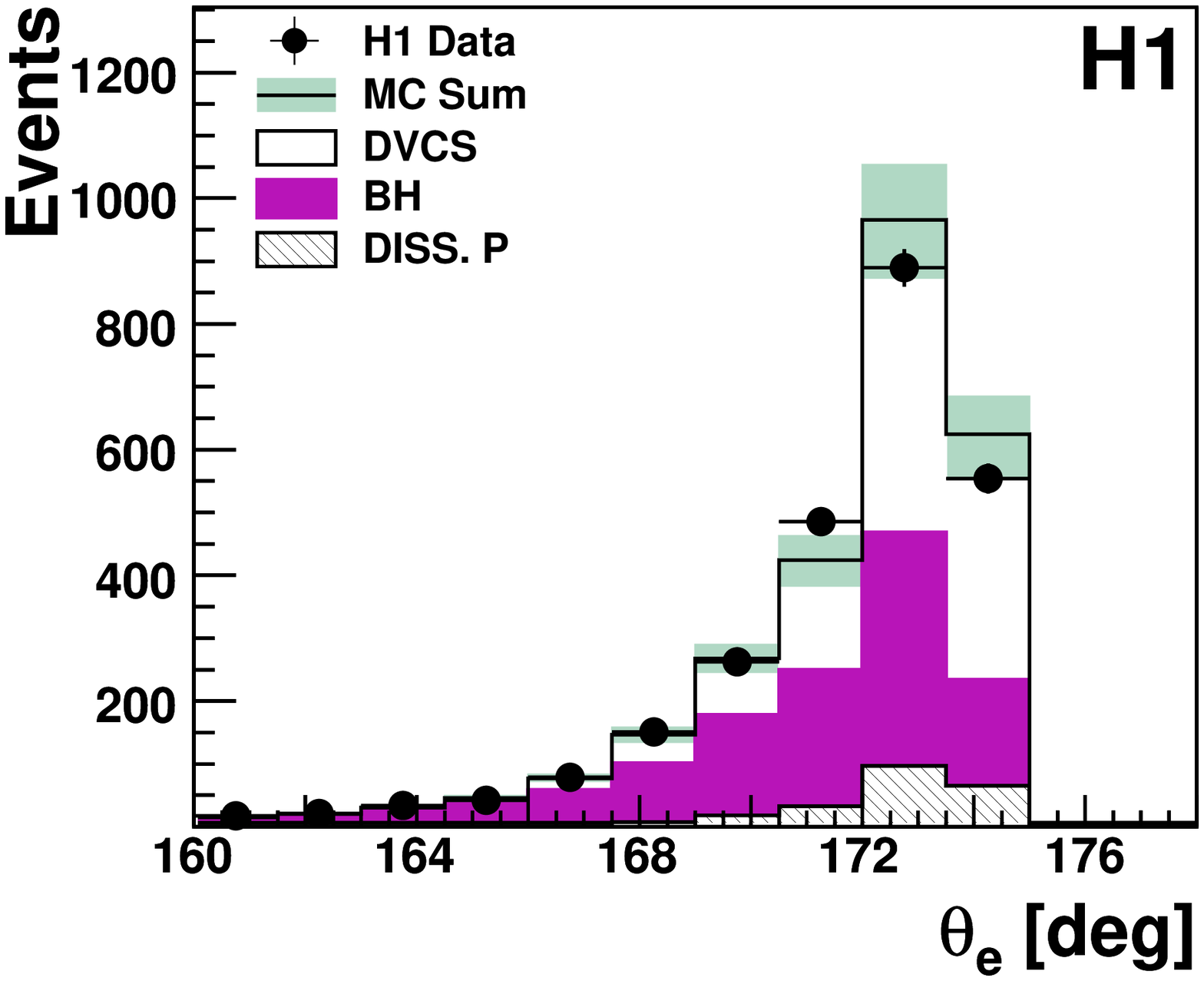}
\end{center}
\caption{\label{figcomp} 
Distributions of the energy  and polar angle of the scattered
electron.
The data are compared with Monte Carlo expectations for 
elastic DVCS, elastic and inelastic BH and inelastic DVCS (labelled DISS. p).
All Monte Carlo simulations are normalized according
to the luminosity of the data.
The open histogram shows the total
prediction and the shaded band its
estimated uncertainty.
}
\end{figure}

Let us notice that the final state for DVCS (QCD process) and BH (QED process)
are identical. This means that both processes interfere, which is of
fundamental interest in the next sections. In this part, we only use the fact that
the BH cross section is precisely calculable in QED
and can be subtracted from the total process rate to extract
the DVCS cross section. Of course, only if the BH contribution is 
not dominating the process rate and if the (integrated) interference 
term is negligible.
Otherwise, the subtraction procedure would be hopeless.
It is the case  at low $x_{Bj}$, and then
for H1 and ZEUS experiments, the DVCS contribution can be measured
directly. Fig. \ref{figcomp} presents the different 
contribution (for the scattered electron variables),
after the experimental analysis
of the reaction $ep \rightarrow ep \gamma$. 

\begin{figure}[t]
\begin{center}
 \includegraphics[totalheight=6.2cm]{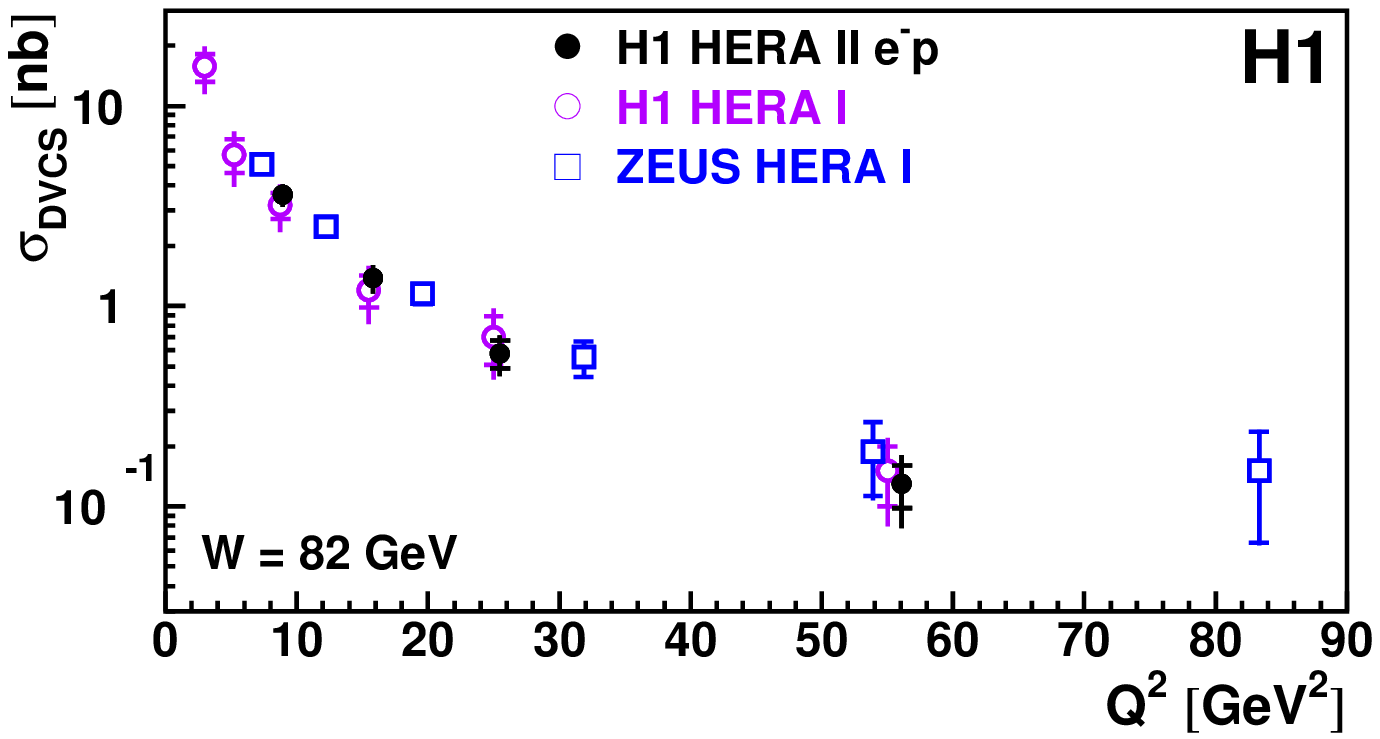}
 \includegraphics[totalheight=6.2cm]{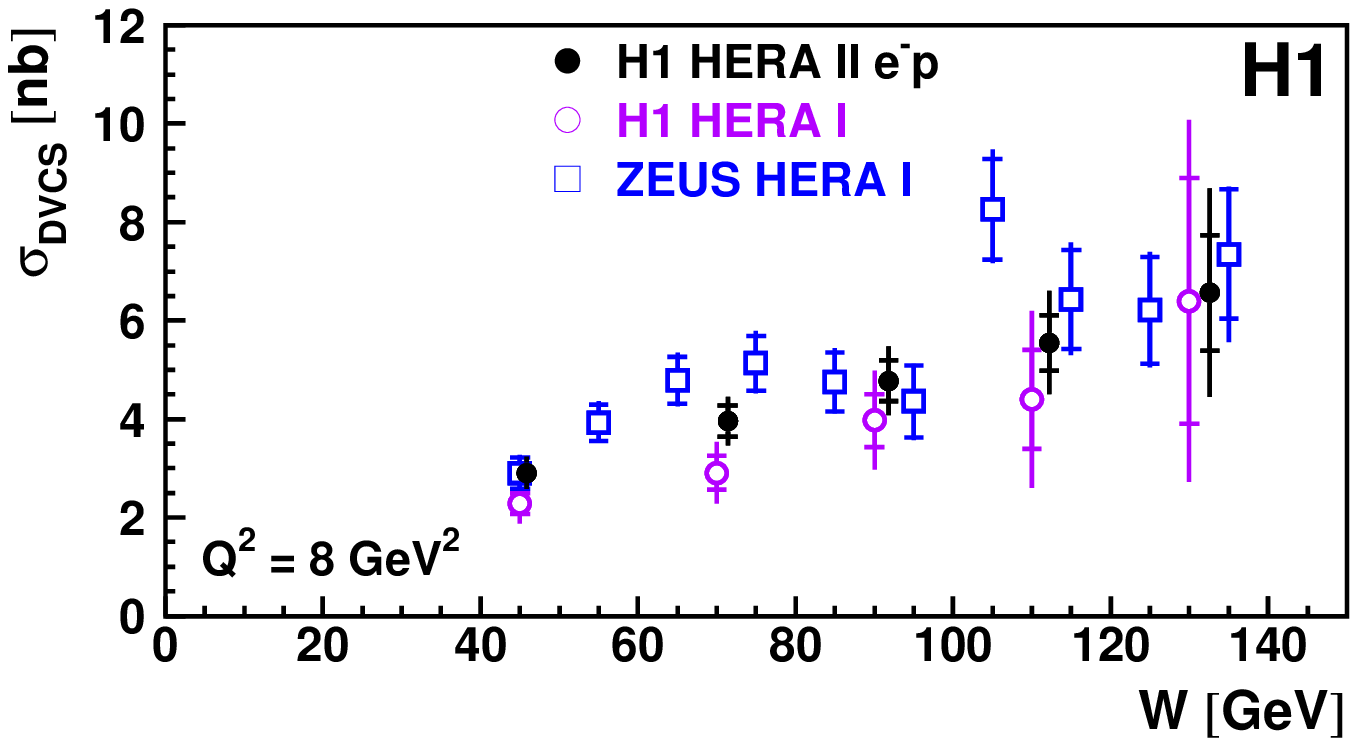}
\end{center}
\caption{\label{fig1d} 
  The DVCS cross section  as a function of
  $Q^2$ at  $W=82$~GeV and as a function of
$W$ at $Q^2=8$~GeV$^2$.
The inner error bars represent the statistical errors, 
the outer error bars the statistical and systematic errors added in quadrature.
}
\end{figure}

We observe that DVCS and BH
contributions are of similar size and thus, the BH contribution
can be subtracted with a systematic uncertainty determined
from a specific experimental study.
In Fig. \ref{fig1d}, we present the DVCS cross sections,
$\gamma^*p \rightarrow \gamma p$, 
obtained
over the full kinematic range of the analysis \cite{dvcsh1,dvcszeus},
as a function of $Q^2$ and $W$. The behavior in $W$ has been 
discussed qualitatively above, it corresponds to the dependence
characteristic for a hard process. The $Q^2$ dependence, measured
to be in $\sim 1/Q^3$ in Fig. \ref{fig1d}, is also understandable
qualitatively.

Indeed, following the discussion of the previous section (see Eq. (\ref{edmond})),
we expect a behavior of the imaginary DVCS amplitude ($\gamma^*p \rightarrow \gamma p$) in
\begin{equation}
Im A \sim \sigma_0 \frac{1}{Q^2}  \int_{1/Q^2}^{1/Q_s^2} \frac{dr^2}{r^4}  (r^2 Q_s^2)
\label{dipoledvcs}
\end{equation}
which leads to a DVCS cross section of the form
$$
\sigma \sim \sigma_0  (\frac{Q_s(x)^2}{Q^2})^2 \sim  \frac{W^{\delta}}{Q^4}
$$
With this expression, we find again the qualitative behavior in $W$.
Interestingly also, the measured $Q^2$ dependence in $\sim 1/Q^3$  
is smaller than expected from this relation.
In fact, to describe qualitatively the observed DVCS cross section,
we must consider a parameterization in 
$$
\sigma \sim \sigma_0  \frac{W^{\delta} [Q^{2}]^\gamma}{Q^4},
$$
after introducing a
term in $[Q^{2}]^\gamma$ in the expression 
of the DVCS cross section.
The term in  $[Q^{2}]^\gamma$ is 
reminiscent from the QCD evolution of the
DVCS amplitude (QCD evolution of the gluon/sea distributions). The experimental 
observation in $\sigma \sim 1/Q^3$ is
compatible with $\gamma \sim 1/2$ (using our notations).
Of course, we do not stay at this qualitative understanding
and
we describe quantitative estimates
of the DVCS cross sections in the following.

A comment is in order concerning the $W$ dependence of DVCS.
It reaches the same value of $\delta$ as in the hard
process of $J/\psi$ electroproduction. Given the fact that the final
state photon is real, and thus transversely polarized, the DVCS
process is produced by transversely polarized virtual photons,
assuming s-channel helicity conservation. 
The steep energy dependence
thus indicates that the large configurations of the virtual transverse
photon are suppressed
and the reaction is dominated by small $q {\bar q}$ configurations
(small dipoles),
leading to the observed perturbative hard behavior.
A similar effect is observed for  $\rho^0$ production \cite{rho}.

\subsection{ Saturation in exclusive processes}

\begin{figure}[t]
\begin{center}

\epsfig{file=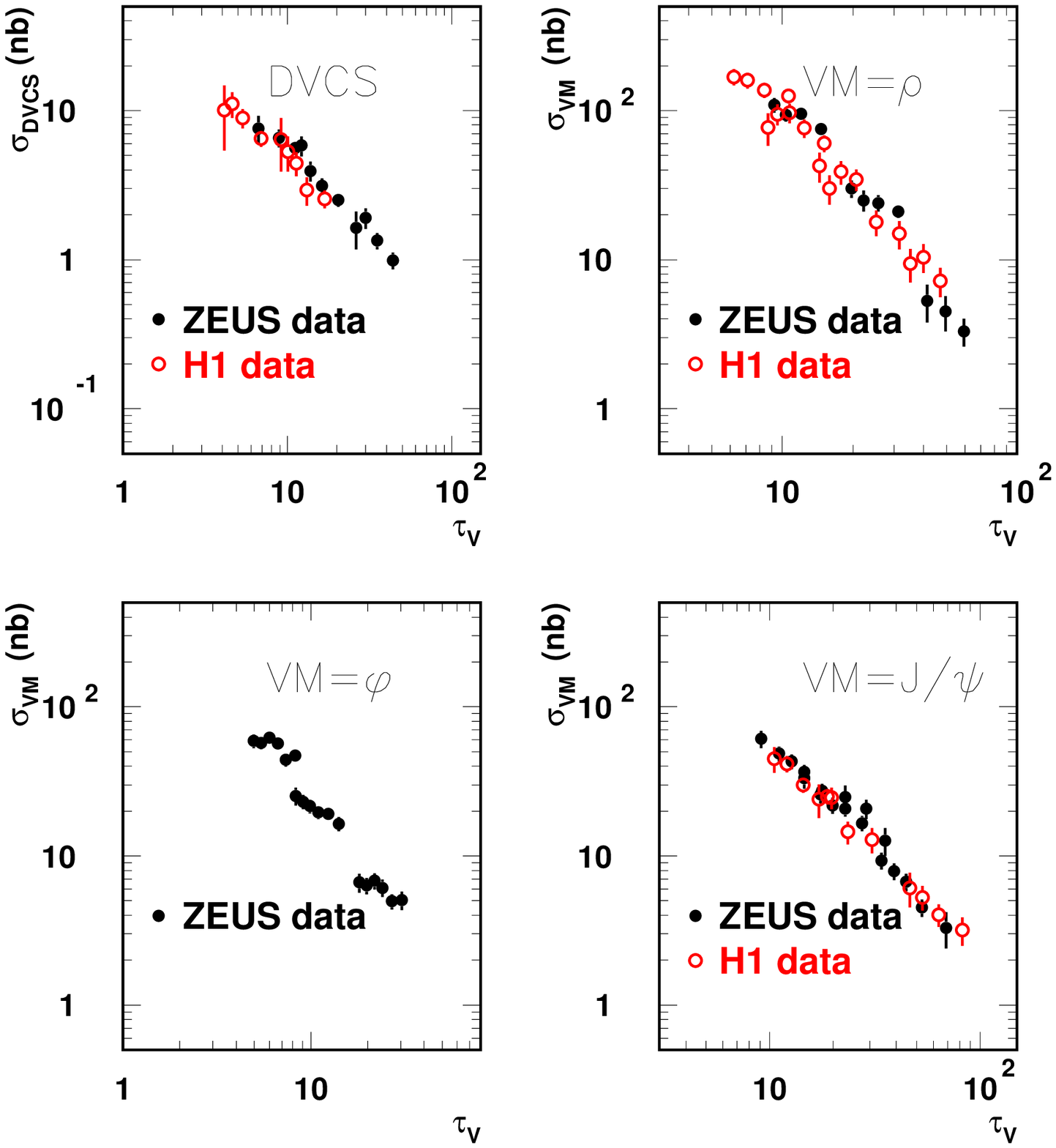,width=10.5cm}
\caption{\label{f3gbwc}
The $\rho$, $J/\Psi$ and $\phi$ production cross-sections 
$\sigma^{\gamma^*p\rightarrow Vp}_{VM}$ and the DVCS cross-section 
$\sigma^{\gamma^*p\rightarrow \gamma p}_{DVCS}$ from H1 and ZEUS 
measurements, as a function of $\tau_V= (Q^2+M_{VM}^2) /Q^2_s(\xpom)$ 
and for $\xpom\!<\!0.01$ \cite{marquet}. Each process verifies the
geometric scaling property.}
\end{center}
\end{figure}

Coming back to the discussion of the previous section about saturation,
we can mention also that
among diffractive interactions, exclusive vector meson production and DVCS
are probably the best processes to study saturation effects 
in DIS since the transverse
size of the $q\bar{q}$ pair forming a meson is controlled 
by the vector meson mass with
$<r> \simeq 1/\sqrt{M_V^2+Q^2}$. Thus we expect saturation 
effects to be more important
for larger (lighter) vector mesons. 
An interesting consequence of this feature  
is illustrated in Fig. \ref{f3gbwc}, where we show that
VM and DVCS process exhibit the property of geometric scaling \cite{marquet}.
This illustrates that this 
qualitative discussion (related to Eq. (\ref{dipoledvcs}))
gives the main elements of understanding of the DVCS and VMs cross sections
dependences.
More generally, as for all other diffractive 
processes presented in this review, it means that the mechanism 
included in the parameterization
of  the dipole cross section of the form written in Eq.
(\ref{eq:7}) is correct and predictive. 

\begin{figure}[ht]
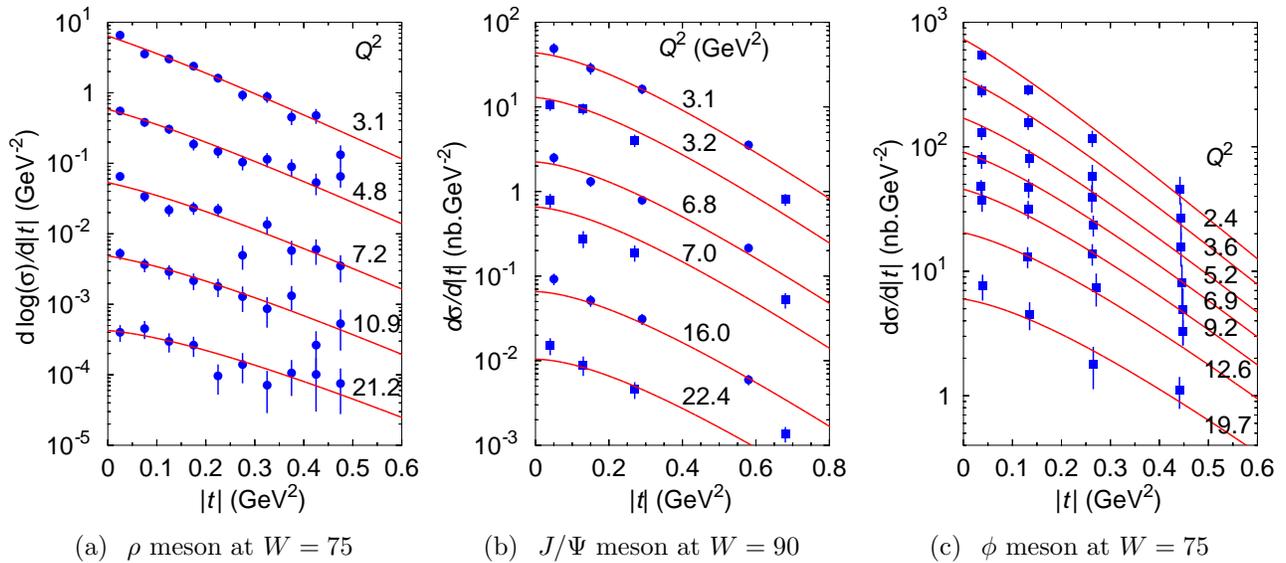

\begin{center}
\subfigure[\ $\rho$ meson at $W=75$]{\includegraphics[width=0.3\textwidth]{dsdt_rho.ps}}
\subfigure[\ $J/\Psi$ meson at $W=90$]{\includegraphics[width=0.3\textwidth]{dsdt_jpsi.ps}}
\subfigure[\ $\phi$ meson at $W=75$]{\includegraphics[width=0.3\textwidth]{dsdt_phi.ps}}
\end{center}
\caption{Fit results for the $\rho$, $\phi$ and $J/\Psi$ 
differential cross-section.}\label{fig:ds_dt}
\end{figure}

In recent works,
it has been shown that dipole models can be extended 
at non-zero $t$ values, with a refined definition
of the saturation scale  \cite{dipole_2bis}.
Then, the geometric scaling property is predicted to
manifest itself in exclusive vector meson production and deeply
virtual Compton scattering (DVCS), also
 at moderate non-zero momentum transfer.
In Fig. \ref{fig:ds_dt}, we  compare data with predictions of Ref. \cite{dipole_2bis}, 
We observe the very good agreement between data
and predictions.

\vfill
\clearpage

%
%
\section{  Nucleon tomography }

\subsection{  $t$ dependence of 
exclusive diffractive processes revisited }

\begin{figure}[htbp]
\begin{center}
\includegraphics[width=8.cm]{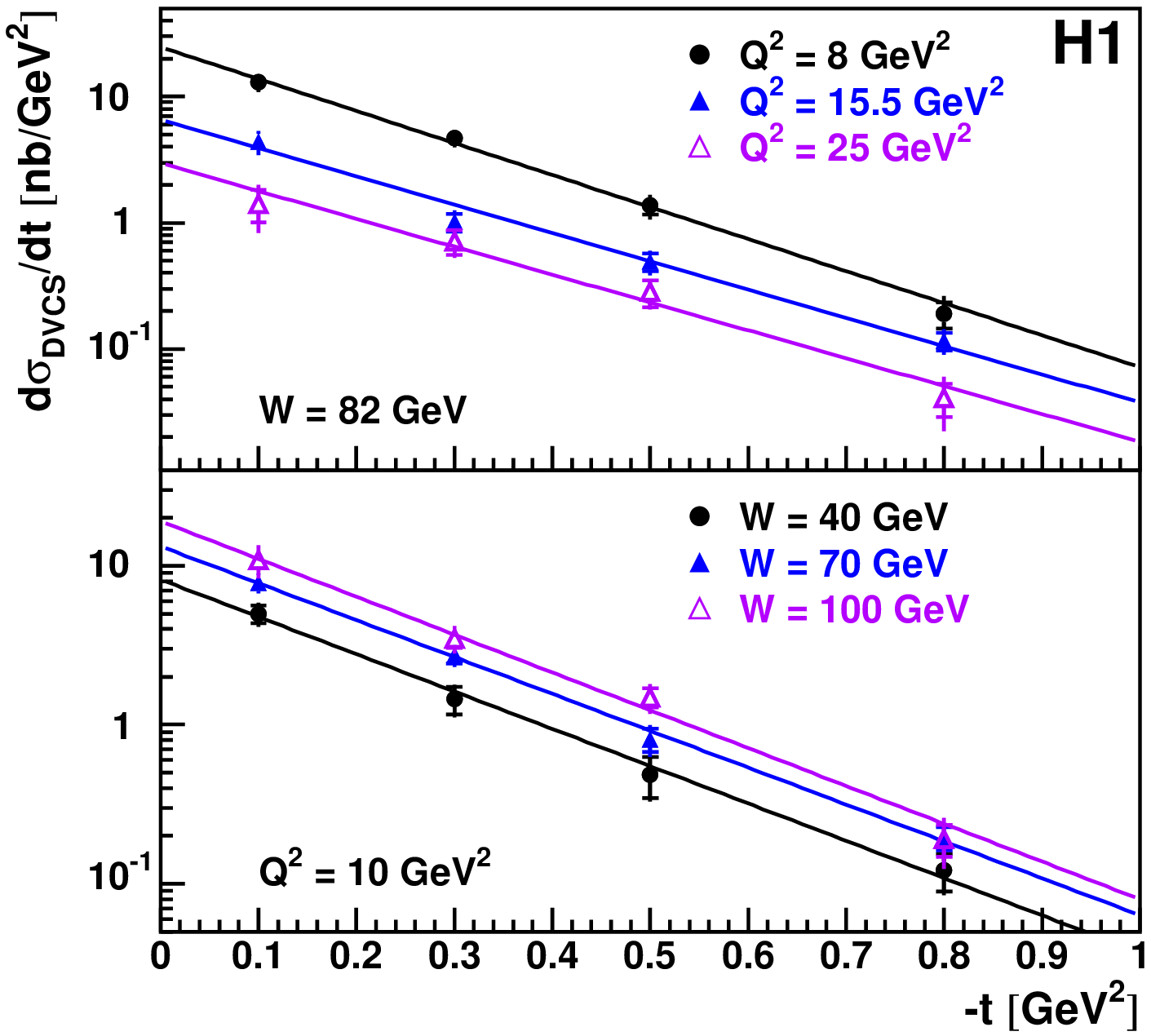}
\end{center}
\vspace*{-0.5cm}
\caption{\label{figbdvcs} 
The DVCS cross section, differential in $t$, 
for three values of $Q^2$ expressed atand 
for three values of $W$.
The solid lines represent the results of fits 
of the form $e^{-b|t|}$.
}
\end{figure}

With $t=(p-p')^2$, the measurement of the 
VM and DVCS cross section, differential in $t$
 is one of the key measurement in exclusive processes.
A parameterization in $d\sigma/dt \sim e^{-b|t|}$, as shown in
 Fig. \ref{figbdvcs}, 
gives a  very good description of measurements.
In addition, in  Fig. \ref{figbdvcs}, we show that 
fits of the form $d\sigma/dt \sim e^{-b|t|}$ can
describe DVCS measurements to a very good accuracy
for different $Q^2$ and $W$ values.
The same conclusions hold in the case of VM production.
That's the reason why we use this parameterization
of the $t$ dependence, with a factorized exponential slope $b$, 
to describe the HERA data on DVCS or VM production
at low $x_{Bj}$. Note that this parameterization
 
\begin{figure}[p]
\vspace{-0.5cm}
\begin{center}
\includegraphics[scale=0.5]{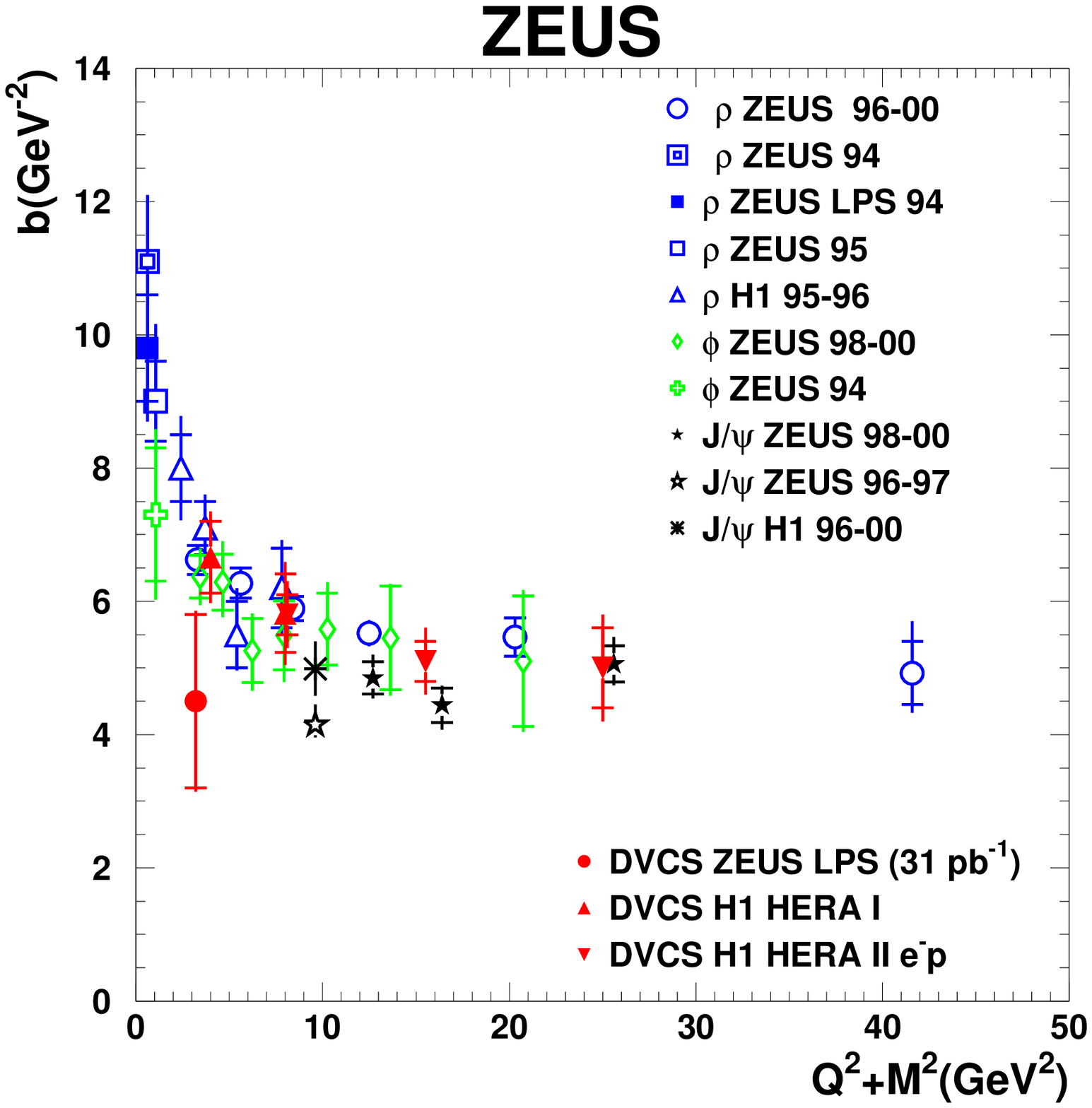}
\end{center}
\vspace{-.5cm}
\caption{
A compilation of the values of the slope b as a function of 
$Q^2+M^2$ for various exclusive 
processes including the present DVCS measurement. 
The inner error bars represent the statistical uncertainty while the outer
error bars the statistical and systematic uncertainties added in quadrature. 
Note that the latest
$t$ slope measurement of DVCS by the ZEUS collaboration
\cite{dvcszeus} is shown. 
It falls below (1 sigma effect)
the H1 measurement at a comparable $Q^2$ value \cite{dvcsh1}.
The main result does not change: at large $Q^2$, exponential
$t$ slopes converge
to a scaling value (see text) ans this is a common
trend for all VM processes.
However, at low $Q^2$ ($Q^2 \simeq 3$ GeV$^2$),
 ZEUS result indicates the absence of effects in $b \sim 1/{Q^2}$
(from the probe) for DVCS, whereas H1 result shows a behavior comparable
to $\rho$ at this $Q^2$, with a clear influence of the probe
to the building of the measured $b$. 
}
\label{fig:bfignew}
\end{figure}

Concerning the interpretation,
we have already briefly mentioned  
the importance of the observation
of the decrease of the exponential slope $b$, 
from a value of about 
10~GeV$^{-2}$ to an asymptotic value of about 5~GeV$^{-2}$, when the
virtuality $Q^2$ of the photon increases
(see Fig. \ref{fig:bfignew}).
The
resulting values of $b$ as a function of the scale $Q^2+M^2$ are
plotted in Fig.~\ref{fig:bfignew}.

A qualitative understanding of this behavior is simple.
Indeed, $b$ is essentially the sum of a component
coming from the probe in $1/\sqrt{Q^2+M_{VM}^2}$ and a 
component related to the target nucleon.
Then, at large $Q^2$ or large $M_{VM}^2$, the $b$ values
decrease to the solely target component.
That's why in Fig.~\ref{fig:bfignew}, we observe that for
large $Q^2$ or for heavy VMs, like $J/\psi$, $b$ is reaching a 
universal value of about $5$ GeV$^{-2}$, scaling with $Q^2$
asymptotically. 
This value is related to the size of the target probed
during the interaction  and we do not 
expect further decrease of $b$ when increasing the scale,
once a certain scale is reached.

To understand this shape of $b(Q^2)$ more quantitatively, 
we need to define a function
that generalizes the gluon density which appears in
Eq. (\ref{jpsieq}) at non-zero $t$ values. That's why, we define a
generalised gluon distribution $F_g$ which depends both on $x$ and $t$
(at given $Q^2$).
From this function, we can compute
a gluon density which also depends on a spatial degree of freedom, 
a transverse size (or impact parameter), labeled $R_\perp$,
in the proton. Both functions are related by a Fourier transform 
$$
g (x, R_\perp; Q^2) 
\;\; \equiv \;\; \int \frac{d^2 \Delta_\perp}{(2 \pi)^2}
\; e^{i ({\Delta}_\perp {R_\perp})}
\; F_g (x, t = -{\Delta}_\perp^2; Q^2).
$$
At this level of the discussion, there is no need to enter into
further details concerning these functions. 
We just need to know that
the functions introduced above define proper (generalized) PDFs,
with gauge invariance
and all the good theoretical properties of PDFs in terms of
operator product expansion.
 In fact, they
 are rooted on fundamental grounds \cite{imaging},
that we develop in further sections
(without heavy formalism). 

\subsection{  Extracting the transverse distribution of the quarks and gluons }

From the Fourier transform relation above, 
the average impact parameter (squared), $\langle r_T^2 \rangle$,
of the distribution of gluons 
$g(x, R_\perp)$ 
is given  by
\begin{equation}
\langle r_T^2 \rangle
\;\; \equiv \;\; \frac{\int d^2 R_\perp \; g(x, R_\perp) \; R_\perp^2}
{\int d^2 R_\perp \; g(x, R_\perp)} 
\;\; = \;\; 4 \; \frac{\partial}{\partial t}
\left[ \frac{F_g (x, t)}{F_g (x, 0)} \right]_{t = 0} = 2 b,
\label{myequation}
\end{equation}
where $b$ is the exponential $t$-slope.
In this expression, $\sqrt{\langle r_T^2 \rangle}$
is  the transverse distance between the struck
parton and the center of momentum of the proton. 
The latter is the average transverse
position of the partons in the proton with weights given by 
the parton momentum fractions.
At low $x_{Bj}$, the transverse
distance defined as $\sqrt{\langle r_T^2 \rangle}$
corresponds also to the relative transverse distance
between the interacting parton (gluon in the 
equation above) and the system defined by spectator partons.
Therefore provides a natural estimate of the transverse
extension of the gluons probed during the hard process.

In other words,
a Fourier transform of momentum
to impact parameter space readily shows that the $t$-slope $b$ is related to the
typical transverse distance in the proton.
This $t$-slope, $b$, corresponds exactly to the slope measured once
the component of the probe itself contributing to $b$ can be
neglected, which means at high scale: $Q^2$ or $M_{VM}^2$.
Indeed,
at high scale, the $q\bar{q}$ dipole is almost
point-like, and the $t$ dependence of the cross section is given by 
the transverse extension 
of the gluons in the  proton for a given $x_{Bj}$ range.

\subsection{  Comments on the physical content of ${\langle r_T^2 \rangle}$}

A short comment is in order concerning the fundamental relation (\ref{myequation})
for DVCS at HERA (at low $x_{Bj}$). Does it make sense to keep only 
the gluon distribution in this expression or do we need to
consider also sea quarks?
This issue can be addressed simply by coming back to Eq. (\ref{dipoledvcs}),
where we have approximated the imaginary DVCS amplitude ($\gamma^*p \rightarrow \gamma p$) in
$$
Im A \sim \sigma_0 \frac{1}{Q^2}  \int_{1/Q^2}^{1/Q_s^2} \frac{dr^2}{r^4}  (r^2 Q_s^2).
$$
Let us give first a more general form to this formula, keeping the 
tracks of the photon wave functions
\begin{equation}
Im A =
\int d^2 r\, dz\,
\Psi^*(r,z,Q_1^2=Q^2) \Psi(r,z,Q_2^2=0) \hat\sigma(x,r),
\label{dipoledvcs2}
\end{equation}
where $\Psi^*(r,z,Q_1^2=Q^2)$ is the wave function for the virtual photon
and $\Psi(r,z,Q_2^2=0)$ for the real photon. Also, 
following the previous discussion on the dipole cross section,
we can write: $\hat\sigma(x,r) \sim \sigma_0 r^2 Q_s(x,r)^2$,
with 
$$
Q_s(x,r)^2 \sim \frac{\alpha_S \ xg(x,1/r^2)}{\pi R_p^2} \sim Q_0^2 (\frac{x_0}{x})^\lambda ,
$$
where $R_p$ is the proton radius.
We conclude immediately that the imaginary part of the DVCS amplitude
is dependent on the gluon density convoluted by the photon (virtual and real) wave functions.
It gives the rationale behind formula (\ref{myequation}).

Of course, this is a matter of representation. In the Eq. (\ref{dipoledvcs2}),
we write the photon-gluon interaction through a quark loop,
with a virtual photon fluctuating in a $q {\bar q}$ pair,
which is exactly the dipole $q {\bar q}$ component entering in $\Psi^*(r,z,Q_1^2=Q^2)$
(see also Fig. \ref{figunified}). In other words,
at low $x_{Bj}$ ($x_{Bj} \simeq 10^{-3}$), the idea is that quarks (sea quarks)
are produced by gluons.

Then, the dipole formalism, summarized in Eq. (\ref{dipoledvcs2})
or Fig. \ref{figunified}, provides a very powerful expression of this behavior.
Of course, in other formalisms, that we present latter, we can express
the cross sections at the level of the photon-quark interaction and thus consider directly
the sea quark distribution.

\subsection{  Experimental results }

\begin{figure}[htbp]
\begin{center}
\includegraphics[width=7cm,height=5cm]{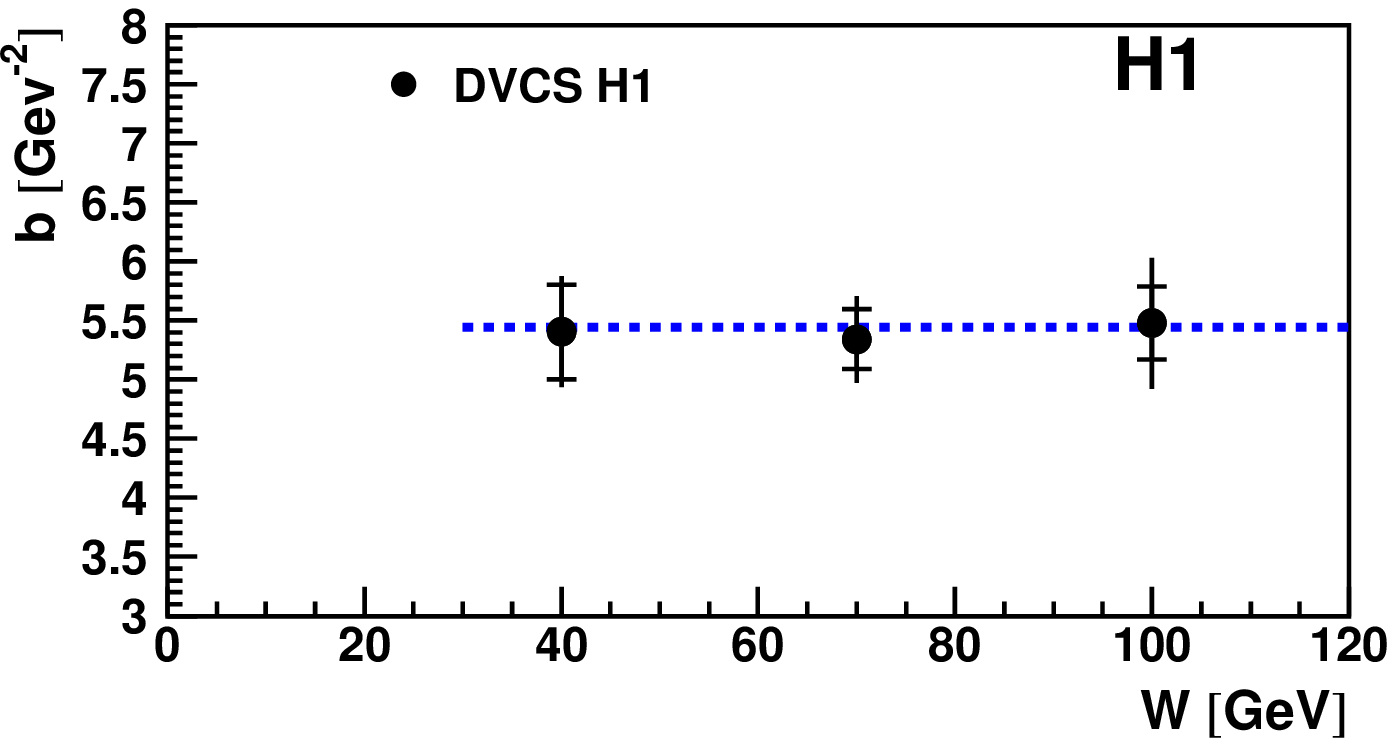}
\caption{  The logarithmic slope of the $t$ dependence
  for DVCS as a function of $W$.}
\label{bslopes}
\end{center}
\end{figure}

DVCS results lead to $\sqrt{r_T^2} = 0.65 \pm 0.02$~fm at large scale $Q^2 > 8$ GeV$^2$ 
for $x_{Bj} \simeq 10^{-3}$ \cite{dvcsh1}.
This value is smaller than the size of a single proton, and, in contrast 
to hadron-hadron scattering, 
it does not expand as energy $W$ increases (see Fig. \ref{bslopes}).
Then, we can parametrize the measured $b$ values displayed in Fig. \ref{bslopes}
in the form of a Pomeron trajectory: $b=b_0+2 \alpha' \ln \frac{1}{x_{Bj}}$. 
We obtain that the $\alpha'$ value,
which is characteristic of the energy dependence
of the trajectory, is close to zero.

This is not useless to recall that this observation
 is extremely challenging on the experimental
analysis side. We are dealing with nano-barn cross sections, that we measure
as a function of $t$, and finally, we measure the energy dependence of this 
behavior in $t$. 
Of course, the gain is  important.
In particular, the great interest of the DVCS is that the $t$ dependence
measured is free of effects that could come from VM wave functions (in case of VMs)
and then spoil (to a certain limit) 
the interpretation of $b$ described above.
Thus,
with DVCS, we have the advantage to work in a  controlled environment
(photon wave functions)
where the generic Eq. (\ref{myequation}) can be applied  to
the measurement (almost directly) and must not be
corrected with effects arising from VMs wave function.

\begin{figure}[htbp]
\centerline{\includegraphics[width=0.5\columnwidth]{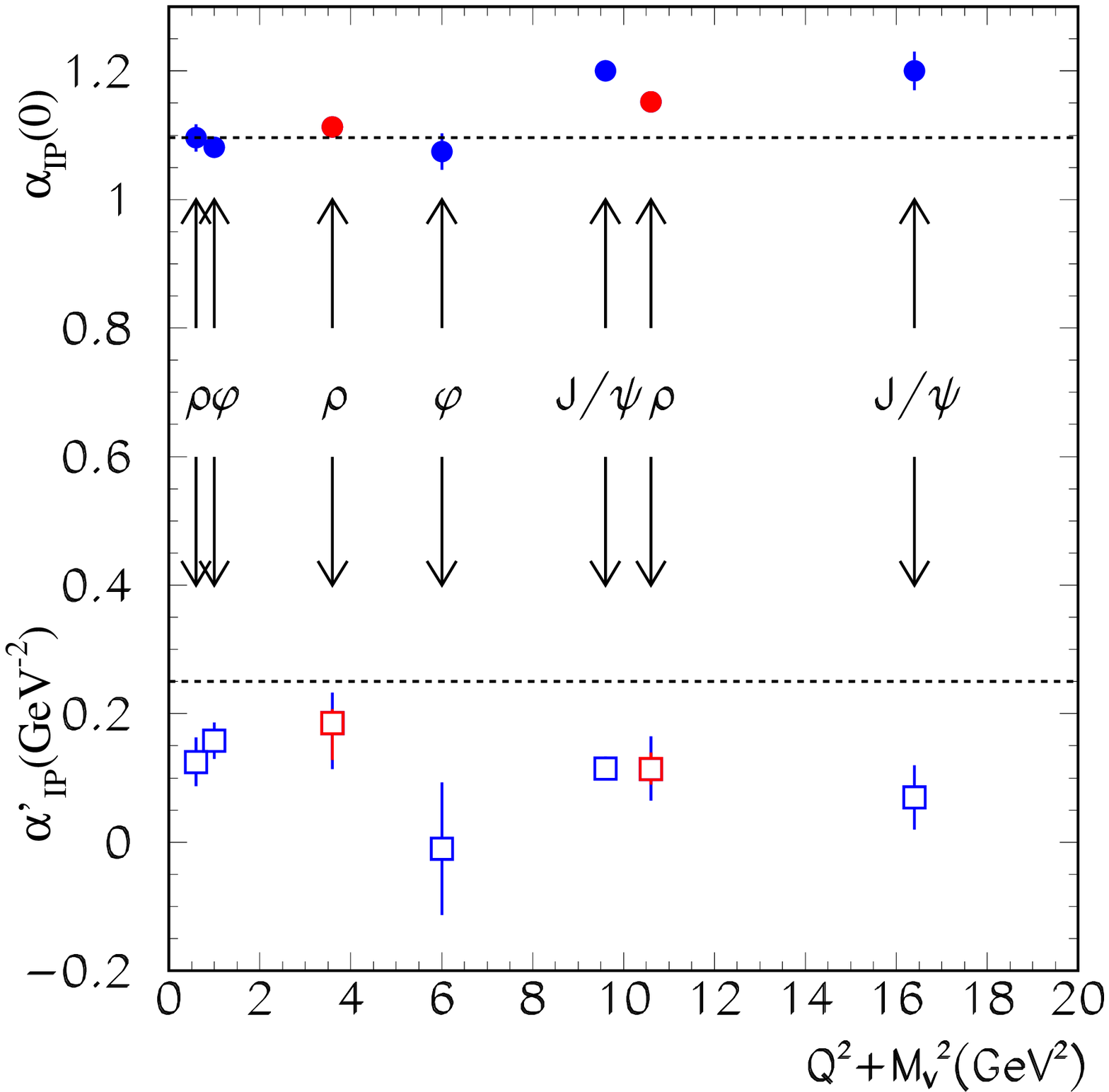}}
\caption{ 
Values of the intercept and slope of the effective Pomeron trajectory
as a function of $Q^2+M^2$, as obtained from measurements of exclusive VM
electroproduction.}
\label{fig:ap-apr-pom}
\end{figure}

\begin{figure}[htbp]
\begin{center}
  \includegraphics [bb= 105 247 487 600,clip,width=0.5\hsize,totalheight=5cm]{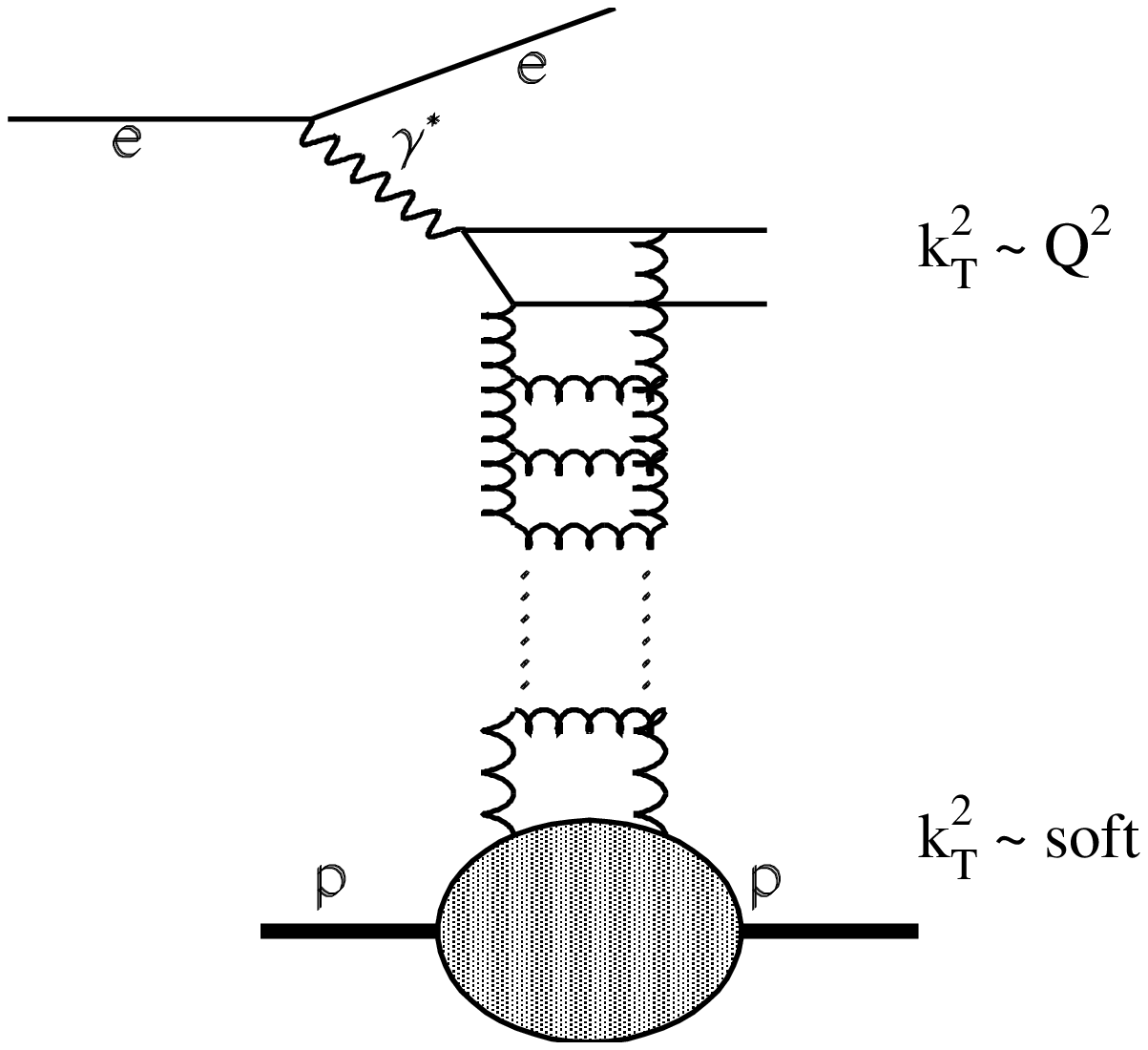}
\caption {
A diagram describing a gluon ladder in a diffractive process.}
\label{fig:ladderb}
\end{center}
\end{figure}

It is obviously very interesting to extend the result presented in Fig. \ref{bslopes}
to all VMs.
Indeed, we can study the $W$ dependence of d$\sigma$/d$t$ 
and extract the energy dependence as done above for all VMs, using
$b=b_0+2 \alpha' \ln \frac{1}{x_{Bj}}$. 
Results are presented in Fig.~\ref{fig:ap-apr-pom} (bottom).
Values are plotted as a function of $Q^2+M^2$. 
We observe that the values of
$\alpha'$ tends to decrease with the scale.
In particular, the measurement of $\alpha'$ done for 
the $J/\Psi$ \cite{jpsi}, leading to a small value for $\alpha'$,
 is well compatible with the DVCS result \cite{dvcsh1}.
 
A short comment can be done qualitatively on such small 
$\alpha'$ value. We can rephrase this observation
as an evidence of
no shrinkage of $d\sigma/dt$ in the process $\gamma^*
p \to J/\Psi p$ or $\gamma^*
p \to \gamma p$. 
Looking at the diagram describing two gluon exchange in
Fig. \ref{fig:ladderb}, the virtual photon fluctuates into two
high $k_T$ quarks. Although in the diagram there are only two gluons
linked to the proton, we actually have a whole ladder due to the
large rapidity range available at these high $W$ energies (see
Fig. \ref{fig:ladderb}). From the virtual photon
vertex down to the proton, the average $k_T$ of the gluons gets
smaller, the configuration larger and we enter the region of low
$k_T$ physics governed by non-perturbative QCD. This process is called
Gribov diffusion. Thus
a process can start as a hard process at the photon vertex but
once it couples to the proton it gets a soft component which makes the
process non calculable in pQCD.  The average $k_T$ of the partons in
the process can be estimated by the slope of the trajectory since
$\alpha^\prime \sim 1/<k_T>$. 

The fact that no shrinkage is observed  
indicates that Gribov diffusion is
not important in this process at the presently available $W$ values, and
the average $k_T$ remains large. Such a behavior is
expected for hard processes, where $\alpha^\prime \ll$ 0.25
GeV$^{-2}$. The experimental results for exclusive $J/\Psi$ production
and DVCS confirm that both processes are fully calculable in perturbative QCD.

\subsection{  Link with LHC issues }

\begin{figure}[htbp]
\begin{center}
  \includegraphics[width=0.34\textwidth]{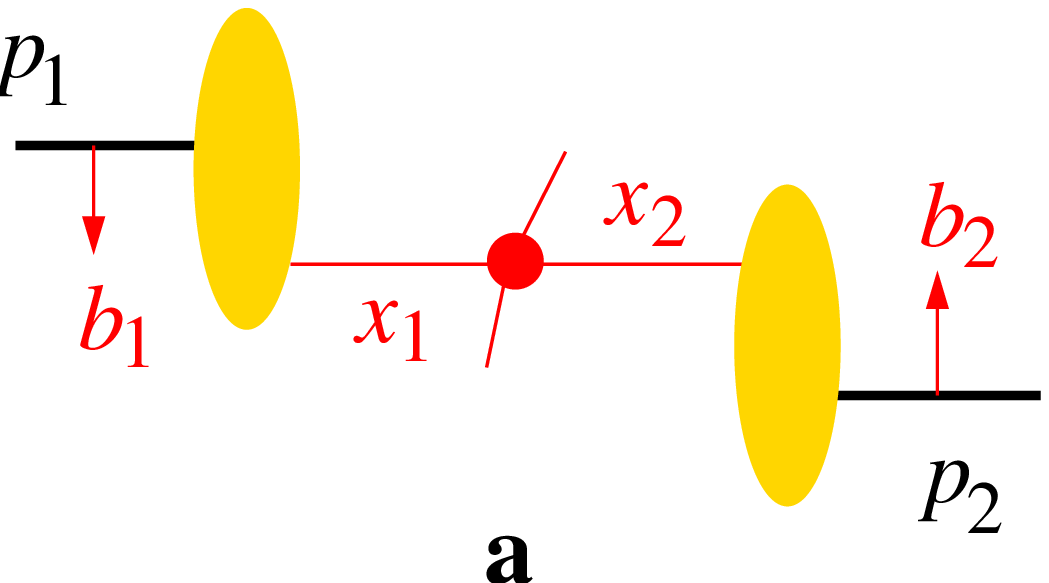}
  \hspace{0.1\textwidth}
  \includegraphics[width=0.34\textwidth]{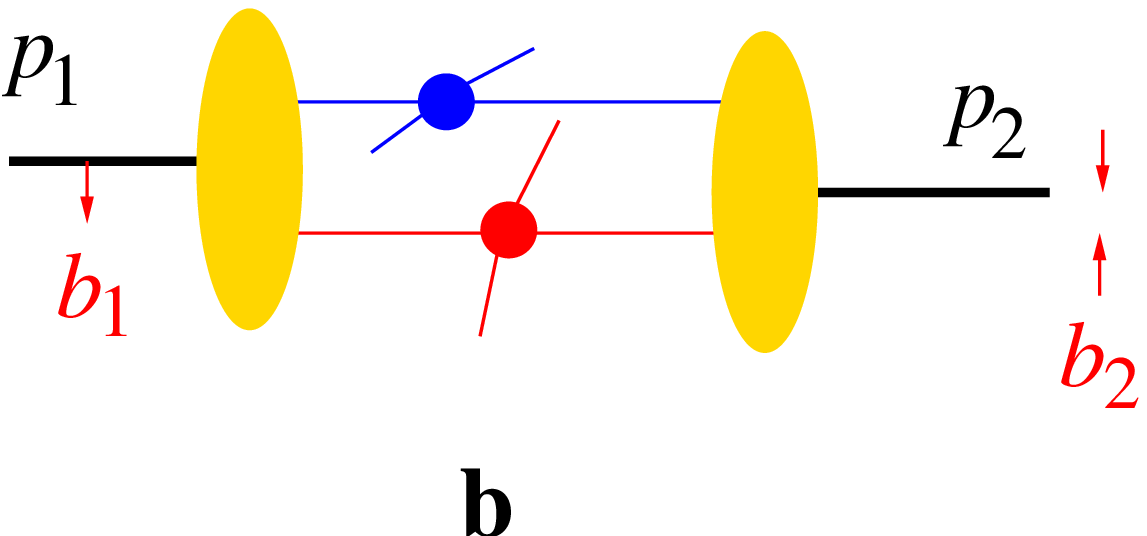}
\caption{\label{fig:multi}  a: Graph with a single hard
  interaction in a hadron-hadron collision.  The impact parameters
  $b_1$ and $b_2$ are integrated over independently. b:
  Graph with a primary and a secondary interaction. }
  \end{center}
\end{figure}

Let us finish this section by a comment making the link with
LHC issues. Indeed,
the  correlation between the transverse distribution of partons
and their momentum fraction is not only interesting from the
perspective of hadron structure, but also has practical consequences
for high-energy hadron-hadron collisions.
Consider the production of a high-mass system (a dijet or a
heavy particle).  For the inclusive production cross section, the
distribution of the colliding partons in impact parameter is not
important: only the parton distributions integrated over impact
parameters are relevant according to standard hard-scattering
factorization (see Fig.~\ref{fig:multi}(a)).  There can however be
additional interactions in the same collision, especially at the high
energies for the Tevatron or the LHC, as shown in
Fig.~\ref{fig:multi}(b).  Their effects cancel in sufficiently
inclusive observables, but it does affect the event
characteristics and can hence be quite relevant in practice.  In this
case, the impact parameter distribution of partons must be
considered. 

The
production of a heavy system requires large momentum fractions for the
colliding partons.  A narrow impact parameter distribution for these
partons forces the collision to be more central, which in turn
increases the probability for multiple parton collisions in the event
(multiple interactions).

%
%
\section{  Generalised parton distributions} 

\subsection{  A brief introduction in simple terms} 

We have already defined in a previous section a
first form for a generalized gluon distribution. In this part,
we move into further details and explain the wide 
experimental field opened in the area of
generalized parton distributions.

First, a short contrarian comment: 
DIS can not be
considered as the continuation of the original Rutherford experiment. 
Indeed, Rutherford measured that
the nucleus is concentrated in a very small part of the atom, and,
as far as we consider only PDFs, we have no possibility to explore the
spatial structure of the nucleon.
The reason is that in the infinite momentum frame picture,
the light-cone description of the Feynman parton model does
not explore the space-time location of partons. In other words, within the
infinite momentum frame description, the variable $x_{Bj}$ has no direct
relation to the space coordinate of a parton but is related to a
combination of the energy and momentum of this parton. 

In the previous section, we have shown that
data on exclusive particle production can give access to the
spatial distribution of quarks and gluons in the proton
at femto-meter scale. 
Then, we have defined functions, which model this property (for gluons)
through the relation
$$
g (x, R_\perp; Q^2) 
\;\; \equiv \;\; \int \frac{d^2 \Delta_\perp}{(2 \pi)^2}
\; e^{i ({\Delta}_\perp {R_\perp})}
\; F_g (x, t = -{\Delta}_\perp^2; Q^2).
$$
Of course, a similar relation holds for quarks, linking
the two functions
$q(x, R_\perp; Q^2) $ and $ F_q (x, t = -{\Delta}_\perp^2; Q^2)$.
The general framework for this physics is encoded in the  so-called
generalized parton distributions (GPDs).

We already know that the reconstruction of spatial images from scattering
experiments by way of Fourier transform of the observed 
scattering pattern is a technique widely used in physics,
for example, in X-rays scattering from crystals.
In simple words, what we have done experimentally is that
we have extended this technique
to the spatial distribution of quarks and gluons
within the proton, using processes that probe the proton at a tiny resolution scale.
Of course, as already mentioned, working at a femto-meter scale
with nano-barn cross sections
is very challenging from the experimental front. We have achieved this
and it immediately opens a way in the ambitious program of
mapping out the GPDs.
We come back below in a more systematic way on different
aspects of that program that requires
a large amount of experimental informations, for which future programs at JLab
and CERN are appealing.

\begin{figure}[htbp]
\begin{center}
  \includegraphics[width=.4\textwidth]{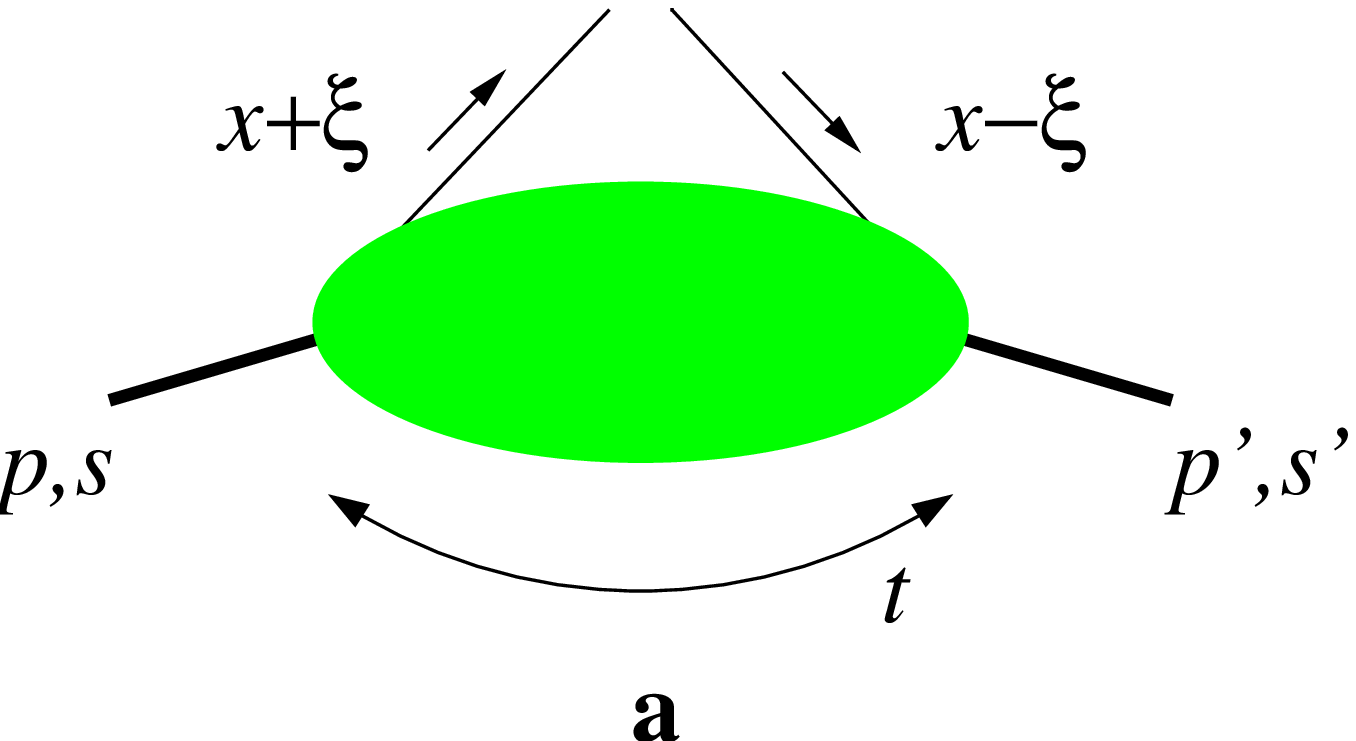}
    \caption{  Picture of a GPD and its variables.  The
  momentum fractions $x$ and $\xi$ refer to the average hadron
  momentum $\frac{1}{2} (p+p')$. Note that $x$ is an internal variable
  and is not equal to $x_{Bj}$. However,
  there is a relation between the skewing variable $\xi$ and 
  $x_{Bj}$, $\xi = x_{Bj}/(2-x_{Bj})$.}
\label{fig:gpd}
\end{center}
\end{figure}

Before coming back to the experimental side,
we can present a short overview of GPDs,
in simple terms. It is interesting, even for 
an experimentalist, as it clarifies the Fourier 
transform relation discussed above and makes
more transparent the goals for the future.
For complete reviews, see Ref. \cite{imaging,gpdsreview,muller}.
GPDs are defined through matrix elements $\langle p' | \mathcal{O} | p
\rangle$ between hadron states $|p'\rangle$ and $|p \rangle$, with
non-local operators $\mathcal{O}$ constructed from quark and gluon
fields.  
From this expression, we understand why GPDs are directly related
to the amplitude for VM or real gamma exclusive production.  
For unpolarized quarks there are two
distributions $H^q(x,\xi,t)$ and $E^q(x,\xi,t)$, where $x$ and
$\xi$ are defined in Fig. \ref{fig:gpd}.  The former is
diagonal in the proton helicity, whereas the latter describes proton
helicity flip.  For $p=p'$ and equal proton helicities, we recover
the diagonal matrix element parameterized by usual quark and antiquark
densities, so that $H^q(x,0,0)=q(x)$ and $H^q(-x,0,0)=-\bar{q}(x)$ for
$x>0$.  
Note that the functions of type $E$ are not accessible in 
standard DIS, as it corresponds to matrix elements
$\langle p',s' | \mathcal{O} | p,s
\rangle$ with $s \ne s'$. Even in DVCS-like analysis, it is 
very difficult to get a sensitivity to these functions,
as in most observables, their contributions are 
damped by kinematic factors of orders $|t|/M_p^2$,
with an average $|t|$ value in general much smaller that $1$ GeV$^2$.
Then, till stated otherwise, our next experimental discussions
are concentrated on the determination of GPDs of type
$H_q$ or $H_g$.
We come back later on this point and show specific cases
where $E$-type functions can be accessed and why this is
an important perspective.

\subsection{  Fundamental relations between GPDs and form factors} 

An interesting property of GPDs, which lightens their
physics content, is that 
their lowest moments give the well-known Dirac and Pauli
form factors
\begin{equation}
\sum_q e_q \int dx\, H^q(x,\xi,t) = F_1(t) 
\qquad\qquad
\sum_q e_q \int dx\, E^q(x,\xi,t) = F_2(t),
\label{ffactors}
\end{equation}
where $e_q$ denotes the fractional quark charge. 
It means that GPDs measure the contribution of quarks/gluons,
with longitudinal momentum fraction $x$, to the corresponding 
form factor. In other words,  GPDs are
like mini-form factors that filter out quark with a 
longitudinal momentum fraction $x$
in the proton.
Therefore, in the same way as Fourier transform of a 
form factor gives the charge
distribution in position space, Fourier transform of GPDs (with
respect to variable $t$)
contains information about the spatial distribution of partons in the proton.

\subsection{  New insights into proton imaging} 

This discussion clarifies also the Fourier transforms,
that can relate $g (x, R_\perp; Q^2) $ and $F_g (x, t = -{\Delta}_\perp^2; Q^2)$
or $q(x, R_\perp; Q^2) $ and $F_q (x, t = -{\Delta}_\perp^2; Q^2)$.
We have already discussed these functions and from their relations,
 it follows that $q(x, R_\perp; Q^2) $ 
is the probability density to find a quark with momentum fraction $x$
at a transverse distance $R_\perp$ from the (transverse) center of momentum
of the proton.
More formal discussions can be found in Ref.
\cite{imaging}.

Exactly, what must be confronted with the proton radius is not
$\sqrt{r_T^2}$ but  $\sqrt{r_T^2}/(1-x_{Bj})$, 
which does not change our result 
with $x_{Bj} \simeq 10^{-3}$ ($\sqrt{r_T^2} = 0.65 \pm 0.02$~fm), 
but must be taken into account for fixed target 
kinematics at larger $x_{Bj}$. 
In particular, at very large $x_{Bj}$ ($x_{Bj} \rightarrow 1$),
the struck quark is carrying almost the entire proton momentum,
thus its relative distance to the center of momentum of the
proton obviously tends to zero.
This means that $\sqrt{r_T^2}$ tends to zero (by definition).
In order to keep finite the ratio
$\sqrt{r_T^2}/(1-x_{Bj})$, we can conclude that the
asymptotic form of $\sqrt{r_T^2}$ at large $x_{Bj}$ is likely
in $(1-x_{Bj})^2$. 

The distance $\sqrt{r_T^2}/(1-x_{Bj})$ is the
associated transverse distance between the struck
  parton (probed during the hard 
  interaction) and the center of momentum of the spectators.
That's why it can be interpreted as
  a typical spatial extension of partons in the proton.

What we have learned so far with the present experimental
situation is already very rich: slow partons (at low $x_{Bj}$) are located at 
the periphery of the proton
whereas fast partons (at large $x_{Bj}$) 
make up the core of the proton (in its center).
This last property is only an indirect observation from fits of form factor
measurements
\cite{gpdsreview} (see below for a short discussion).

We need to get more information. How large can be the spread in space of slow partons? 
Could it be larger that 1 fm? 
Also, what is the spread for the large $x$ (constituent) partons? 
Where is the transition between the large $x$ partons
and the peripheric partons? 
We need more experimental results and then 
more experiments with different setups to address these questions 
from all possible angles.

For example, if we would observe a gradual increase of the $t$ 
dependence of  the GPD 
$H(x,0,t)$ (quarks or gluons) when varying $x_{Bj}$ from large to 
small values, it would mean
exactly that quarks at large $x_{Bj}$ come from the more localized 
valence core of the proton,
while the small $x_{Bj}$ region receives contribution from the periphery 
or, in other words, from the
wider meson cloud. 
This is a very nice perspective for the future to expect 
direct measurements of $\sqrt{r_T^2}$ from many experiments in the world.

\subsection{  An elegant application} 

Let us come back briefly to form factors and their essential role
in the interplay between $x$ and $t$ kinematic variables.
A complete analysis is presented by Diehl et al. in Ref. \cite{diehlff}.
Indeed, it is clear that
indirect information on impact parameter distributions can be obtained
by using the sum rules presented in Eq. (\ref{ffactors}), which 
provides a natural link between the GPDs dependences in $x$ and $t$.
We can exemplify the structure of the link on the Dirac form factor
for proton and neutron
$$
F_1^p(t)=\int_0^1 dx [\frac{2}{3} H^u_v(x,t)- \frac{1}{3} H^d_v(x,t)] 
$$
$$
F_1^n(t)=\int_0^1 dx [\frac{2}{3} H^d_v(x,t)- \frac{1}{3} H^u_v(x,t)]
$$
where we have neglected the contribution from the $s$ quarks.
Note that only valence type distributions appear in these relations,
since the electromagnetic
current is only sensitive to the difference of quark and antiquark
distributions.
Then, from an ansatz for the functional dependence of $H^q_v(x,0,t)$
and measurements of the Dirac form factor $F_1(t)$ (and $F_2(t)$),
a fit of some GPDs parameters can be performed \cite{diehlff}.

\begin{figure}[p]
\begin{center}
\includegraphics[height=.37\textwidth,
  bb=120 75 370 280,clip=true]{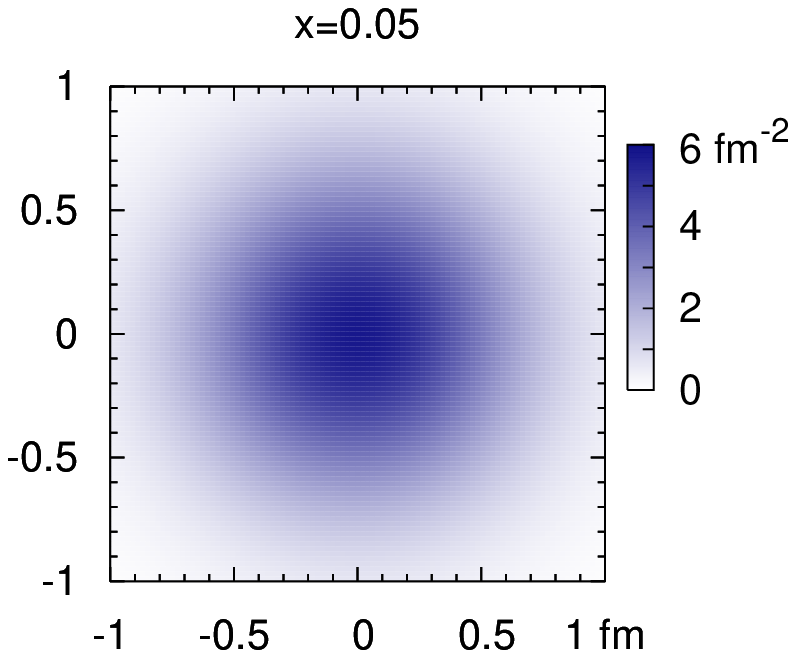}
\includegraphics[height=.37\textwidth,
  bb=120 75 320 280,clip=true]{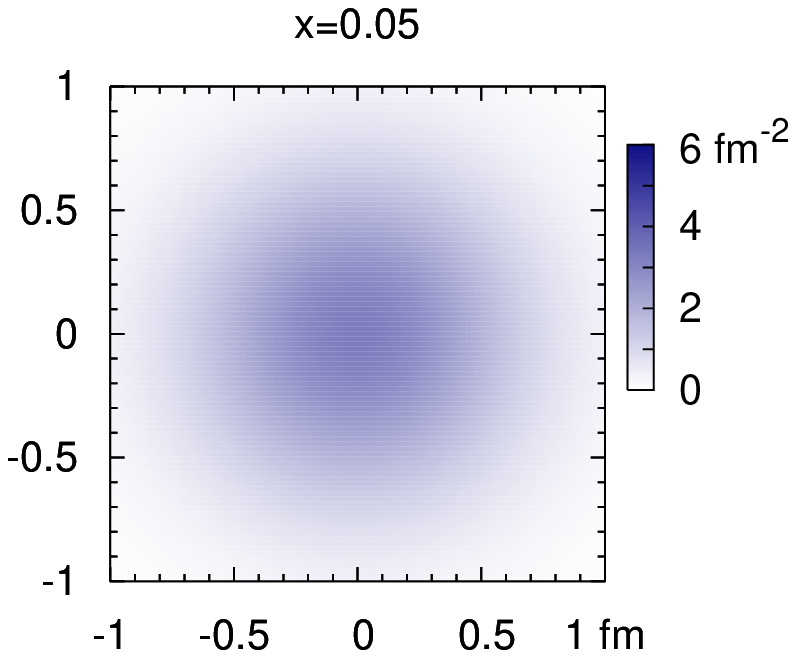}
\\[2em]
\includegraphics[height=.37\textwidth,
  bb=120 75 370 280,clip=true]{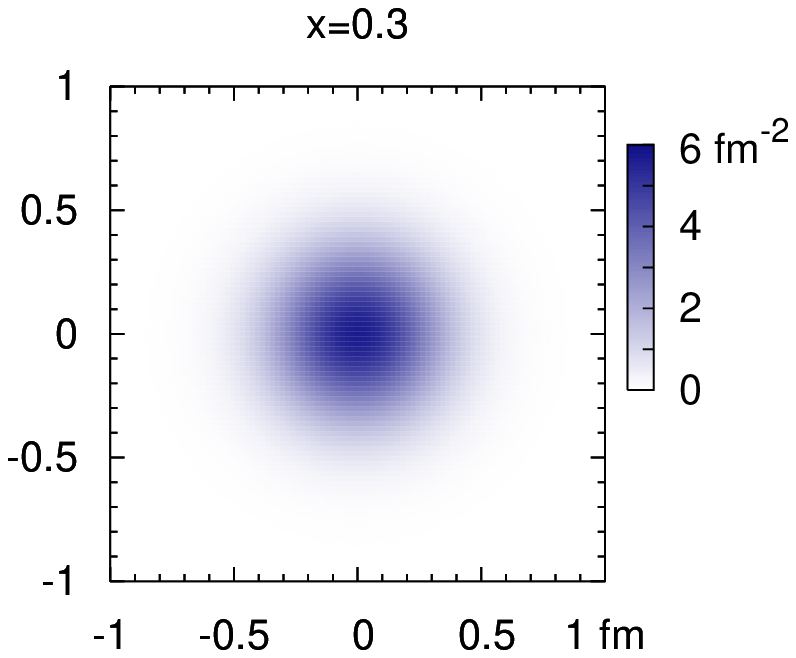}
\includegraphics[height=.37\textwidth,
  bb=120 75 320 280,clip=true]{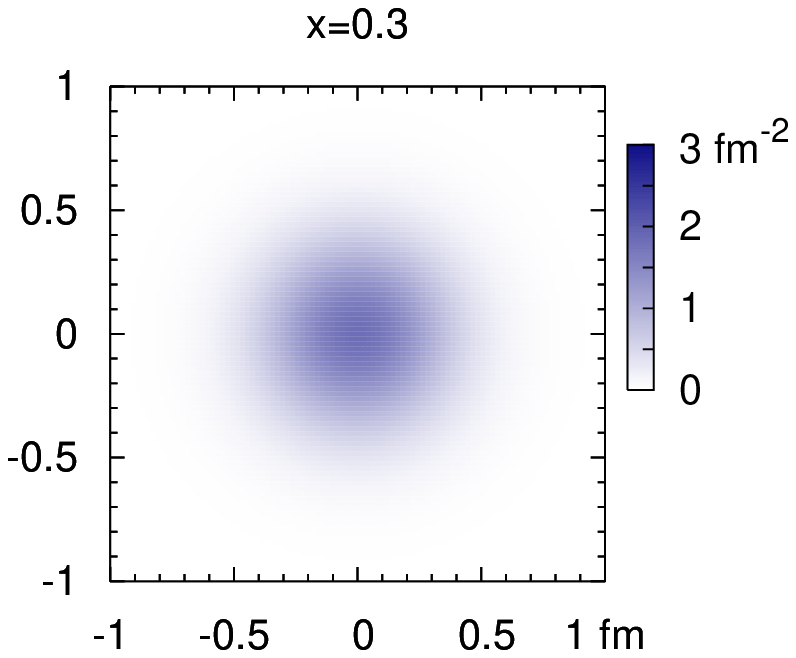}
\\[2em]
\includegraphics[height=.37\textwidth,
  bb=120 75 370 280,clip=true]{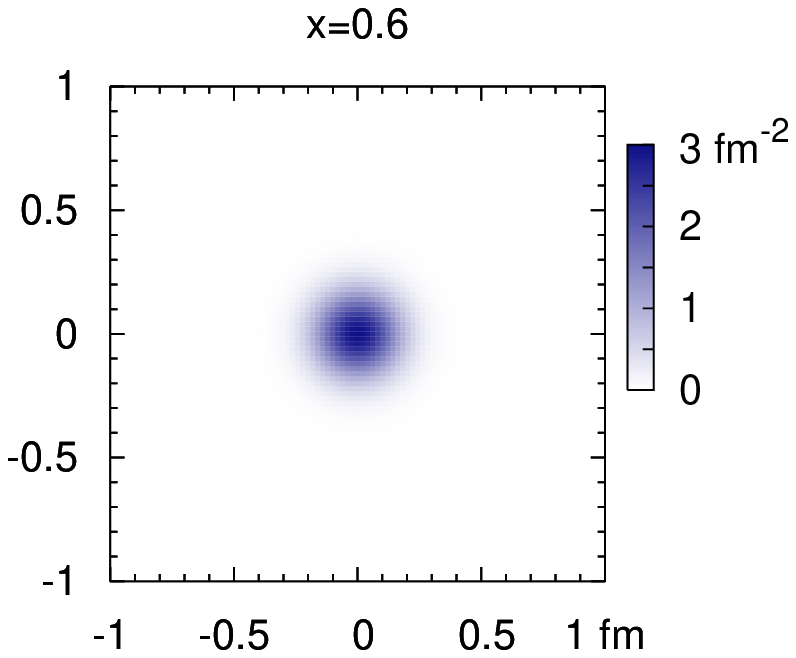}
\includegraphics[height=.37\textwidth,
  bb=120 75 320 280,clip=true]{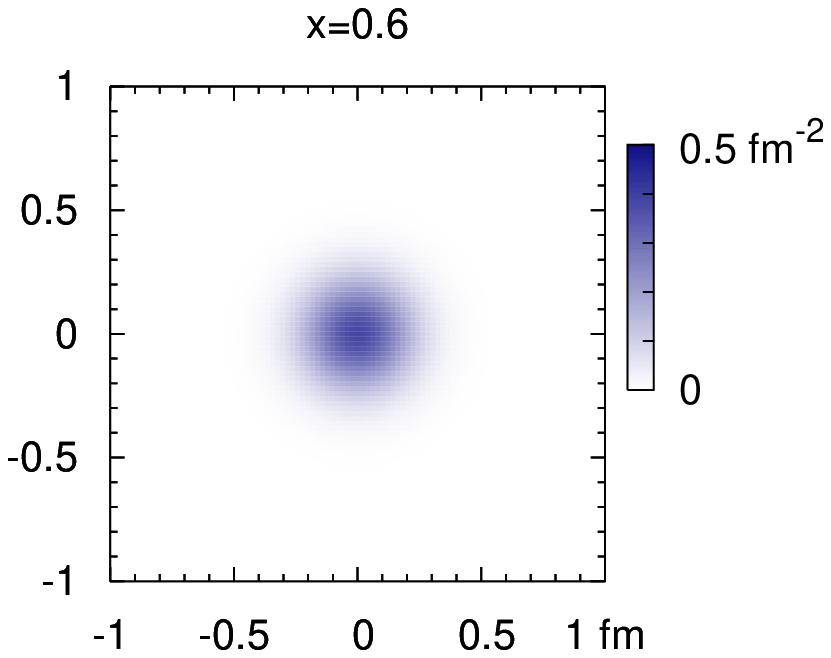}
\end{center}
\caption{\label{fig:tomo_u} Tomography plots of $u_v(x,\bf{r_T})$
(left) and $d_v(x,\bf{r_T})$ (right) in the transverse $r_T^x$--$r_T^y$
plane.  Note that the scale of intensity for longitudinal momentum
fraction $x=0.6$ differs from the one for $x=0.3$ and $x=0.05$ \cite{diehlff}.}
\end{figure}

Obviously, the sensitivity of such a fit is governed by the parameters
building the interplay of $x$ and $t$ dependences
(for valence distributions), which is the purpose of this
approach.
In Fig.~\ref{fig:tomo_u} the
default results for GPDs as tomography plots in impact parameter space
is illustrated
for fixed longitudinal momentum fraction $x$ \cite{diehlff}. 

This confirms the results on $\sqrt{r_T^2}$ discussed above:
 low $x_{Bj}$ partons are located at 
the periphery of the proton
whereas valence like partons 
make up the core of the proton (in its center).

%
%
\section{  Quantifying skewing effects on DVCS at low $x_{Bj}$} 

\subsection{  DVCS in the context of GPDs}

After this short overview of GPDs physics, 
we understand clearly why DVCS is the typical (and cleanest)
process to extract GPDs, or at least to extract
informations on GPDs. Then,
we  can
come back on the DVCS cross section measurements and their interpretation
in terms of GPDs.
In order to quantify the magnitude of 
skewing effects, and thus the impact of
GPDs on the DVCS process
($\gamma^*p \to \gamma p$), we need to derive for example 
the following ratio
from measured cross sections \cite{dvcsh1}
$$
R \equiv 
{{ I}m \,{{ A}}\,(\gamma^* p \to \gamma  p)_{t=0}} /
{{ I}m\, {{  A}}\,(\gamma^* p \to \gamma^*  p)_{t=0}}.
$$
In this expression,
 ${ I}m {A}(\gamma^*p \to \gamma p)_{t=0}(Q^2,W)$ 
is the imaginary part of the DVCS process and is
directly proportional to the GPDs.
Also, the diagonal term 
${{ I}m\, {{  A}}\,(\gamma^* p \to \gamma^*  p)_{t=0}}$
is directly proportinal to the total cross section.
The ratio $R$ is then is equivalent to
the ratio of the GPDs to the PDFs.
That's why its measurement can provide directly the impact of GPDs,
when compared to pure PDFs predictions.

\begin{figure}[!]
\begin{center}
 \includegraphics[totalheight=8cm]{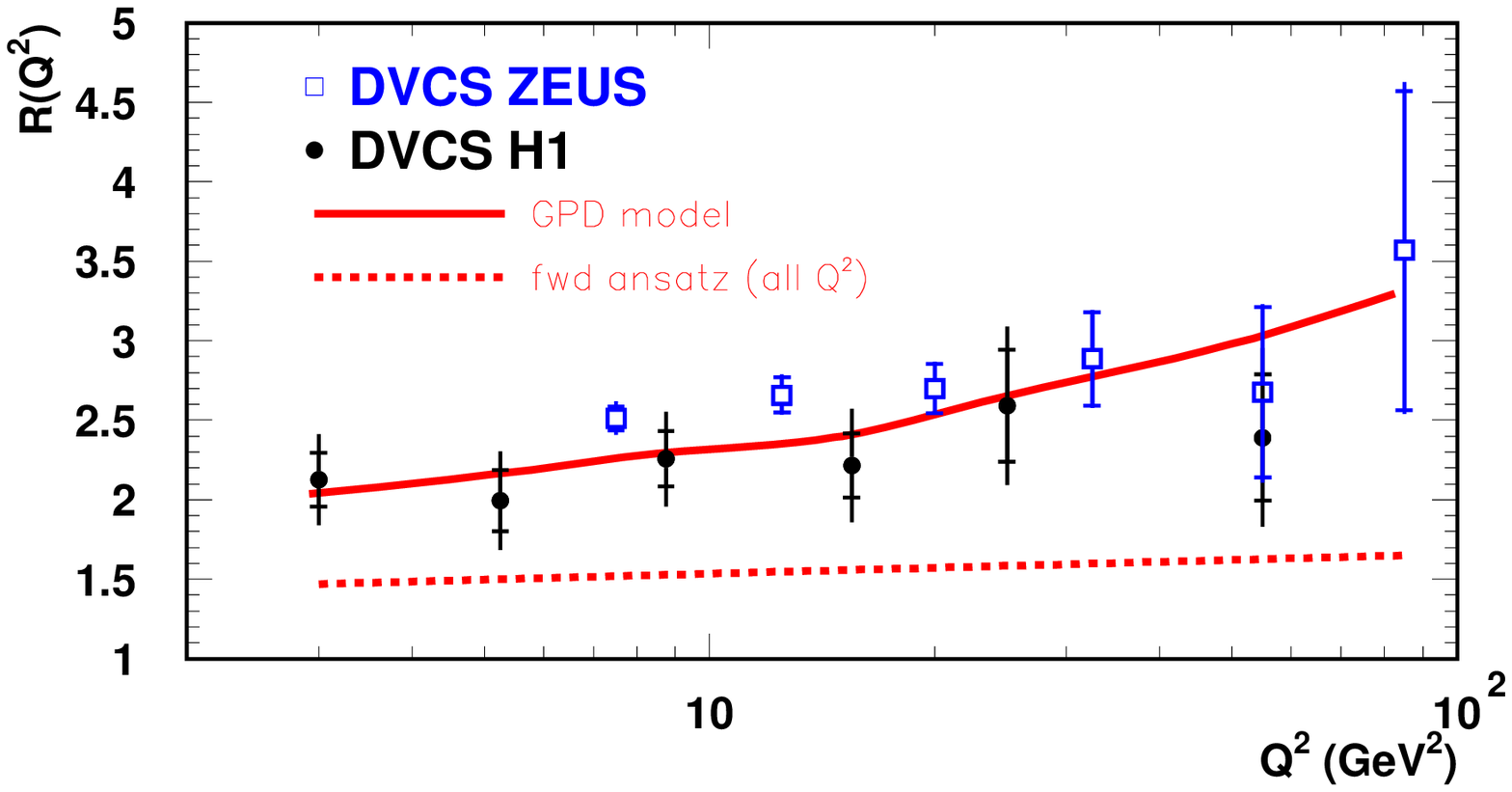}
\end{center}
\vspace*{-1.7cm}
\caption{\label{figssr}
Skewing factor
$R \equiv 
{{\cal I}m \,{{ A}}\,(\gamma^* p \to \gamma  p)_{t=0}} /
{{\cal I}m\, {{  A}}\,(\gamma^* p \to \gamma^*  p)_{t=0}} $
extracted from DVCS and DIS cross sections  \cite{compasslolo}
The GPD model is also displayed and gives a good agreement of the data (full line).
The forward ansatz model \cite{Frankfurt:1998et}, used
at all values of $Q^2$, fails to reproduce the total skewing effects generated by
the QCD evolution  (dashed line). 
 }
\end{figure}

\subsection{  Experimental results}
 
In Ref. \cite{compasslolo}, we have shown how to extract this ratio
from the DVCS and DIS cross sections.
Results are presented in 
Fig. \ref{figssr}. 
The typical values of $R$ are found around $2$, 
whereas in a model without skewing $R$ would be equal to unity. 
Therefore, the present measurements confirm the large effect of skewing.

Values of $R$ are also compared with a
 GPDs model based on a forward ansatz at low scale ($Q_0=1.3$ GeV)
 \cite{Frankfurt:1998et}.
Namely, the singlet GPD is parametrized as follows: $H_S(x,\xi)=Q_S(x)$,
where $Q_S(x)$ is the singlet PDFs and $x$ and $\xi$ are the variables
used in the previous part for the definition of GPDs (see Fig. \ref{fig:gpd}).
It does not mean that the GPD is taken to be exactly the PDF. Indeed, 
at $x=\xi$, we get  $H_S(\xi,\xi)=Q_S(\xi)=Q_S(x_{Bj}/2)$ and not
$Q_S(x_{Bj})$. 
In other words, in this forward ansatz parameterization of the GPDs,
we simply consider that at a low scale $Q_0$, we can
forget the profile function and take directly the parameterization of
the GPD from a PDF like form. The same is done with non-singlet and
gluon distributions.

If the GPDs are parametrized in such a way at initial scale,
then we have two possibilities. Either, we evolve the GPDs
using skewed QCD evolution equations, which  naturally generates the
skewing dependence (in $\xi$)  along the $Q^2$ evolution,
 or we forget about the
skewed evolution and we consider only the standard QCD evolution equations
like for PDFs \cite{compasslolo,Frankfurt:1998et}. 
This corresponds to the two curves presented in
Fig. \ref{figssr}. The full line represents the complete GPDs model,
with skewed evolution equations and the dashed curve, labeled forward ansatz (all $Q^2$),
represents the 
case where initial distributions are evolved with standard QCD evolution 
equations.
Then,  Fig. \ref{figssr} demonstrates that we need the full GPDs model
to describe our data on DVCS cross sections (converted in $R$ values). 
If we forget about the skewing generated during the
QCD evolution, we miss the data by about 30\%.
This is clearly a deep impact of the skewing effects present 
in the data \cite{compasslolo,Frankfurt:1998et}.

Another influence of GPDs that we can check on  data concerns the $t$ dependence.
We have already shown that in the kinematic domain of H1 and ZEUS measurements,
DVCS cross section ($d\sigma / dt$) can be factorized and 
approximated to a good accuracy by an exponential form $e^{-b|t|}$,
which implies a factorized dependence in $e^{-b/2 |t|}$ for GPDs.
However, we can think of taking into account 
a non-factorized form in $|x|^{-\alpha'/2 t}$ as well.
With the small $\alpha'$ value determined previously, we know
that this term can only be small (negligible) correction to the
dominant (factorized) $t$ dependence in $e^{-b/2|t|}$ for GPDs.

%
%
\section{  On the way of mapping out the GPDs} 

\subsection{  Prospects for the COMPASS experiment at CERN} 

As we have shown, the mapping of the GPDs is certainly a difficult work
due to the flexibility of these functions. However, we have already illustrated
some elements that can be constrained with the present 
DVCS data at low $x_{Bj}$. Concerning the  $t$ dependence of the GPDs
in this kinematic domain,
we have shown that the impact of a potential non-factorized term in
$|x|^{-\alpha't}$ is small, 
due to the small value of $\alpha'$ observed
at low $x_{Bj}$.

This is one important element of the experimental project to measure DVCS at COMPASS 
in the future,as we need to check this kind of effects at larger $x_{Bj}$. 
DVCS at COMPASS (located at CERN) can be  measured  with muon beams on 
fixed target,
$\mu p\rightarrow \mu p \gamma$. If the muon energy is large enough,
for example $E_{\mu } = 190$ GeV,  DVCS  dominates
over the BH contribution (as for H1 and ZEUS) so that DVCS  
cross section can be measured directly.

At smaller lepton energy,  $E_{\mu }
= 100$ GeV, the DVCS signal is not dominant and can not be measured directly.
Then, we need to use the property that DVCS and BH, having identical final state,
can interfere. When the DVCS cross section itself can not be measured,
the interference can be observed.
The strong interest is that the $x_{Bj}$ kinematic 
domain of COMPASS follows the
one of H1 and ZEUS at larger $x_{Bj}$, with $x_{Bj} \sim [0.05-0.15]$, thus
much larger values than in the kinematic domain of H1 and ZEUS.
A project is ongoing to install a proton recoil detector in the COMPASS setup
  and then 
measure DVCS cross section or DVCS/BH interference \cite{dhose}. 
Some tests have already been done to show the technical
feasibility of the proposed experiment.

Let us discuss how we can access an interference
between DVCS and BH reactions.
In fact,
since these two processes have an identical final state, they can 
obviously interfere.
The squared photon production amplitude is then given by 
\begin{equation} \label {eqn:tau}
\left| A \right|^2 
= \left| A_{{\scriptscriptstyle BH}} \right|^2 + 
\left| A_{{\scriptscriptstyle DVCS}} \right|^2 + \underbrace{
A_{{\scriptscriptstyle DVCS}} \, A_{{\scriptscriptstyle BH}}^* 
+ A_{{\scriptscriptstyle DVCS}}^* \, A_{{\scriptscriptstyle BH}}}_I,
\end{equation}
where $A_{\scriptscriptstyle BH}$ is the BH amplitude,
$A_{\scriptscriptstyle DVCS}$ represents the DVCS amplitude and
$I$ denotes the interference term.

For  unpolarized proton target and lepton beam, the interference term can be written
quite generally as a linear combination of harmonics of the azimuthal angle
$\phi$, which is the angle between the 
plane containing the incoming and outgoing leptons 
and the plane defined by the virtual and real photons. 
In the leading twist approximation (at sufficiently high $Q^2$),
if only the first term in $\cos\phi$ and $\sin\phi$ are considered, 
it can be written as:

\begin{equation}
I \propto 
-C \, 
[  a \cos \phi \, \mathrm{Re}  A_{DVCS}  
+b P_l \sin \phi \, \mathrm{Im}  A_{DVCS}.  
]
\label{toto}
\end{equation}
 In this expression,
 $C = \pm 1$ is the lepton beam charge, $P_l$
its longitudinal polarization 
and $a$ and $b$ are 
functions of the ratio of longitudinal to 
transverse virtual photon flux \cite{gpdsreview}.

\begin{figure}[t]
{\centering \resizebox*{0.6\textwidth}{!}{\includegraphics{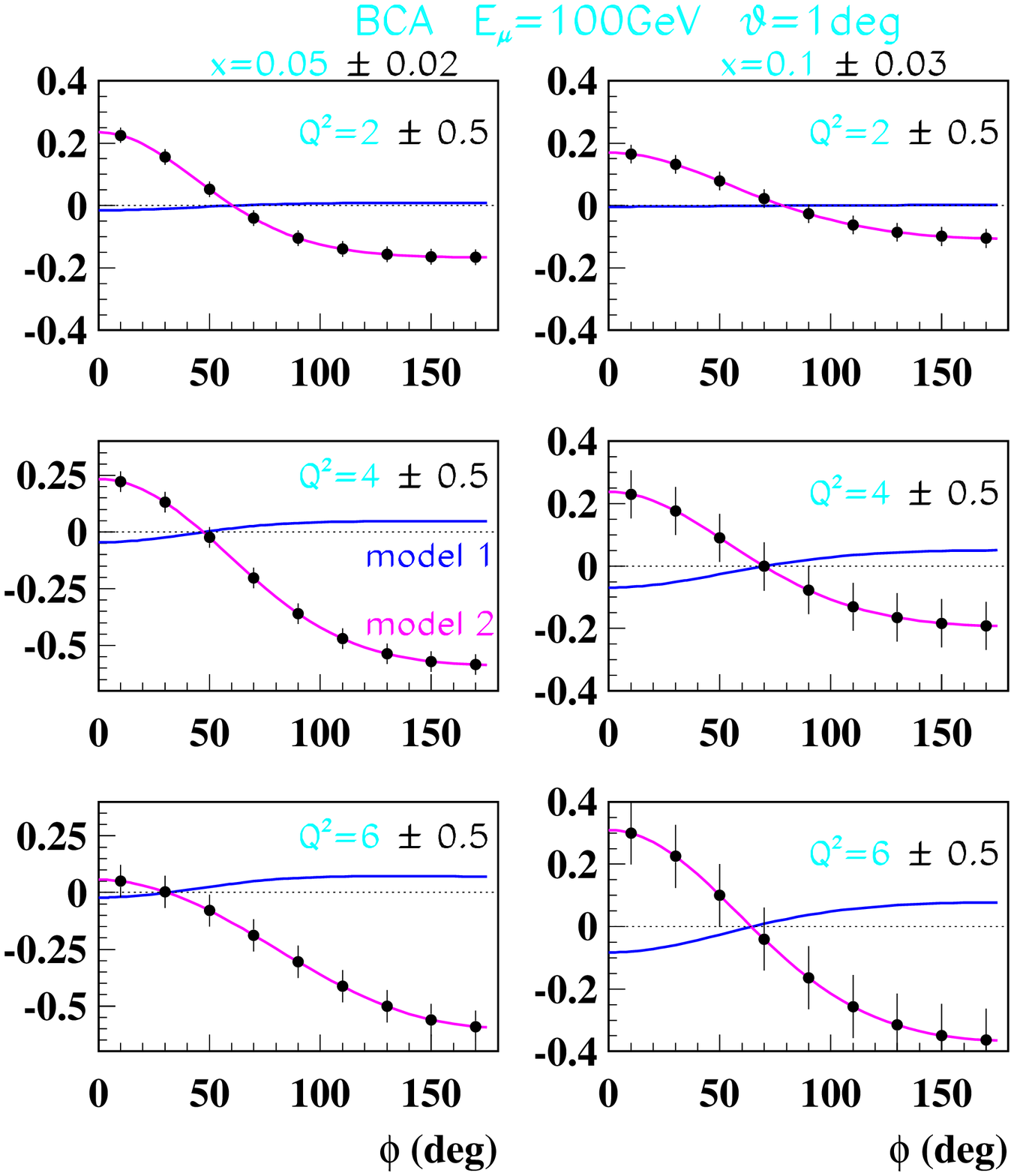}} \par}
\caption{\label{compass2}  Azimuthal distribution of the beam charge asymmetry
measured at COMPASS at \protect\( E_{\mu }\protect \)= 100 GeV and
\protect\( |t|\leq 0.6\protect \) GeV\protect\( ^{2}\protect \)
for 2 domains of \protect\( x_{Bj}\protect \) (\protect\( x_{Bj}=0.05\pm 0.02\protect \)
and \protect\( x_{Bj}=0.10\pm 0.03\protect \)) and 3 domains of \protect\( Q^{2}\protect \)
(\protect\( Q^{2}=2\pm 0.5\protect \) GeV\protect\( ^{2}\protect \),
\protect\( Q^{2}=4\pm 0.5\protect \) GeV\protect\( ^{2}\protect \)
and \protect\( Q^{2}=6\pm 0.5\protect \) GeV\protect\( ^{2}\protect \))
obtained in 6 months of data taking with a global efficiency of 25\%
and with \protect\( 2\cdot 10^{8}\protect \) \protect\( \mu \protect \)
per SPS spill (\protect\( P_{\mu ^{+}}=-0.8\protect \) and 
\protect\( P_{\mu ^{-}}=+0.8\protect \)) \cite{dhose}.}
\end{figure}

\begin{figure}[!htbp]
\begin{center}
 \includegraphics[totalheight=8cm]{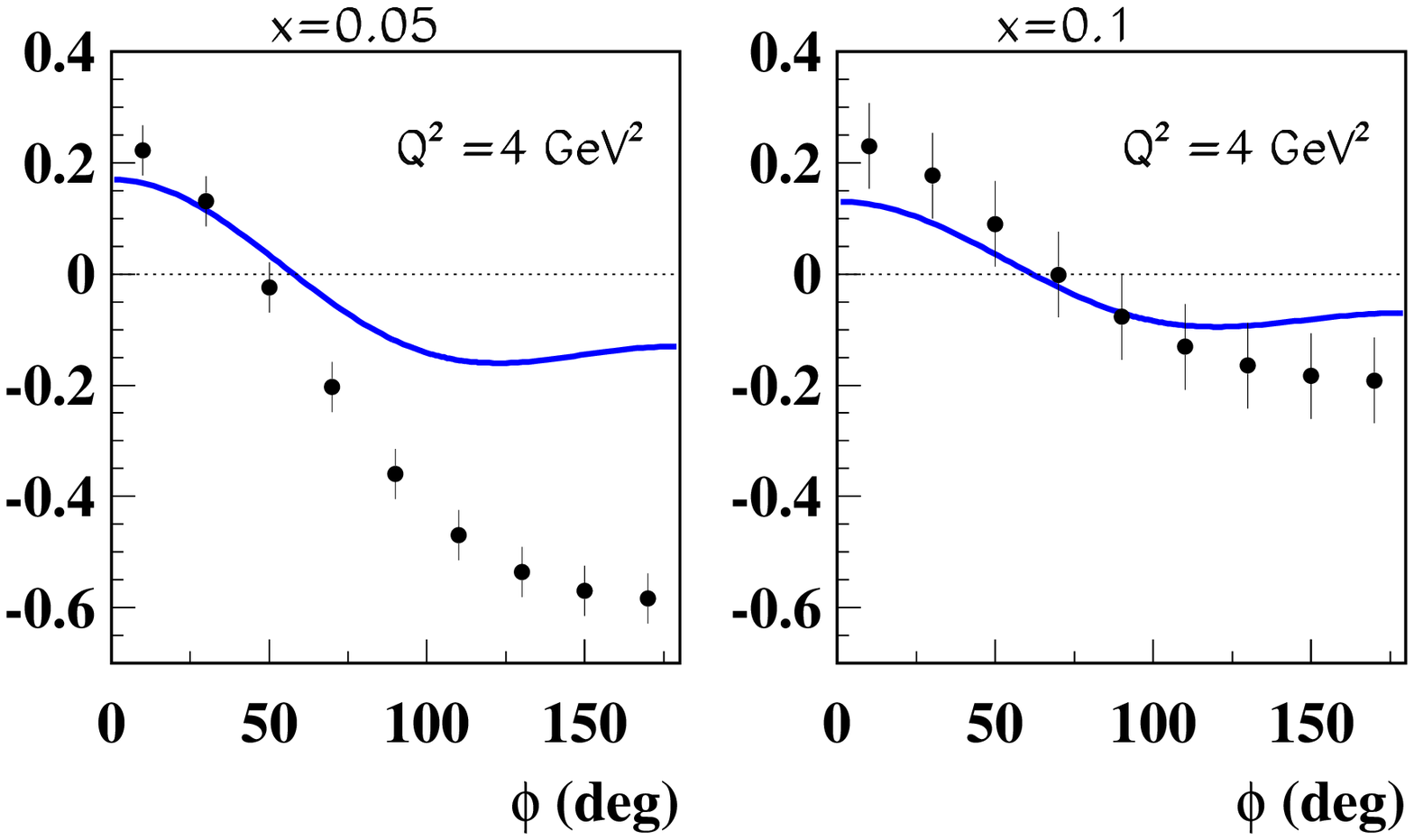}
\end{center}
\vspace*{-0.4cm}
\caption{\label{lolo}
Simulation of the azimuthal angular distribution of the beam charge
asymmetry measurable at COMPASS at $E_\mu=100$ GeV.
We present the projected values and error bars in the range
$|t|<$ 0.6 GeV$^2$
for 2 values of $x_{Bj}$ ($0.05$ and $0.1$) at $Q^2=4$ GeV$^2$ (see Ref. \cite{dhose}).
The prediction of the GPD model with a non-factorized $t$ dependence 
is shown (full line). The case of a factorized $t$ dependence would lead to
a prediction of the BCA compatible with zero and is not displayed.
 }
\end{figure}

At COMPASS, if we measure a beam charge asymmetry (BCA), the polarization
of the muon beam flips with the charge and so, the $\sin \phi$ 
terms disappears. Then, the BCA reads
\begin{equation}
A_C = \frac{d\sigma^+/d\phi -d\sigma^-/d\phi} 
{d\sigma^+/d\phi + d\sigma^-/d\phi}  \sim p_1 \cos \phi = 2 A_{BH} \frac{\mathrm{Re}  A_{DVCS} } 
{ |A_{DVCS}|^2+ |A_{BH}|^2 } \cos \phi.
\label{bcadef}
\end{equation} 

Note that DVCS cross section measurements which are integrated over $\phi$ are 
not sensitive  to the interference term (see Eq. \ref{toto}). 
Simulations done for COMPASS \cite{dhose} are shown in Fig.
\ref{compass2} for BCA  in a  setup described in the legend of the figure.
Two models of GPDs, with 
a factorized and non-factorized $t$ dependence, 
are shown in Fig. \ref{compass2} and we can observe easily the great discrimination
power offered by  COMPASS, with the proton recoil detector fully operational \cite{dhose}.
Of course, the discrimination is large in  Fig.  \ref{compass2} due to the fact that
$\alpha'$ is taken to be large ($\alpha' \sim 0.8$ GeV$^{-2}$) in 
 simulations.
If it happens to be much smaller, as measured at low $x_{Bj}$ by H1 \cite{dvcsh1}
(see previous section),
both predictions for BCA in Fig.  \ref{compass2} would be of similar shape,
as both curves would converge to the factorized case.

In Fig. \ref{lolo}, we compare predictions of the GPD model used in the
previous section for H1 data to simulations of 
the BCA extraction at COMPASS using a muon beam
of 100 GeV \cite{compasslolo}. 
We present the comparison for one value of $Q^2$ ($4$ GeV$^2$) and two values of
$x_{Bj}$ ($0.05$ and $0.1$).
When we compute the BCA in the factorized exponential $t$ dependence approximation,
we find  values compatible with zero, which are not represented in Fig. \ref{lolo}. 

Therefore,
we  display only the predictions of the model obtained in the non-factorized case
using the same $\alpha' \sim 0.8$  GeV$^{-2}$ value than in Ref. \cite{dhose}.
Both the $\cos(\phi)$ and $\cos(2\phi)$ terms contribute 
to a significant level to the BCA at COMPASS,
as illustrated in Fig. \ref{lolo}.
We notice that our predictions do not match with the COMPASS simulation done with the model
described in Ref. \cite{dhose}. 
This is another illustration of the large discriminative power of this observable
on GPDs parameterizations, even for identical $t$ dependence input.

\begin{figure}[htbp]
\centering
 \includegraphics[width=0.7\columnwidth]{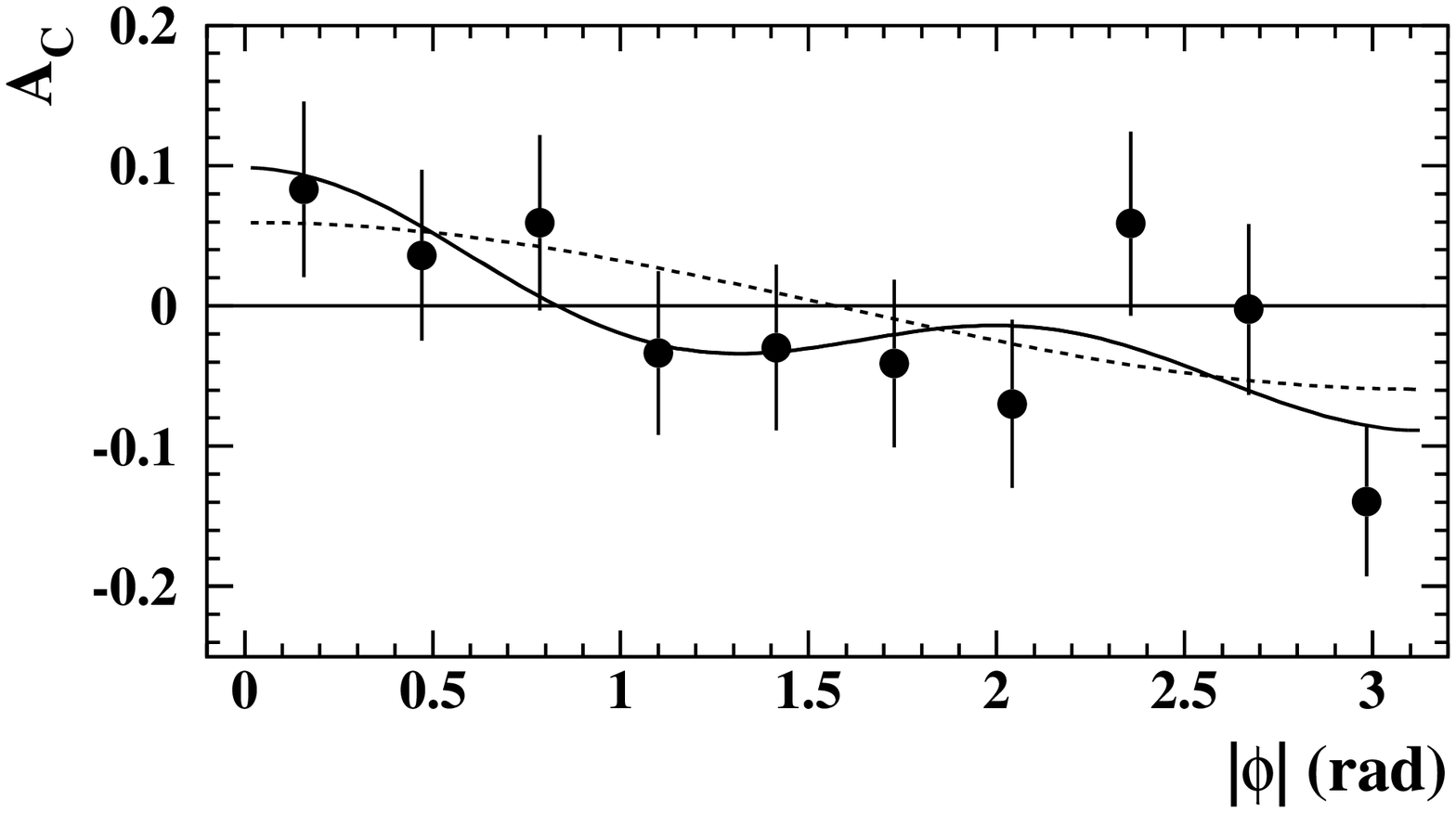}
  \caption{\label{fig:bcahermes0} 
  Beam charge asymmetry $A_C$  from HERMES data
  as a function of $|\phi|$.
  Statistical uncertainties are shown.
The solid curve represents the four--parameter fit:
$(-0.011\pm0.019) + (0.060\pm0.027) \cos \phi + (0.016\pm0.026) 
\cos 2 \phi + (0.034\pm0.027) \cos 3 \phi$. 
The dashed line shows the pure $\cos \phi$ dependence.}
\end{figure}

\begin{figure}[htbp]
  \begin{center}
\includegraphics[width=11cm]{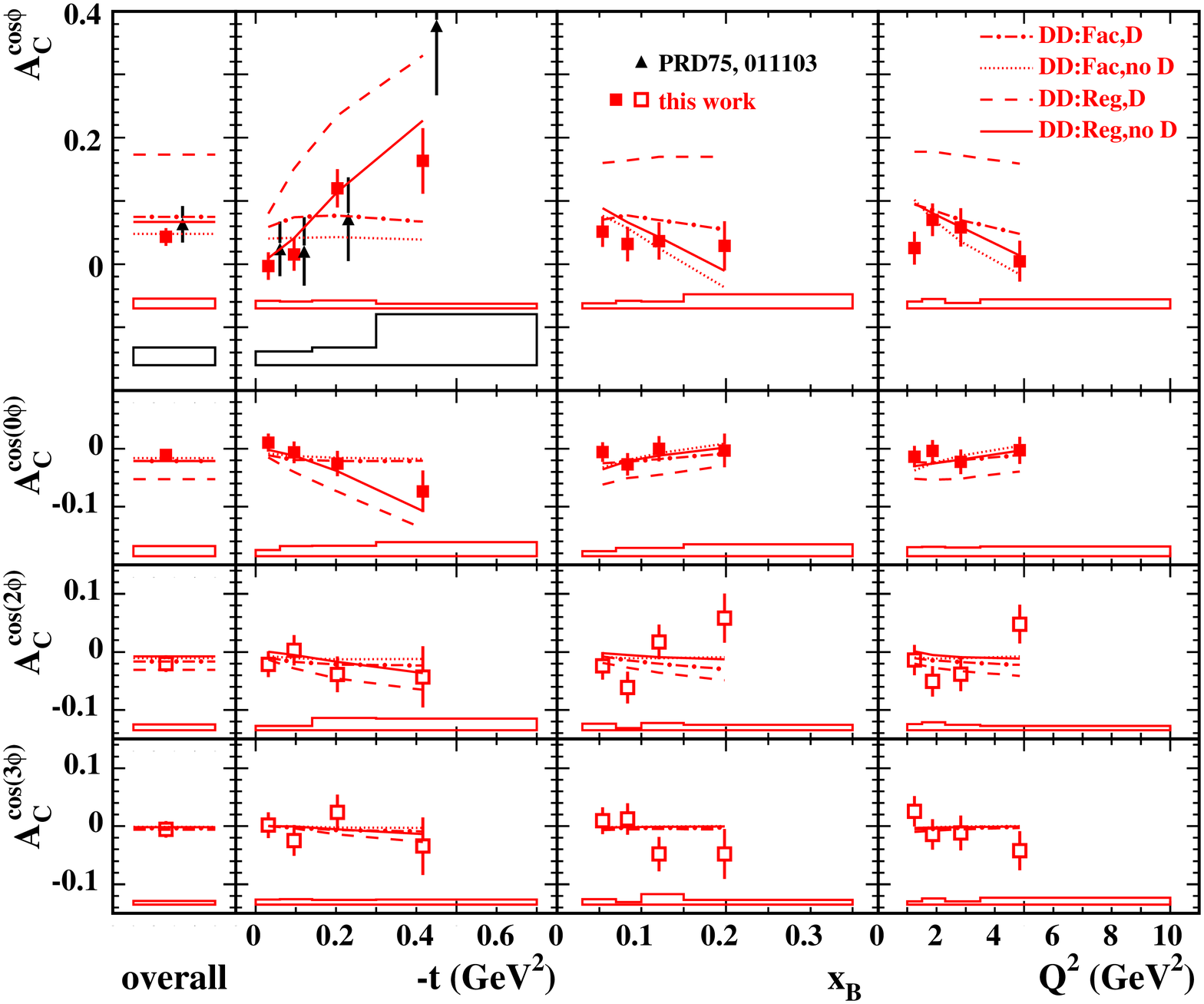}
\caption{ Moments of the Beam charge asymmetry from HERMES data.}
\label{fig:bcahermes}   
\end{center}
\end{figure}

\begin{figure}[htbp] 
  \begin{center}
    \includegraphics[width=10cm]{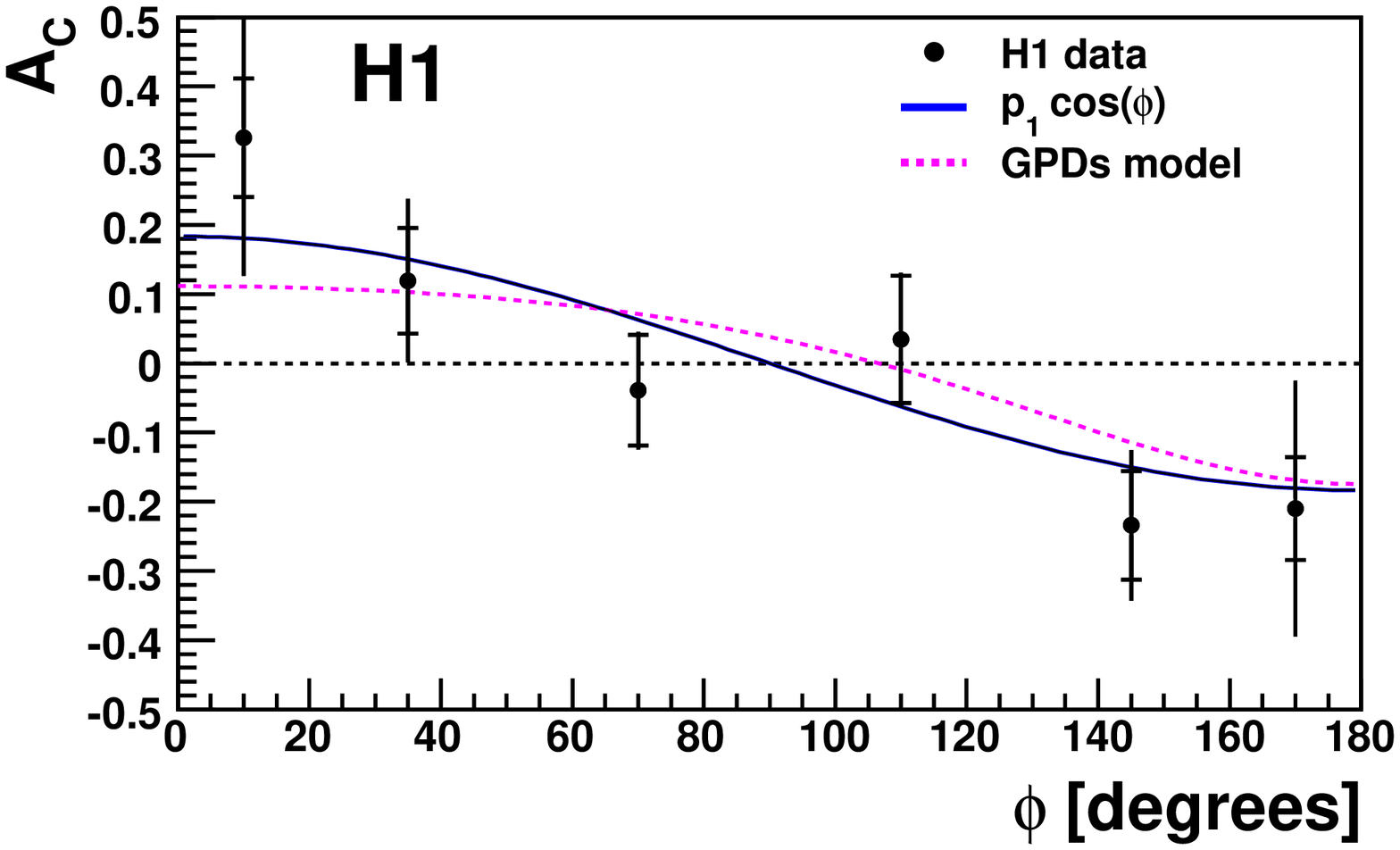}
  \end{center}
  \caption{
Beam charge asymmetry as a function of $\phi$ from H1 data
\cite{dvcsh1}.
Note that the $\phi$ convention chosen in H1 is different
from HERMES (see text).
The inner error bars represent the statistical
errors,
the outer error bars the statistical and systematic errors added in
quadrature.
A fit in $p_1\cos\phi$ is presented, where $p_1$ is a free parameter (full line),
together with the GPDs model prediction (dashed line).
}
\label{fig:bcah1} 
\end{figure}

\subsection{  Recent results on azimuthal asymmetries at HERA}

At HERA, we have also samples of data with electron and positron beams. Therefore,
it has been possible to extract the beam charge asymmetry,
$
A_C = \frac{d\sigma^+/d\phi -d\sigma^-/d\phi} 
{d\sigma^+/d\phi + d\sigma^-/d\phi}.  
$
A former pioneering measurement of the BCA at HERMES \cite{hermes}
is  shown in Fig. \ref{fig:bcahermes0}. 
HERMES was a fixed target experiment located at DESY operating with  the 
electrons or positrons beams of  27.6 GeV.
Recent results from HERMES \cite{hermes} and H1 \cite{dvcsh1} 
are presented in Fig. \ref{fig:bcahermes} and \ref{fig:bcah1}, which
correspond to $x_{Bj} \sim 0.1$ for HERMES and $x_{Bj} \sim 10^{-3}$ for H1.
Note that for H1 results, we have kept a different convention in the definition
of $\phi$ than in fixed target experiments, namely $\phi_{H1}=\pi-\phi_{HERMES}$.
The advantage of the convention we have considered in H1 is that,
a positive $p_1$ (with $BCA = p_1 \cos\phi$) means a positive real part of the
DVCS amplitude. 

Both experiments show that the present status of GPD models can
correctly described the BCA measurements. 
In case of H1, factorized parameterizations of GPDs 
(in $t$) are the most simple
choices compatible with measurements (see above), and for HERMES, the
sensitivity of the hypothesis 
of the $t$-dependence is illustrated in Fig. \ref{fig:bcahermes}.

If we consider the overall description
in Fig. \ref{fig:bcahermes},
the factorized ansatz (without D-term) is favored
by HERMES BCA measurements. 
The so-called D-term is part of some parameterizations of GPDs
 \cite{hermes}. That's why BCA, which provides a sensitivity
the  real part of the DVCS amplitude, gets a sensitivity to this 
(unknown) term.
In general also,
the factorized ansatz is much more stable with respect to the D-term
contribution, when compared to the non-factorized (Regge) ansatz.
Indeed, the spread between Regge with/without D-term predictions
is huge, whereas
the  D-term has only a small impact on the factorized
predictions. 
As the D-term is almost completely unknown, it is interesting to
make choice of parameterizations (if possible) that can reduce their
sensitivity to it.

In Ref. \cite{hermes}, it is mentioned that the 
Regge (without D-term) is favored, based on the observation of
the $t$ dependence. However, it is not that clear when considering all
data points. 

\begin{figure}[htbp] 
  \begin{center}
    \includegraphics[width=8cm]{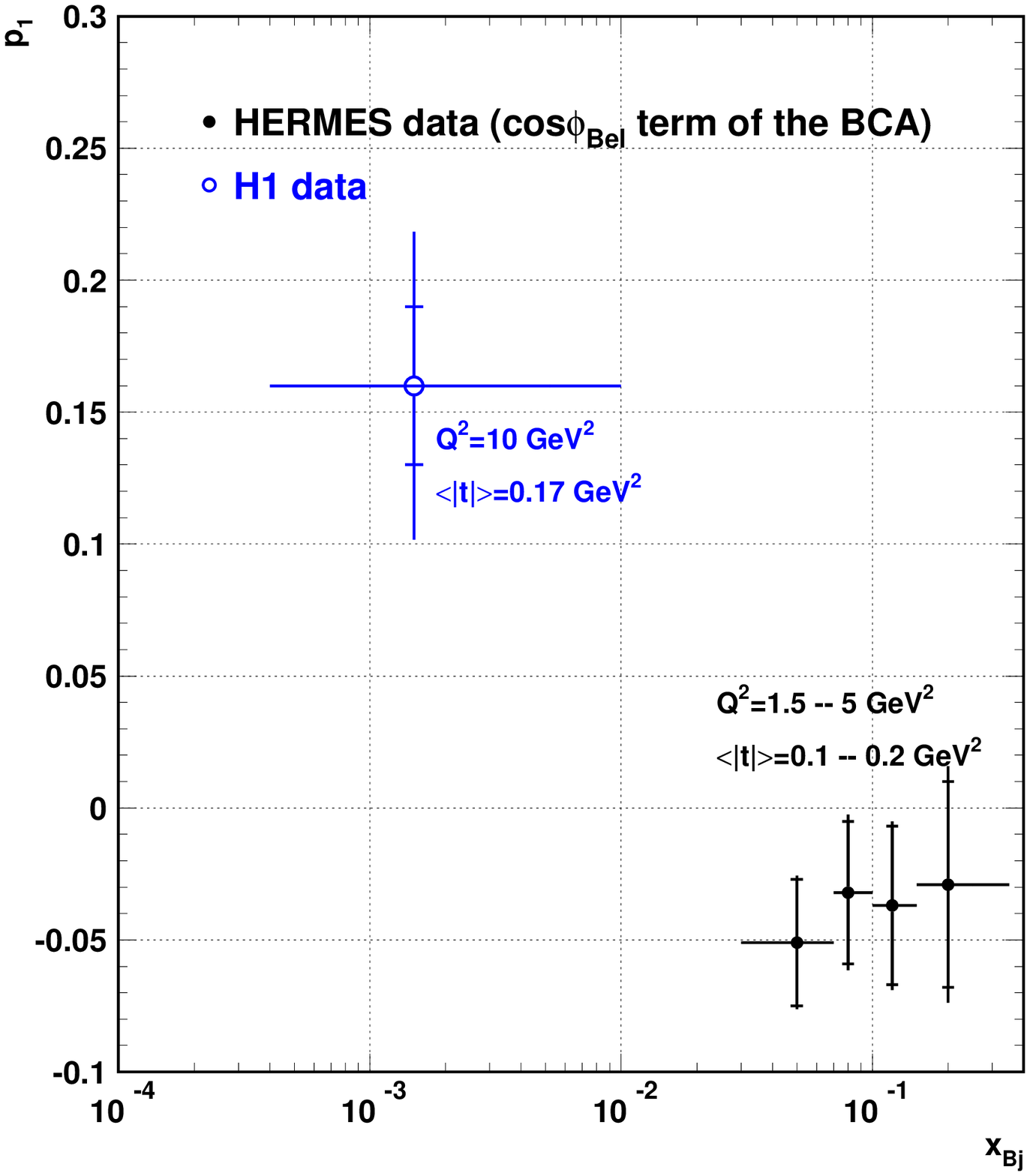}
  \end{center}
  \caption{
The beam charge asymmetry for H1 and HERMES is fitted by a the dominant harmonic
in $p_1\cos\phi$, where $p_1$ is a free parameter (see text).
Results for $p_1$ are presented for both experiments as a function of $x_{Bj}$.
This plot can be interpreted as a reflection of the different values of $\alpha'$
in both kinematic ranges, negligible value in the H1 domain and large value for HERMES.
We understand immediately the great interest of COMPASS kinematical domain,
lying at intermediate  $x_{Bj}$.
}
\label{fig:bcah1hermes}  
\end{figure}

In any case, the experimental results presented in Fig. 
 \ref{fig:bcahermes} and \ref{fig:bcah1} are the first obtained on 
 BCA and then important pieces to provide constraints on the real
 part of the amplitude in future developments of GPDs phenomenology. 
A compilation of H1 and HERMES results is presented in Fig. \ref{fig:bcah1hermes}.

\subsection{  Experimental analysis of dispersion relations} 

A specific analysis has been done in the H1 experiment concerning the real part of the
DVCS amplitude \cite{dvcsh1}. From Eq. (\ref{bcadef}) and  measurements
of  BCA and DVCS cross section, it is possible to extract the ratio
of the real to imaginary parts of the DVCS amplitude
$$
\rho = \mathrm{Re}  A_{DVCS} / \mathrm{Im}  A_{DVCS}. 
$$

This ratio is a key quantity which can also be derived through a 
dispersion relation, which takes a simple form in the high energy limit. 
Indeed, at low $x_{Bj}$, when the $W$ dependence of the DVCS cross section
is dominated by a single term in $W^\delta$ (with $\delta>0.3$), 
the dispersion relation can be written as
\begin{equation}
\mathrm{Re}  A_{DVCS} / \mathrm{Im}  A_{DVCS} 
= \tan \left( \frac{\pi \delta(Q^2)} {8} \right),
\label{dispersion}
\end{equation}
where $\delta(Q^2)$ is the power governing the $W$ dependence of the DVCS cross 
section at a given $Q^2$.
As we have measured  $\delta$ independently from DVCS cross sections only
\cite{dvcsh1} (see previous section), we can compute this ratio,
with the very reasonable assumption that the dispersion relation are correct.
We obtain: $\rho=0.25 \pm 0.06$.
To be compared to the value extracted from  BCA measurement and the
subsequent extraction of $p_1$, which gives $\rho=0.23 \pm 0.08$.
Both values are found in good agreement.

After this brief discussion, we can also understand simply 
how the sensitivity of the
beam charge asymmetry observable is built with $\alpha'$.
The BCA is proportional to the ratio of real to imaginary part
of the DVCS amplitude and this ratio can be expressed with
respect to $t$ as
$$
\rho[t] = \mathrm{Re}  A_{DVCS} / \mathrm{Im}  A_{DVCS}[t] \simeq
\mathrm{Re}  A_{DVCS} / \mathrm{Im}  A_{DVCS}[0](1-\alpha' |t|).
$$
Then, trivially, for small values of $\alpha'$ at low 
$|t|$ values, we do not expect much sensitivity
(on $\alpha'$)
of this ratio and thus of the BCA.
This is  what is illustrated for HERMES results 
in Fig. \ref{fig:bcah1hermes}.

\subsection{ Jefferson Laboratory experiments } 

\begin{figure}[htb]
\centering
\includegraphics[width=8.5cm]{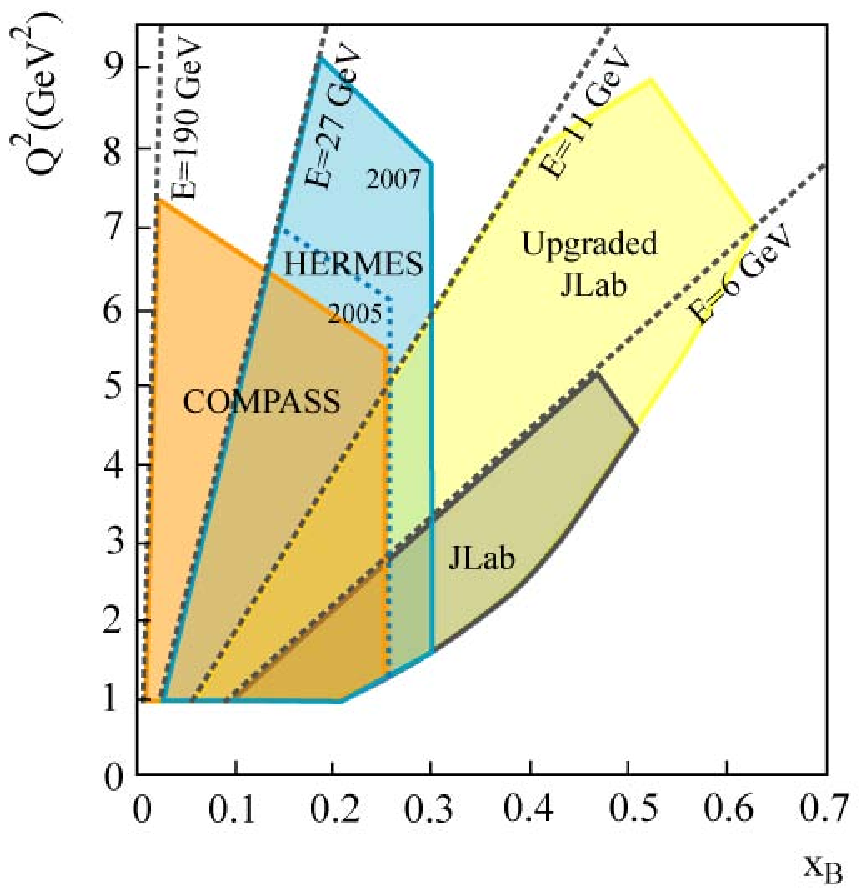}
\centering
\caption{ Kinematic coverage for fixed-target experiments:
(i) {\sc Compass} at 190 GeV; (ii) {\sc Hermes} at 27.6 GeV, dotted line for 
existing data ($\leq$ 2005), solid line for future (2005-2007) data 
with an integrated luminosity higher by about one order of magnitude;
(iii) {\sc JLab} experiments at 6 GeV, and at 11 GeV (after upgrade).}
\label{compass1}
\end{figure}

Regarding the kinematic coverage of fixed-target 
experiments (see Fig. \ref{compass1}),
the Jefferson Lab (JLab) experiments play a major role in the field,
exploring the large $x_{Bj}$
and low $Q^2$ kinematic domain.
JLab experiments, colliding an electron beam 
in the energy range of 6 GeV on a fixed target,  
can measure beam spin or target spin asymmetries \cite{jlaball}
and then access directly the imaginary part of the DVCS amplitude
in the valence domain.

Of course, we can not exclude a priori that higher twists effects would 
completely spoil any perturbative treatment of the experimental results
in this area. 
Below, we describe briefly few characteristic  measurements at JLab
 related to GPDs physics.

First, let us recall that in this kinematic domain, the BH cross section
is completely dominating the $ep \rightarrow ep \gamma$ cross section.
Then, the DVCS signal can hardly be observed and only the BH/DVCS interference
can be accessed through different observables with different
sensitivities to GPDs \cite{jlaball}.

\begin{figure}
\centering
\includegraphics[width=8.cm]{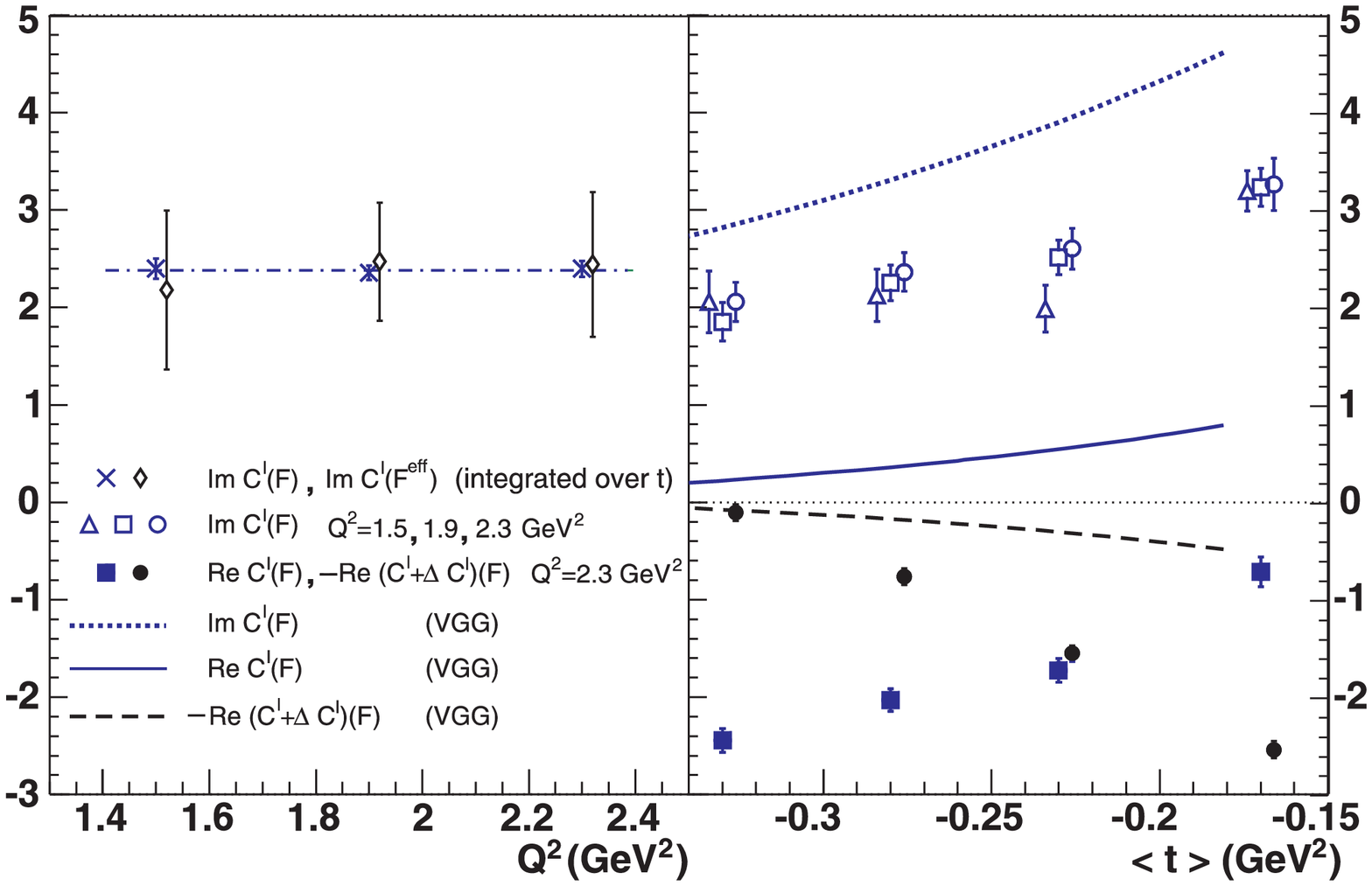}
\caption{
$Q^2$ dependence of imaginary part of the
DVCS amplitude (left). 
We observe a scaling over the range in $Q^2$ covered by the analysis.
}
\label{fig:coef}
\end{figure}

An important recent 
result has been obtained by the Hall A E-00-110 experiment \cite{jlaball},
which  demonstrates that
measurements at JLab are  dominated by leading twists 
contributions. It is shown in Fig. \ref{fig:coef}.
From the observed $Q^2$ scaling of the imaginary part of 
the DVCS amplitude (see Fig. \ref{fig:coef}) this result provides
an indication that the 
measurement of the imaginary part of the DVCS amplitude follows
a typical Bjorken scaling, observed over the $Q^2$ range
covered by the experiment. Which means that leading twists terms are 
likely to dominate.
Of course, the range in $Q^2$ accessible is not large but the
the high precision of the data makes this last
statement quite reasonable.
An upgrade at larger energies of the lepton beam is obviously an
important issue to get a sensitivity to higher $Q^2$ values (and
larger $W$). 

\begin{figure}[htbp]
\begin{center}
\epsfxsize=4.cm
\epsfysize=7.cm
\hspace*{-2cm}
\epsffile{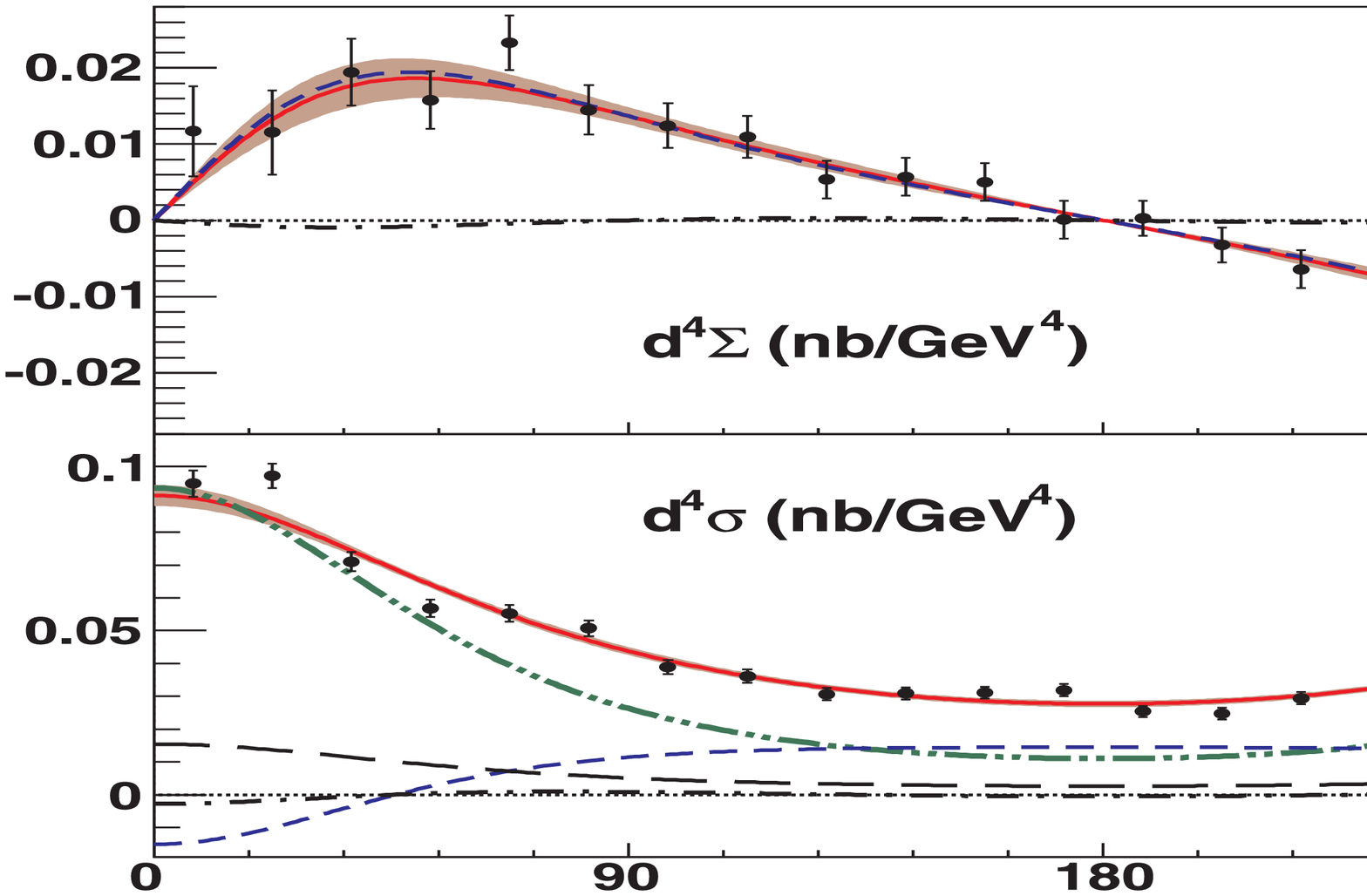}
\caption{Difference of (beam) polarized
cross sections for DVCS on the proton, as a function of the $\Phi$ angle, measured 
by the JLab Hall A collaboration. The average kinematics is
$<x_{Bj}>$=0.36, $<Q^2>$=2.3~GeV$^2$ and $<-t>$=0.28~GeV$^2$. The figure on 
the bottom shows the total (i.e unpolarized) cross section as a function of 
$\Phi$. The BH contribution is
represented by the dot-dot-dashed  curve. }
\label{fig:halla}
\end{center}
\end{figure}

Apart from DVCS/BH interference measurements,
a separation between BH and DVCS processes has been
obtained with the Hall A E-00-110 experiment.
The measurement of the 
4-fold (polarized and unpolarized) 
differential cross sections $\frac{d\sigma}{dx_BdQ^2dtd\phi}$
(for the real photon production process) has been done
 for three values of $Q^2$(in the kinematic domain
 $W\approx$ 2~GeV and $x>0.1$). Results are shown
in Fig.~\ref{fig:halla} for $<Q^2>$=2.3~GeV$^2$ \cite{jlaball}.
The particular shape in $\phi$ of the unpolarized cross section (Fig.~\ref{fig:halla}) 
is typical of the BH process. 
The dot-dot-dashed  curve in
 Fig.~\ref{fig:halla} shows its precise shape and contribution. It can be seen that it
dominates most of the cross sections and, only around $\Phi$= 180$^o$, there is
a large discrepancy (a factor $\approx$ 2) between the BH and the data which could
 be attributed to the DVCS process itself.
 It opens a possibility in a future analysis to extract directly
 a DVCS signal, which would be a first time measurement in this kinematic range.

\begin{figure}[htbp]
\centering
 \includegraphics[width=100mm]{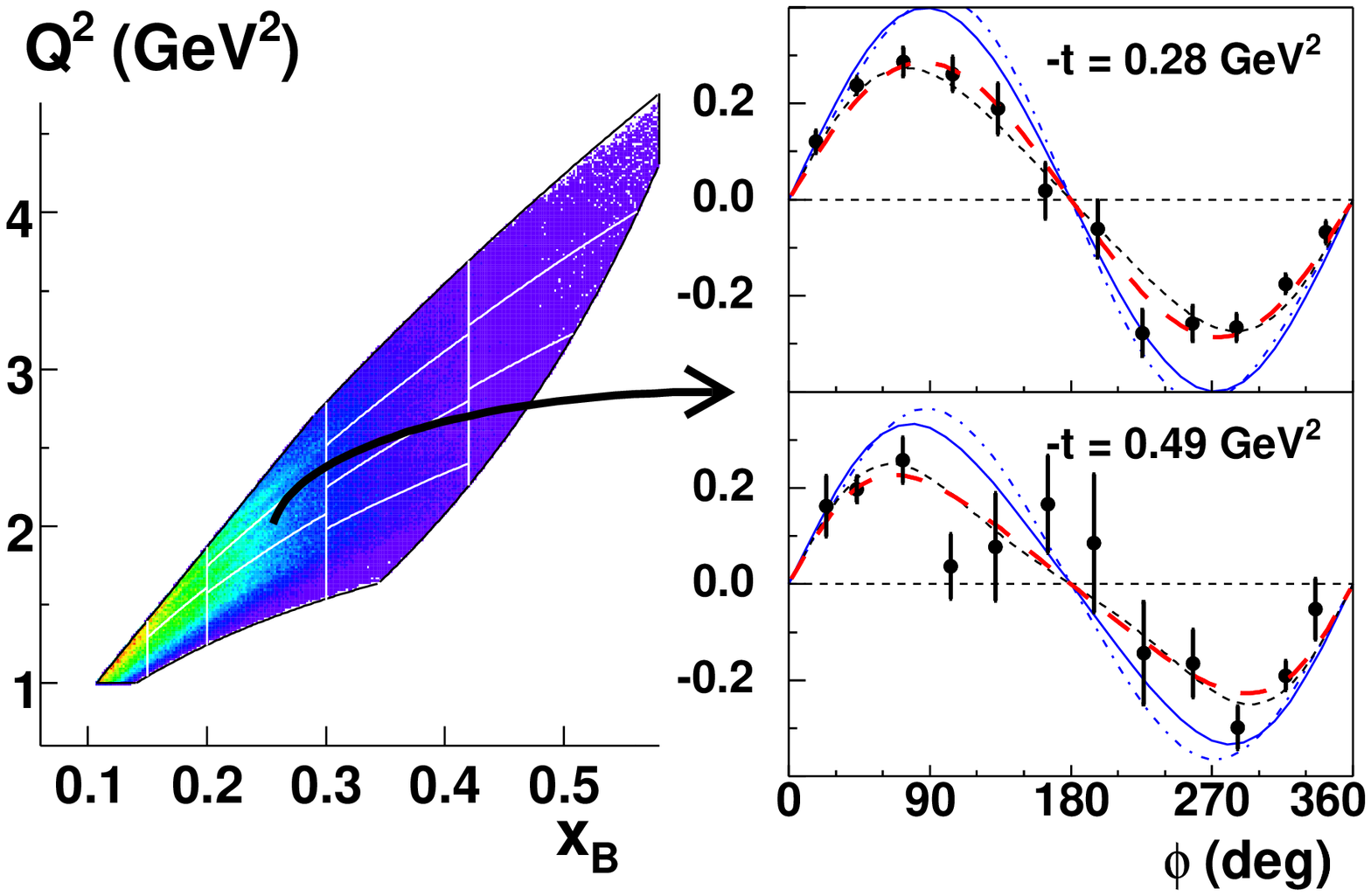}
 \caption{\label{fig:result1}
 Left: kinematic coverage and binning in the ($x_B$, $Q^2$) space. 
 Right: Beam spin asymmetry $A(\phi)$ for 2 of the 62 ($x_B$, $Q^2$, $t$) bins,
     corresponding to
     $\langle x_B\rangle =0.249$, 
     $\langle Q^2\rangle =1.95$ GeV$^2$, 
      and two values of $\langle t\rangle$.
      The  long-dashed curves correspond to fits with
      $A = \frac{a\sin\phi}
       		{1+c\cos\phi}$.
      The  dashed curves correspond
      to a Regge calculation.
      GPDs calculations are also shown as full lines.
		}
\end{figure}

\begin{figure}[htbp]
\begin{center}
\includegraphics[width=90mm]{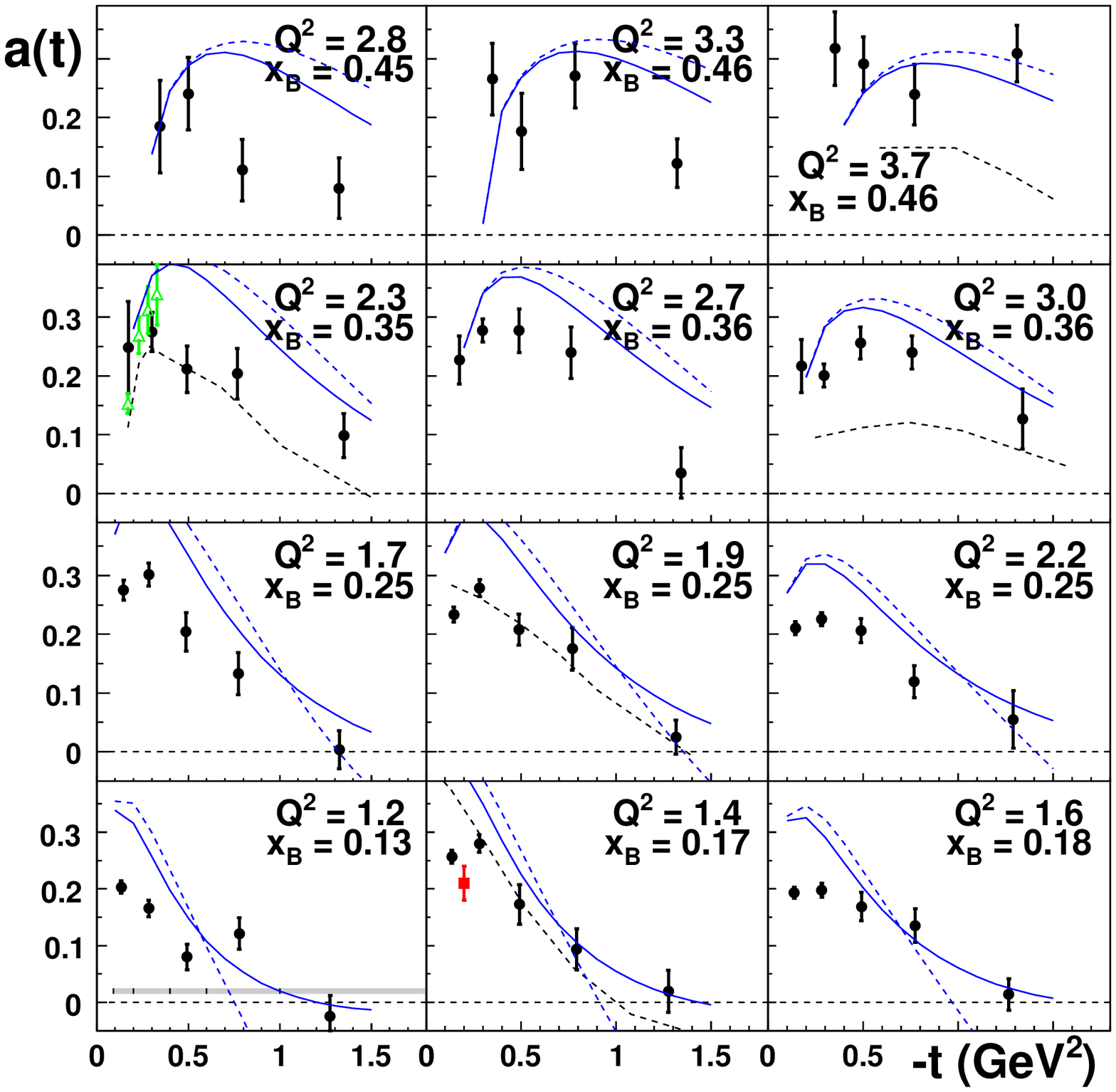}
 \caption{\label{fig:result2}          
	  Extracted values $a$ 
	  of the DVCS-BH interference as measured
	  in a spin asymmetry, with
$
	A = \frac{a\sin\phi}
       		{1+c\cos\phi+d\cos2\phi}.
$
	  They are shown as a function of $-t$
	  for different bins in $(x_B, Q^2)$.
	  The solid curves represent a GPD model and the dashed curves
	  a Regge approach.
 	 }
\end{center}	 
\end{figure}

Let us present a final measurement from the (JLab) Hall B E-00-113 experiment,
concerning beam spin asymmetries (BSAs) \cite{jlaball}, which shows (again) clearly 
the interest for an upgrade at larger energies.
Results are presented in
Fig.~\ref{fig:result1} with 
GPDs or Regge (non-perturbative) models. 
The asymmetries are fitted according to the relation
\begin{equation}
	A = \frac{a\sin\phi}
       		{1+c\cos\phi+d\cos2\phi}\ 
       	\label{eq:A_alpha}
\end{equation}
and extracted values of $a$ are displayed in Fig.~\ref{fig:result2}.
As can be seen in Fig. \ref{fig:result2}, the discrimination of Regge (soft) or
 GPDs (hard) approaches is not conclusive from the present data. Therefore, the
upgrade with 12 GeV electrons is very interesting to address this separation
between soft and hard physics at JLab.

\subsection{ Experimental prospects on the orbital angular momentum of partons }

A final comment is in order concerning the measurement 
of asymmetries (from DVCS/BH interference) in fixed target experiments.
Experiments at JLab and data collected by HERMES allow to 
determine transverse target-spin asymmetries, by controlling the
polarization of the target.
This would  be also a possibility of the future COMPASS project described above.
Experimentally, we need to introduce another 
azimuthal angle $\phi_S$ to characterize completely 
the events measured in such configurations, where $\phi_S$ 
represents the direction of the
spin of the target with respect to the plane of the leptons
(incident and scattered).

The great interest is then that the $\cos\phi$ moment of the asymmetry
$d\sigma(\phi,\phi_S)-d\sigma(\phi,\phi_S+\pi)
$ 
is proportional to the imaginary part of GPDs of types $H$ and $E$.
Remind the short note we have written in the last section:
the contribution of GPDs of type $E$
are damped by kinematic factors of orders $|t|/M_p^2$
in all the observables we have discussed till now.
This is not the case for  transverse target-spin asymmetries.

\begin{figure}[t]
\centering
\includegraphics[width=0.6\textwidth]{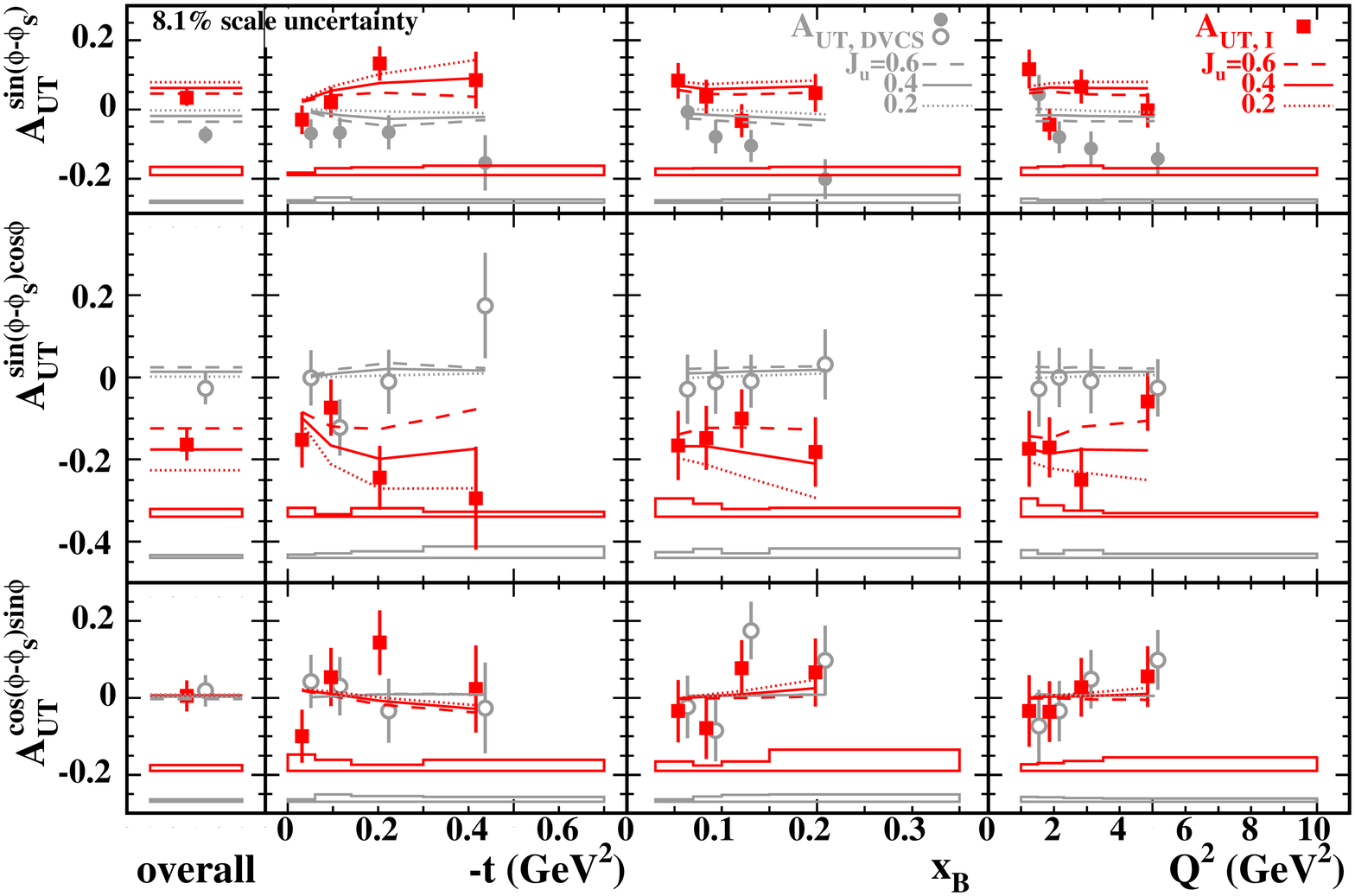}
\caption{\label{hermestsa}
Target-spin asymmetry amplitudes describing the 
dependence of the squared DVCS amplitude 
(circles, $\AUTDVCS$) and the interference term (squares, $\AUTI$) on the transverse target polarisation. In the notations, $U$ refers to Unpolarized
beam and $T$ to Transversely polarized target.
The circles (squares) are shifted right (left) for visibility. 
The curves are predictions  of a
GPD model  with three different values for the $u$-quark total angular momentum $J_u$ 
and fixed $d$-quark total angular momentum $J_d=0$ (see \cite{hermes}). This is a first important (model dependent) check 
of the
sensitivity these data to the Ji relation.}
\end{figure}

Therefore, these measurements are particularly interesting in the quest
for GPDs. The strong interest in determining GPDs of type $E$
is that these functions appear in a 
fundamental relation between GPDs and angular momenta of partons.
Indeed,
GPDs have been shown to be related directly to the total 
angular momenta carried by partons in the nucleon, via the Ji relation~\cite{gpdsreview}
\begin{equation}
\frac{1}{2} \int_{-1}^1 dx x\left(H_q(x,\xi,t) + E_q(x,\xi,t)\right) = J_q .
\label{eq:JiRelation}
\end{equation} 
As GPDs of type $E$ are essentially unknown apart
from basic sum rules, any improvement of their knowledge is essential.
From Eq. (\ref{eq:JiRelation}), it is clear that we could access directly to the
orbital momentum of quarks if we had a good knowledge of GPDs $H$ and $E$.
Indeed,  $J_q$ is the sum of the longitudinal angular momenta of quarks
and their orbital angular momenta. The first one is relatively well known
through global fits of polarized structure functions.
It follows that a determination of $J_q$ can provide an estimate of
the orbital part of its expression. 
In Ji relation (Eq. (\ref{eq:JiRelation})), the function
$H$ is not a problem as we can take its limit at $\xi=0$,
where $H$ merges with the PDFs, which are well known.
But we need definitely to get a better understanding of $E$.

First measurements of transverse target-spin asymmetries have been 
realized at JLab \cite{jlaball} and HERMES \cite{hermes}.
We present results obtained by HERMES \cite{hermes} in Fig. \ref{hermestsa}.
The typical sensitivity to hypothesis on $J_q$ values 
is also illustrated in Fig. \ref{hermestsa},
with the reserve that in this analysis, the observed sensitivity to $J_q$
is  model dependent. It is already a first step, very challenging
from the experimental side. Certainly, global fits of GPDs (if possible) would give
a much more solid (less model dependent) sensitivity to $J_q$
(see next section).

\subsection{ A few comments on the Ji relation } 

In order to give more intuitive content to the Ji relation (\ref{eq:JiRelation}),
we can comment further its dependence in the function $E$.
From our short presentation of GPDs, we know that
functions of type $E$ are 
related to matrix elements of the form
$\langle p',s' | \mathcal{O} | p,s
\rangle$ for $s \ne s'$, which means helicity flip at the proton vertex ($s \ne s'$).
That's why their contribution vanish in standard DIS or in processes where
$t$ tends to zero. More generally, their contribution would vanish
if the proton had only configurations where
helicities of the partons add up to the helicity of the proton.
In practice, this is not the case due to angular momentum of partons.
This is what is reflected in a very condensed way in the Ji relation (Eq. (\ref{eq:JiRelation})).

Then, we get the intuitive interpretation of this formula: it connects
$E$ with the angular momentum of quarks in the proton.
A similar relation holds for gluons \cite{gpdsreview}, linking $J_g$
to $H_g$ and $E_g$ and
both formulae, for quarks and gluons, add up to build the proton spin  
$$
J_q+J_g = 1/2.
$$
This last equality must be put in perspective with the
asymptotic limits for $J_q$ and $J_g$ at large scale $Q^2$,
which read $J_q \rightarrow \frac{1}{2} \frac{3 n_f}{16+3n_f}$ and
$J_g \rightarrow \frac{1}{2} \frac{16}{16+3n_f}$, 
where $n_f$ is the number of active flavors of quarks
at that scale (typically $n_f=5$ at large scale $Q^2$) \cite{gpdsreview}.

In words, half of the angular momentum of the proton is carried by gluons
(asymptotically). It is not trivial  to make quantitative estimates
at medium scales, but it is a clear indication that orbital angular
momentum plays a major role in building the angular momentum of the proton.
It implies that all experimental physics issues
that intend to access directly or indirectly to GPDs of type $E$
are essential in the understanding of the proton structure,
beyond what is relatively well known concerning 
its longitudinal momentum structure in $x_{Bj}$.
And that's also why first transverse target-spin asymmetries
(which can provide the best sensitivity to $E$)
are so important and the fact that such measurements have already
been done is promising for the future.

Clearly, we understand at this level the major interest of GPDs
and we get a better intuition on their physics content.
They simultaneously probe
the transverse and the longitudinal distribution of quarks and gluons
in a hadron state and the possibility to flip helicity in GPDs
  makes these functions sensitive to orbital angular momentum
in an essential way.
This is possible because they generalize the
purely collinear kinematics describing the familiar twist-two
quantities of the parton model. This is obviously
illustrating a fundamental 
feature of non-forward exclusive processes.

\subsection{ Towards global fits in the GPDs context } 

A direct continuation of the analysis exposed in the previous section
is to perform a global fit of all previous experimental results.
In the same spirit as it is done for global QCD fits of 
proton structure function data, obtained in DIS scattering, a 
global fit can be done of observables measured for
exclusive processes, like exclusive real gamma production.
Instead of defining initial conditions on PDFs (at a low scale $Q_0$),
 initial conditions on GPDs must be assumed.
An important step in this direction is presented in Ref.\cite{muller}.

\begin{figure}[t]
\includegraphics[scale=0.48]{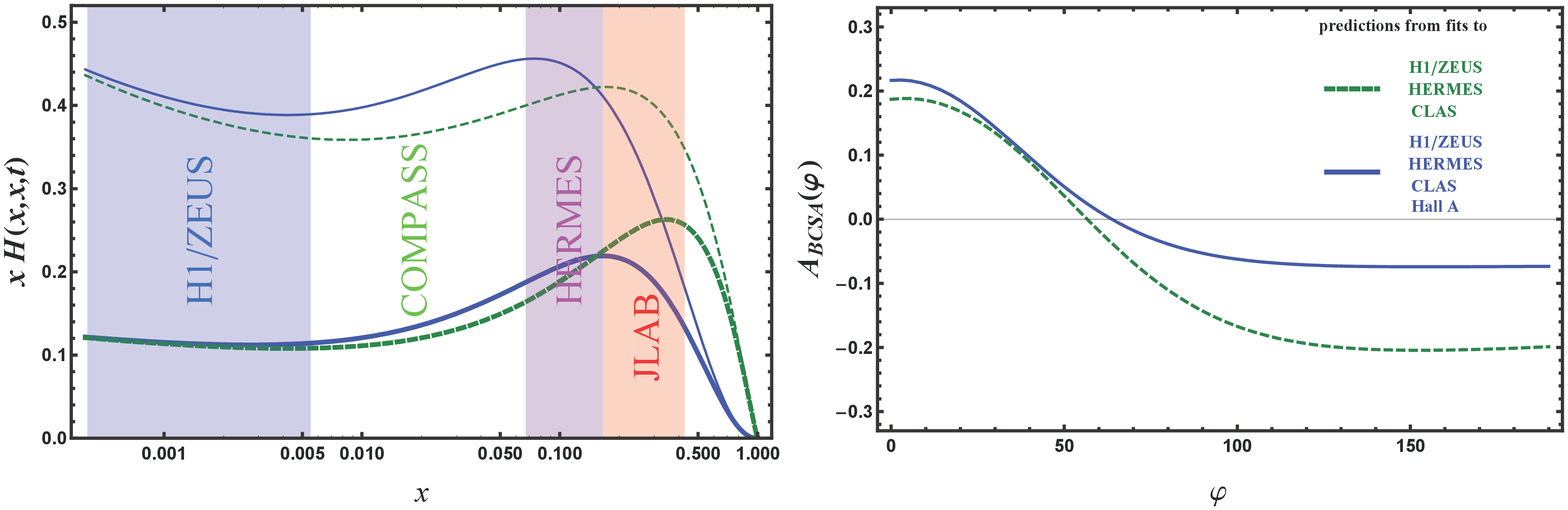}
\caption{\small
(a) Global GPD fits at  $t=-0.3$ GeV$^2$ (thick)
and $t=0$ (thin) are displayed as dashed and solid lines, respectively.
(b) Prediction of the BCA  for COMPASS kinematics
($E_{\mu} =160$ GeV,  ${\cal Q}^2=2$ GeV$^2$, $t= -0.2$ GeV$^2$) 
\cite{muller}.
}
\label{fig:Glo}
\end{figure}

A typical result derived in this work is displayed in Fig. \ref{fig:Glo}.
The GPD of type $H$ is shown 
for two values of $t$  (see Fig.\ref{fig:Glo}-left-)
and the influence
of JLab results is illustrated on the prediction of the BCA in
the COMPASS kinematics (see Fig.\ref{fig:Glo}-right-).

From these global GPDs fits \cite{muller}, the impact 
parameter space distribution can be extracted with
\begin{eqnarray}
\label{eq:TraProFun}
\rho(r_T,x,{\cal Q}^2) =
\frac{
\int_{-\infty}^{\infty}\! d^2\vec{\Delta}\;
e^{i \vec{\Delta} \vec{r_T}} H(x,\eta=0,t=-\vec{\Delta}^2,{\cal Q}^2)
}{
\int_{-\infty}^{\infty}\! d^2\vec{\Delta}\;
H(x,\eta=0,t=-\vec{\Delta}^2,{\cal Q}^2).
}
\end{eqnarray}
Results obtained in Ref.\cite{muller} confirms what we have
already discussed in previous sections \cite{diehlff}.
This theoretical framework to analyze GPDs is a promising
trend for the future, in parallel to the production of 
new experimental measurements.

\section{  Outlook }

We have reviewed the most recent experimental results from
hard diffractive scattering at HERA and Tevatron.
We have shown that
many aspects of diffraction in $ep$ collisions can be successfully 
described in QCD if a hard scale is
present. A key to this success are factorization theorems, which render parts of the dynamics 
accessible to calculation in perturbation theory. The remaining 
non-perturbative quantities, namely diffractive
PDFs and generalized parton distributions, can be extracted from 
measurements and contain specific
information about small-$x_{Bj}$ partons in the 
proton that can only be obtained in diffractive processes. To
describe hard diffractive hadron-hadron collisions is more 
challenging since factorization is broken by
re-scattering between spectator partons. 
These re-scattering effects are of interest in their own right 
because of their intimate relation with multiple scattering effects, which at 
LHC energies are expected to be
crucial for understanding the structure of events in hard collisions. 

A combination of data on inclusive
and diffractive $ep$ scattering hints at the onset of parton saturation at 
HERA, and the phenomenology
developed there is a helpful step towards understanding high-density 
effects in hadron-hadron collisions.
In this respect,
we have discussed  
a very important aspect that makes diffraction in DIS  so
interesting at low $x_{Bj}$. Its interpretation in the
dipole formalism and its connection
to saturation effects. Indeed, diffraction in DIS has appeared as
a well suited process to analyze saturation effects 
at large gluon density in the proton. In the dipole model,
it takes a simple and luminous form, with the introduction of
the so-called saturation scale $Q_s$. Diffraction is then dominated
by dipoles of size $r \sim 1/Q_s$. In particular, it provides a simple 
 explanation of the
constance of the ratio of diffractive to total cross sections
as a function of $W$ (at fixed $Q^2$ values).

Then, exclusive processes in DIS, like VMs production or DVCS, have appeared
as key reactions to trigger the generic mechanism of diffractive scattering.
Decisive measurements have been performed recently, in particular 
concerning  dependences of exclusive processes
cross section within the momentum exchange (squared) at the proton vertex, $t$.
This allows to extract first  experimental features concerning proton 
tomography, on how partons are localized in the proton.
It provides 
a completely new information on the spatial extension of partons
inside the proton (or more generally hadrons), as well as on the 
correlations of longitudinal momenta.
A unified picture of this physics is encoded in the GPDs formalism.
We have shown that
Jefferson laboratory experiments or prospects at COMPASS are essential,
to gain relevant information on GPDs.
Of course, we do not forget that the dependence of GPDs on three
kinematical variables, and the number of distributions describing
different helicity combinations present a considerable complexity.  In
a sense this is the price to pay for the amount of physics information
encoded in these quantities.  It is however crucial to realize that
for many important aspects we need not fully disentangle this
complexity.  The relation of longitudinal and transverse structure
of partons in a nucleon, or of nucleons in a nucleus, can be studied
quantitatively from the distribution in the two external kinematical
variables $x_{Bj}$ and $t$.  

 \begin{figure}[th]
      \centering
          \includegraphics[scale=0.4,clip=true,angle=0]{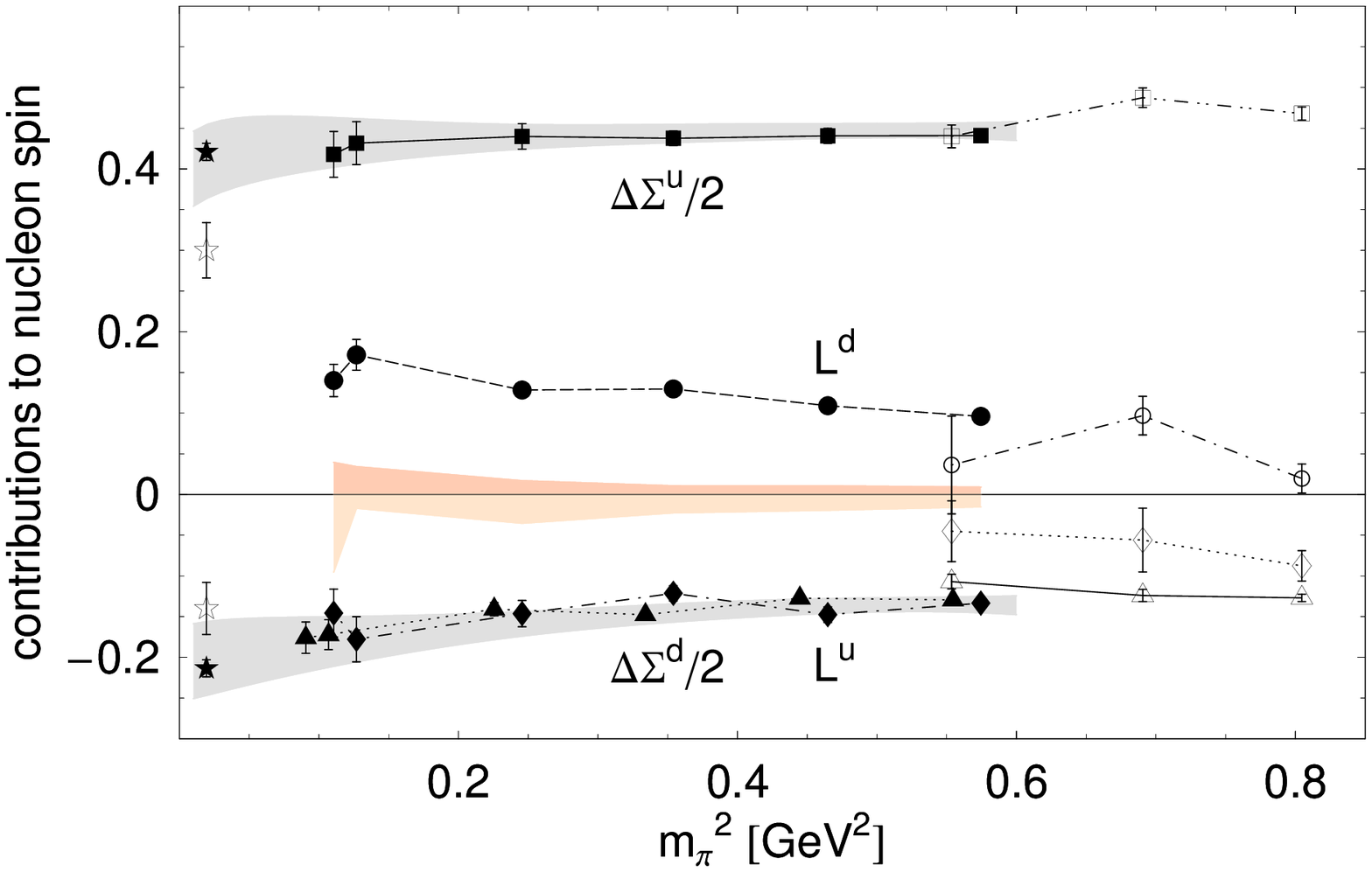}
       \caption{\label{thelast}
       Quark spin and orbital angular momentum contributions to 
       the spin of the nucleon for up and down quarks.
       Squares and triangles denote $\Delta \Sigma^u$ and $\Delta \Sigma^d$ respectively, 
       and diamonds and circles denote $L^u$ and $L^d$ respectively \cite{lattice}.}
    \end{figure}
To conclude, we can illustrate these issues with results
from lattice QCD \cite{lattice}.
In Ref. \cite{lattice}, lattice QCD calculations are performed. They show 
two remarkable features of the quark contributions to the nucleon spin. The first is that the
magnitude of the orbital angular momentum contributions of the up and down quarks, 
$L^u$ and $L^d$, are separately
quite substantial, and yet they cancel
nearly completely (see Fig. \ref{thelast}). The second is the close cancellation between the orbital and spin contributions
of the $d$ quarks, $L^d$ and $\Delta \Sigma^d/2$. 
Of course, we can not take these results as granted 
but calculations are solid. It would be obviously valuable to understand the physical origin of both
features, with more data.



\end{document}